\documentclass{aa}
\pdfoutput=1 
\usepackage{appendix}
\usepackage{latexsym}
\usepackage{graphicx}
\usepackage[normalem]{ulem}
\usepackage{lscape}
\usepackage{afterpage}
\usepackage{booktabs}
\usepackage{multirow}
\usepackage{arydshln}
\usepackage{ulem}

\usepackage{ulem}
\usepackage[citecolor =blue]{hyperref}
\usepackage{psfrag,color,ulem,gensymb}
\definecolor{deepblue}{rgb}{0,0,0.5}
\definecolor{deepred}{rgb}{0.6,0,0}
\definecolor{deepgreen}{rgb}{0,0.5,0}
\definecolor{darkgreen}{rgb}{0,0.6,0}
\definecolor{darkspringgreen}{rgb}{0.09, 0.45, 0.27}
\definecolor{debianred}{rgb}{0.84, 0.04, 0.33}
\definecolor{deepgreen}{rgb}{0,0.5,0}
\definecolor{amethyst}{rgb}{0.6, 0.4, 0.8}

\usepackage{amstext}
\hypersetup{draft}
\begin{document}

\title{Interactions between the large-scale radio structures    \\
and the gas in a sample of optically selected type 2 quasars}
\author{M. Villar Mart\'{i}n$^{1}$, B.\,H.\,C. Emonts$^{2}$, A. Cabrera Lavers$^{3,4}$, E. Bellocchi$^5$, A. Alonso Herrero$^5$ \\ A. Humphrey$^6$, B. Dall’Agnol de Oliveira$^{7}$, T. Storchi-Bergmann$^{7,8}$}
\institute{$^{1}$Centro de Astrobiolog\'{i}a, CSIC-INTA, Ctra. de Torrej\'{o}n a Ajalvir, km 4, 28850 Torrej\'{o}n de Ardoz, Madrid, Spain\\
$^{2}$National Radio Astronomy Observatory, 520 Edgemont Road, Charlottesville, VA 22903, USA \\
$^3$GRANTECAN, Cuesta de San Jos\'e s/n, E-38712 , Bre\~na Baja, La Palma, Spain \\
$^4$Instituto de Astrof\'\i sica de Canarias, V\'\i a L\'actea s/n, E-38200 La Laguna, Tenerife, Spain \\
$^5$Centro de Astrobiolog\'\i a, CSIC-INTA, ESAC Campus, E-28692, Villanueva de la Ca\~nada, Madrid, Spain \\
$^6$Instituto de Astrof\'{i}sica e Ci\^encias do Espa\c{c}o, Universidade do Porto, CAUP, Rua das Estrelas, PT4150-762 Porto, Portugal \\
$^7$Departamento de Astronomia, Universidade Federal do Rio Grande do Sul, IF, CP 15051, 91501-970 Porto Alegre, RS, Brazil \\
$^8$Harvard-Smithsonian Center for Astrophysics, 60 Garden St., Cambridge, MA 02138, USA \\
 \email{villarmm@cab.inta-csic.es}}        
\date{}
\abstract
{The role of   radio mode feedback in non radio-loud quasars  needs to  be explored in depth to determine its true importance. Its effects can be identified based on the evidence of interactions between the radio structures and the ambient ionised gas.}
{We  investigate this  in a sample of 13 optically selected  type-2 quasars (QSO2) at $z<$0.2 with Very Large Array (VLA) FIRST Survey  radio detections. None are radio loud. The ranges of  [OIII]$\lambda$5007 and monochromatic  radio  luminosities are   log($L_{\rm[OIII]}/\rm erg ~s^{-1}$)$\sim$42.08-42.79   and  log($P_{\rm 1.4~GHz}/\rm erg ~s^{-1}~Hz^{-1})\sim$30.08-31.76. All show  complex optical morphologies, with signs of   distortion  across tens  of kpc due to  mergers/interactions.}
 {We have searched for evidence of  interactions between the radio structures and the ionised gas by characterising  and comparing their morphologies. The first is traced by narrow band H$\alpha$ images obtained with the GTC 10.4m Spanish telescope and the Osiris instrument. The second is traced by VLA radio maps obtained with A and B configurations to achieve both high resolution  and brightness sensitivity.}
{The radio luminosity has an active galatic nucleus (AGN) component  in 11/13 QSO2, which is spatially extended in our radio data in   9 of them (jets/lobes/other). The relative contribution of the extended radio emission to the total $P_{\rm 1.4~GHz}$ is in most cases in the range 30\% to 90\%. The maximum  sizes are  in the range $d_{\rm max}^{\rm R}\sim$few-500 kpc.  

 QSO2 undergoing interaction/merger events appear to be invariably associated with ionised gas spread over large spatial scales with maximum distances from the AGN in the range $r_{\rm max}\sim$12-90  kpc.   The  morphology of the ionised gas at $<$30 kpc   is strongly influenced by  AGN related processes.

 Evidence for radio-gas interactions exist in 10/13  QSO2;  that is, all but one  with confirmed AGN radio components.  The interactions are identified across different spatial scales, from the nuclear narrow line region   up to tens  of kpc.}
{Although this sample cannot be considered representative of the general population of QSO2, it supports the idea  that  large scale low/modest power radio sources can exist in radio-quiet QSO2, which  can provide a source of feedback  on scales  of the spheroidal component of  galaxies and  well into the circumgalactic medium, in  systems where radiative mode feedback is expected to dominate.}


\titlerunning{Radio mode feedback in SDSS QSO2}
\authorrunning{Villar-Mart\'\i n et al.}

\maketitle

\section{Introduction}
\label{intro}

Type 2 quasars (QSO2) are very interesting objects for investigating  feedback in  the general population of quasars (QSO). The accretion disk and the broad line region (BLR) are occulted  by obscuring material,  allowing a detailed study  of the surrounding medium. This is more complex in the unobscured counterparts, type 1 QSO or QSO1, due to the dominant contribution of the nuclear point spread function. 

During  the last ten years it has become clear that ionised outflows are  ubiquitous  in QSO2 at different redshift ($z$) 
 (e.g. Villar-Mart\'\i n et al. \citeyear{Villar2011,Villar2014,Villar2016}; \citealt{Greene2011,Mullaney2013,Zakamska2014,Bellocchi2019}).    They  appear to be triggered in general by  processes related to the nuclear activity (e.g. \citealt{Greene2011,Mullaney2013,Villar2014,Zakamska2014,Jarvis2019}) but the dominant specific mechanism is uncertain. 
 The quasar mode, in which the intense flux of photons and particles produced by the  transfer of energy and momentum from the active galactic nucleus (AGN) to the surrounding environment (\citealt{Fabian2012,King2015}), is often assumed to be  dominant   because of their high AGN luminosities and  accretion rates and the fact that only  $\sim$10-15\% QSO2 are  radio loud (RL) (\citealt{Lal2010,Zakamska2014}). 
For this same reason, the role of radio mode feedback, where the bulk of the energy is ejected in kinetic form through  jets  coupled to the galaxies' gaseous environment, has been often considered irrelevant.

The nature of the radio emission in radio quiet quasars (RQQ) is still a matter of debate, whether it is dominated by star formation  in the host galaxy  (\citealt{Padovani2011,Bonzini2013,Condon2013,Kellermann2016}) or by non thermal emission due to mechanisms driven by the nuclear activity (e.g. \citealt{Maini2016,Herrera2016,Zakamska2004,Zakamska2016}).   The relative contribution of AGN related processes   appears to increase with radio luminosity (\citealt{Kimball2011,Kellermann2016}).

\cite{Zakamska2014} studied the radio luminosity of 
568   objects out of the entire sample of   QSO2 at $z\la$0.83 in the Sloan Digital Sky Survey (SDSS, \citealt{York2000}) selected by \cite{Reyes2008}. The median redshift is $z_{\rm med}$=0.397. The authors conclude that the origin of the radio emission for those with FIRST (Faint Images of the Radio Sky at Twenty-Centimeters)  detections ($\sim$65\%) is unlikely due to star formation, but to AGN related processes instead (see also \citealt{Lal2010}).  

The authors inferred a median monochromatic 1.4 GHz luminosity    log($P_{\rm 1.4}$)=log($P_{\rm 1.4~GHz}/\rm erg ~s^{-1}~Hz^{-1}$)=30.85\footnote{This is based on the assumption that the non detections are close to the FIRST survey detection limit}. When  compared  with the general population of RL AGN, this value is   surprisingly high for an object class often referred to as predominantly radio-quiet (see also \citealt{Kellermann2016}). \cite{Best2005} identified $\sim$2100 RL AGN  in the second data release of the SDSS ($z_{\rm med}\sim$0.1).  At least 30$\%$ have log($P_{\rm 1.4}$) below the QSO2 median monochromatic luminosity. Radio AGN  with  log($P_{\rm 1.4}$)$\la$31.5 (the range spanned by most QSO2) currently attract great interest concerning the role of mechanical feedback in galaxies since, as shown by \cite{Best2005},  they greatly outnumber their high luminosity counterparts.

Therefore,  independently of the  classification in terms of radio-loudness, and the fact that  the radio emission of optically selected QSO2 is low or modest relative to the optical output, many  host AGN driven radio sources  of significant power in comparison with the general population of radio AGN. Whether these  interact with the ambient gas and provide a mechanism of efficient feedback is an open issue and, thus, the role of   radio mode feedback in QSO2 (non RL quasars in general) needs to  be explored in depth to determine its true importance.

The frequency of jets and related structures (hot spots, lobes) in QSO2 and, more generally, non radio loud quasars (RQ and RI)  is uncertain.  Large-area radio surveys typically lack
the sensitivity to detect and/or resolve the radio structures and for this reason  the precise mechanism that produces the AGN radio emission   is often difficult to discern   (e.g. jets/lobes/hot spots vs. relativistic  particles  accelerated in the shocks produced by  quasar-driven outflows  (\citealt{Zakamska2014,Panessa2019}). \cite{Jarvis2019}  have recently suggested   that  radio jets/lobes with modest radio luminosities and  sizes of up to 25 kpc may be common in QSO2 and may provide a crucial feedback mechanism for massive galaxies during a quasar phase. 
Moreover, even in systems where the radio emission appears to be dominated by star formation,  this does not exclude the existence of jets. 
  
Radio mode feedback  has been observed in  Seyferts and RQQ  for decades (e. g. \citealt{Wilson1983,Whittle1992,Leipski2006,Mullaney2013,Husemann2013,Tadhunter2014,Alatalo2015,Aalto2020}). 
 Radio jets have now been identified even in galaxies which host weakly active or silent super-massive black holes (SMBH), including the Milky Way (\citealt{Baldi2018,Issaoun2019}).   SMBH driven jets may thus exist in many galaxies, even with such low levels of nuclear activity that they are considered inactive. 

Different works suggest that    the most kinematically extreme  nuclear ionised outflows  in AGN with low/modest radio luminosities (log($P_{\rm 1.4}$)$\la$31.0),  including QSO2, are triggered by compact ($\la$few kpc) radio jets  (\citealt{Mullaney2013,Villar2014,Molyneux2019}). Studies of large scale ($>$several kpc from the AGN, well beyond the nuclear region) radio induced feedback   in non-radio loud QSO, including QSO2, are very scarce (\citealt{Husemann2013,Villar2017,Jarvis2019}).  Recently, we discovered radio-induced feedback across large scales in the Beetle galaxy, which is a radio-quiet QSO2 at z=0.1 \citep{Villar2017}. Even though the radio emission in the Beetle was classified as compact based on archival survey data, deep VLA imaging revealed an extended radio structure with hot-spots stretching almost 50 kpc. The extended radio emission was detected only at the mJy level, and thus too faint to be detected in surveys. Still, this faint radio source interacts with the circumgalactic gas far outside the galaxy. The Beetle galaxy revealed that radio jets of modest power can be a relevant feedback mechanism acting across large scales, even in non-radio loud  QSOs.

The purpose of our current work is to investigate whether extended, low-power radio sources, analogue to what we previously observed in the Beetle galaxy, can be hiding in radio-quiet QSO2, and provide a source of feedback across large scales. We investigate this in a sample of optically selected QSO2 at z <0.2, based on the morphological characterisation of the ionised gas and the radio structures and the relative comparison.
The first is traced by narrow band H$\alpha$ images obtained with the GTC 10.4m Spanish telescope. The second is traced by radio maps obtained with the Karl G. Jansky Very Large Array (VLA).

The paper is organised as follows: we describe the sample  and observations in Sect. \ref{sample} and \ref{observations} respectively.  General results are presented in Sec. \ref{results-global} and discussed in Sect. \ref{discussion}.  Summary and conclusions are in Sect. \ref{conclusions}. Results on individual objects  are explained in  the Appendix \ref{results-objects2}.

We assume the cosmological parameters H$_{0}$\,=\,71, $\Omega_{\rm M}$\,=\,0.27, and $\Omega_{\Lambda}$\,=\,0.73 and calculate distances following \cite{Wright2006}.

\section{Sample selection}
\label{sample}

The VLA sample consists of 12   objects  from the SDSS catalogue of QSO2  at $z<$0.2 by \cite{Reyes2008} (Table \ref{table-sample}).

 Eight of the above objects were  observed with GTC to map the H$\alpha$ morphology.  Hubble Space Telescope (HST) or SDSS images were used for objects that could not be observed with GTC.  Although not in our VLA program, we  also included the Teacup in this  sample  to map the morphology of the giant ($\sim$110 kpc) nebula discovered by \cite{Villar2018}. The extended radio emission of this QSO2 has been studied in detail by \cite{Harrison2015} and \cite{Jarvis2019}.

\subsection{Optical properties}

  Our main interest is to investigate radio induced feedback
across galactic and extragalactic scales. For this mechanism to operate and be detectable, the radio source must interact with widely spread gas. For this reason, the $z<$0.2 QSO2  were selected to  show  complex optical morphologies, with signs of  morphological distortion  across tens  of kpc due to  mergers/interactions.  This suggests that the objects may be associated with a large scale gaseous environment.  The identification of the optical distortion was based on SDSS and/or HST available images.

No constraints were applied regarding prior evidence and/or strength of nuclear ionised outflows.

 The range of  [OIII] luminosities is log($L_{\rm[OIII]}$)$\sim$42.08-42.79 in erg s$^{-1}$  (the log of the median is 42.42), which is the same spanned by QSO2 at similar $z$  (Fig. \ref{xu}).

\subsection{Radio properties}
\label{radioprop}

\subsubsection{Radio-loudness}

The objects were selected to  have FIRST  and NRAO VLA Sky Survey  (NVSS) 1.4 GHz   detections and, in some cases, to show solid or tentative evidence for extended AGN driven radio sources (see below). The range of monochromatic  radio  luminosities is    log($P_{\rm 1.4})\sim$30.08-31.76 (Table \ref{table-sample}). The log of the median is 31.15.

\begin{figure}
\centering
\includegraphics[width=0.5\textwidth]{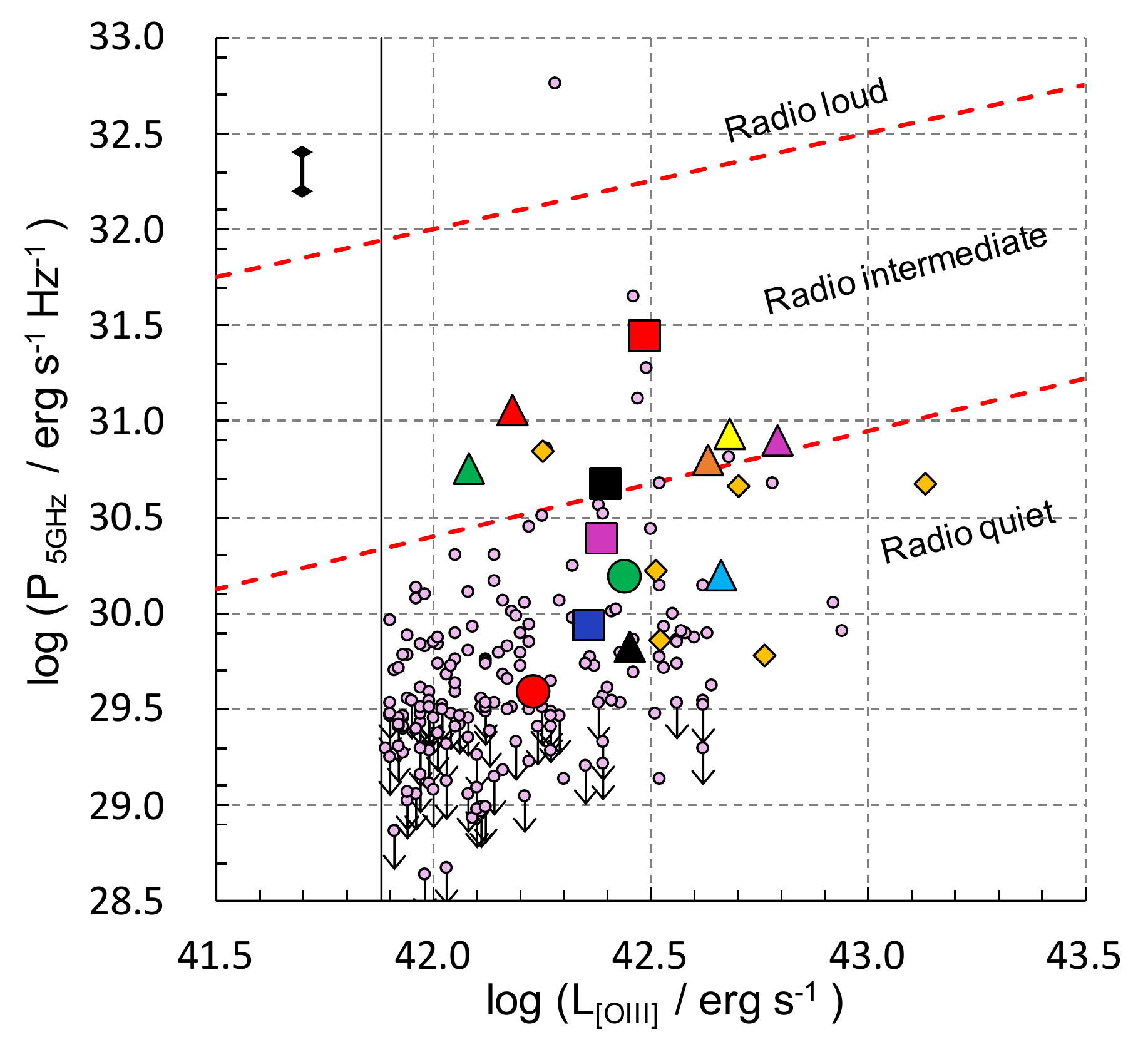}
\includegraphics[width=0.5\textwidth]{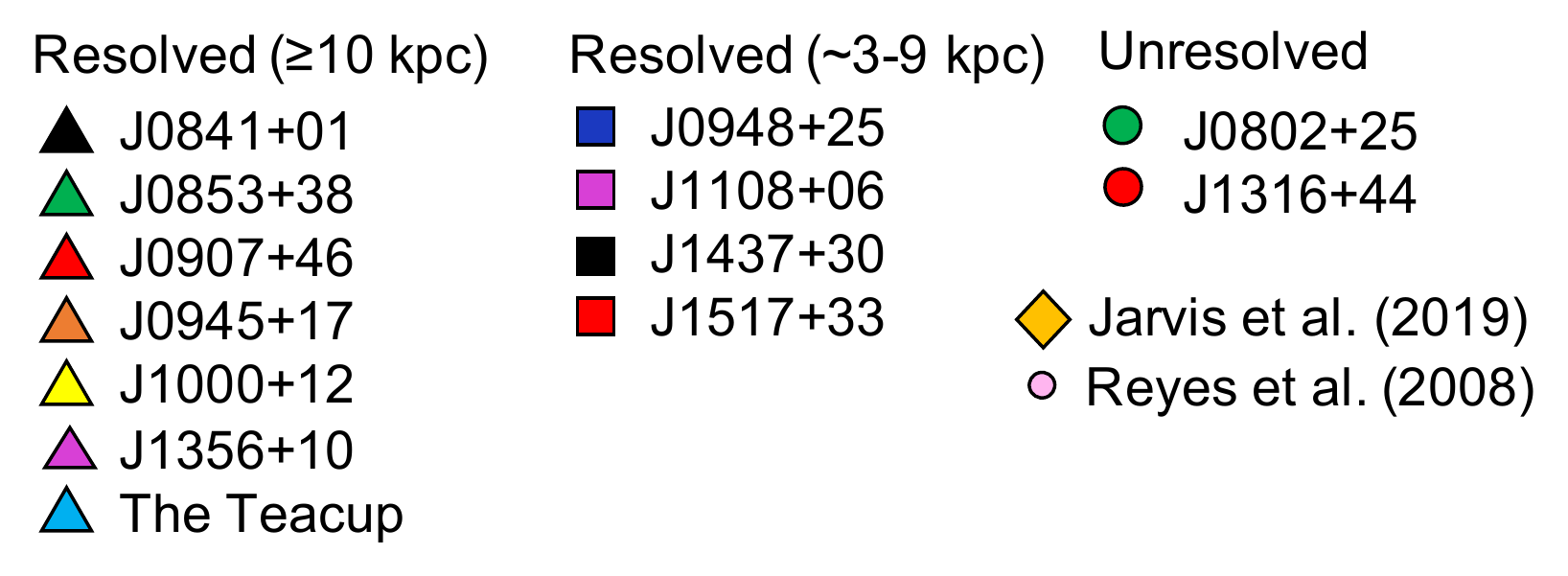}
\caption{Location of the object sample (large symbols) in the  [OIII]$\lambda$5007 vs. monochromatic 5.0 GHz  luminosity plane (\citealt{Xu1999}). Triangles are used for radio sources with total sizes $d_{\rm max}^R\ge$10 kpc (see Table \ref{radprop}), squares for   3$\le d_{\rm max}^R\le$9 kpc and circles for spatially unresolved or marginally resolved objects. The  vertical small bar shows the expected uncertainty of the radio luminosities due to the unknown spectral index $\alpha$. \cite{Jarvis2019} objects not included in our sample  are also shown for comparison.  The small pink circles are  QSO2 at $z<$0.2 in \cite{Reyes2008} catalogue.  The vertical solid line marks the lower [OIII] luminosity these authors assumed  to select QSO2. Arrows  indicate upper limits for objects with no FIRST detections.  The areas above  and below  the red dashed lines are filled with AGN classified as RL  and RQ respectively (\citealt{Xu1999}).  The gap is populated much more sparsely by  radio intermediate (RI) AGN. Our sample includes several RI sources.}
\label{xu}
\end{figure}

 To classify them according to the radio-loudness, we show in Fig. \ref{xu} their location    in the log($P_{\rm 5GHz}$) (rest frame value) vs. log($L_{\rm[OIII]}$) plane following \cite{Xu1999}. $P_{\rm 5GHz}$ was calculated using the 5 GHz  fluxes,  $S_{\rm 5GHz}$, when available (\citealt{Rosario2010,Jarvis2019,Bondi2016}). When not,  we computed  $S_{\rm 5GHz}$ with the NVSS 1.4 GHz fluxes  and assumed a spectral index $\alpha$=-0.864$\pm$0.222  (with $S_{\nu}\propto \nu^{\alpha}$). This is the median  inferred for the range of $\alpha$ values measured by \cite{Jarvis2019} for their QSO2 sample. The K-correction is negligible.
We have also plotted in  Fig. \ref{xu} all QSO2 at $z<$0.2 \cite{Reyes2008} catalogue. For objects with no NVSS or FIRST  detections,  a 5$\sigma$  upper limit from FIRST of 1 mJy has been used. 

The classification is shown in (Table \ref{table-sample}).   Seven  have high radio powers that place them  in  or near the RI region.  Most importantly, most QSO2 in our sample are at the high end of the radio luminosities spanned by QSO2 at similar $z$.

A more quantitative approach to the ``radio-loudness'' can be taken by looking at the $q$ parameter, which is a measure of the FIR/radio flux-density ratio (\citealt{Helou1985}): 

\begin{equation} 
q=log \left(\frac{S_{\rm FIR}}{3.75\times 10^{12}\rm ~ W ~m^{-2}}\right ) - log(\frac{S_{\rm 1.4~GHz}}{\rm W ~m^{-2} ~Hz^{-1} })
\label{ec:qeq}
\end{equation}

 where $S_{\rm 1.4~GHz}$ is in units of W m$^{-2}$ Hz$^{-1}$ and   $S_{\rm FIR}$=1.26$\times$10$^{-14}$ (2.58 $S_{\rm 60~\mu m}$ + $S_{\rm 100~\mu m}$) W m$^{-2}$. $S_{\rm 60~\mu m}$ and  $S_{\rm 100~\mu m}$)  are the IRAS fluxes at 60 and 100 $\mu$m in W m$^{-2}$. The $q$ values are shown in Table \ref{table-sample} when available.  Objects with $q\le$1.8 show an excess of radio emission above that expected from star formation and,  therefore, they have a significant AGN contribution  (\citealt{Villar2014}).
This is the case for  seven of the nine objects with measured $q$, while the radio emission is consistent with star formation in two (J1108+06 and J1316+44). Notice that this does not discard the existence of AGN driven radio structures. As an example, the Beetle QSO2  has $q$=1.89$\pm$0.10 and it is associated with a $\sim$4 kpc jet and a  large scale ($\sim$46  kpc) radio source (\citealt{Villar2017}). 

\subsubsection{Extended radio emission}
\label{extended}

As part of our sample selection, we compared the flux density values from FIRST (peak and integrated, $S^{\rm FIRST}_{\rm peak}$ and $S^{\rm FIRST}_{\rm int}$) and NVSS (integrated, $S^{\rm NVSS}$) 1.4 GHz fluxes. Our sample consists of sources for which $S^{\rm FIRST}_{\rm peak}$ $<$ $S^{\rm NVSS}$, and where this difference could indicate radio continuum emission at the mJy level on scales larger than the 5$^{\prime\prime}$ beam of FIRST. This would mimic the case of the Beetle Galaxy \citep{Villar2017}. We note, however, that differences between $S^{\rm FIRST}_{\rm peak}$ and $S^{\rm NVSS}$ in some cases are close to or within the quoted uncertainties in the flux density estimates, and could also be the result of source variability. 

In order to obtain a rough prior indication about the likely existence of radio emission extended on scales of $\ga$5$\arcsec$ in our sample sources, we follow \cite{Kimball2008}, who calculated log$(\theta^2)$, with $\theta$=$\sqrt\frac{S^{\rm FIRST}_{\rm int}}{S^{\rm FIRST}_{\rm peak}}$, and  $\Delta t= -2.5 \times {\rm log}\frac{S^{\rm FIRST}_{\rm int}}{S^{\rm NVSS}}$.  $\theta$ gives a dimensionless source concentration on $\sim$5$\arcsec$ scale.  Sources with log$(\theta^2)<$0.05 and $\ge$0.05 are classified as highly concentrated (``unresolved'', using the authors terminology) and  extended (``resolved'') respectively.  $\Delta t$ provides a measurement of source morphology that indicates angular extent and complexity.  \cite{Kimball2008} find a bimodal distribution  so that sources with $\Delta t\sim$0 are single component sources, while those  with $\Delta t\sim$0.7  are multiple-component or extended. Following their  approach, we classified roughly sources with $\Delta t<$0.35 as ``simple'' and those with  $\Delta t\ge$0.35  as ``complex'' (Table \ref{rad-ext}).

At the moment of  the observations there was no evidence for extended radio emission on scales $\ga$5$\arcsec$  for 4 out of 12 objects (J0802+25,  J1316+44, J1356+10 and J1437+30), while radio emission  was suspected  or confirmed to be extended for 7  objects (J0841+01, J0853+38, J0907+46, J0945+17, J1000+06, J1108+06, and J1517+33), as well as the Teacup. J0948+25 was uncertain due to the possible contamination of the NVSS flux by a nearby source. We note, however, that this a-priori classification is only approximate. For instance, sources with very bright compact cores and faint extended structures will be classified as ``unresolved''  and ``simple''.  This is the case of the Beetle (log$(\theta^2)=$0.033 and $\Delta t=$0.204). Also J1108+06, classified as ``unresolved'' and ``simple'' using the above method, is known to be associated with a multicomponent $\sim$3$\arcsec$ radio source (\citealt{Bondi2016}). Therefore, to reach our goal of investigating how common extended radio structures at the mJy level are in radio-quiet and -intermediate QSO2, our sample covers a wide range of log$(\theta^2)$ and $\Delta t$.

Nevertheless,with all the above restrictions, the sample cannot be considered representative of the general population of QSO2. The objects are characterised by complex optical morphologies. They were selected with FIRST detections, with many having relatively high radio luminosities  in comparison with  SDSS QSO2 in the same $z$ range, and some having prior indications for the presence of extended radio emission.

\begin{table*}
\centering
\tiny
\caption{The sample. General properties. (6) $L_{\rm [OIII]}$ is  the [OIII]$\lambda$5007 luminosity; (7) P$_{\rm 1.4\,GHz}$ is the monochromatic luminosity at 1.4 GHz inferred from the NVSS flux; (8) $q$  is a quantitative  measure of the FIR/radio flux-density ratio;  $q\le$1.8, suggests that the  the radio flux   has a significant AGN contribution (see text); (9) shows the classification of the object in radio quiet (RQ) or radio intermediate (RI)  (there are no radio loud objects) based on Fig. \ref{xu}. The ambiguous classification of several targets (RQ/I) reflects the vague division between  classes. `-' is shown when the information  is not available.}
\label{table-sample}
\begin{tabular}{lcccccccccccc}
\hline
Source & RA & Dec & $z$ & Scale  & log($L_{\rm [OIII]}$) & log(P$_{\rm 1.4\,GHz})$  & $q$ &  RQ/I/L \\ 
    &     &  & &  kpc/$\arcsec$ & erg\,s$^{-1}$ &  erg\,s$^{-1}$  Hz$^{-1}$\\ 
(1)  &  (2) &  (3)  &  (4)  &  (5) &  (6)  &  (7)   & (8)  & (9)  \\ \hline
J0802+25 & 08:02:52.93 & +25:52:55.6 & 0.080 &  1.491 & 42.44   & 30.67 & 1.39$\pm$0.06 &  RQ  \\ 
J0841+01 & 08:41:35.09 & +01:01:56.3 & 0.110 &  1.982 &  42.45   & 30.32 &  -&  RQ  &   \\
J0853+38 & 08:53:19.48 & +38:52:39.0 & 0.127 & 2.246 & 42.08 & 31.24 &  -  &  RI &  \\ 
J0907+46 & 09:07:22.36 & +46:20:18.1 & 0.167 &  2.827 & 42.18 & 31.55 &  -   &  RI &    \\
J0945+17 &	09:45:21.34 & +17:37:53.3 &	 0.128 &  2.261 & 42.63  & 31.28 & 1.35$\pm$0.10 &  RQ/I   \\
J0948+25 & 09:48:25.24  & +25:06:58.0  &  0.179 &  2.991 & 42.36  & 30.42 &    - &  RQ  &  \\
J1000+12 & 10:00:13.14 & +12:42:26.2 & 0.148 & 2.556 &  42.68 & 31.30 & 1.08$\pm$0.09 &   RQ/I  &   \\
J1108+06 &   11:08:51.03 & +06:59:00.5 & 0.182 &   3.031 &  42.39 &  31.01 & 1.97$\pm$0.07&  RQ    \\ 
J1316+44 & 13:16:39.75 & +44:52:35.1 & 0.091  & 1.675 & 42.23 &  30.08 & 2.35$\pm$0.04 &   RQ \\ 
J1356+10 & 13:56:46.11 & +10:26:09.1 & 0.123 & 2.185 &  42.79 &   31.39 & 1.23$\pm$0.04 &    RQ/I  \\
Teacup &  	14:30:29.88 	& +13:39:12.0& 0.085 & 1.576 &  42.66 & 30.67 & 1.18$\pm$0.11  &  RQ   \\
J1437+30 & 14:37:37.85 & +30:11:01.1 & 0.092 & 1.692 & 42.40 & 31.15 & 0.43$\pm$0.02 & RQ/I    \\ 
J1517+33 &    	15:17:09.21 & +33:53:24.7 & 0.135 & 2.366 & 42.49 & 31.76 &  0.36$\pm$0.18 &  RI     \\ 
\hline
\end{tabular} 
\end{table*} 

\begin{table*}
\centering
\footnotesize
\caption{Classification of the QSO2  VLA sample according to the possible  existence of extended (5$\arcsec$) radio emission based on \cite{Kimball2008}. (1) Source name; (2) FIRST peak flux density. The quoted error is the RMS is the noise in the FIRST map (mJy/beam); (3)  FIRST integrated flux density for the source; (4) NVSS integrated  flux density with the  mean error. N/A means that there is no NVSS flux available; 
(5) log$(\theta^2)$, with $\theta$=$\sqrt\frac{S^{\rm FIRST}_{\rm int}}{S^{\rm FIRST}_{\rm peak}}$. Sources with log$(\theta^2)<$0.05 and $\ge$0.05 are classified as ``unresolved' and  ``resolved'' respectively (see text) in column (6); (7) $d_{\rm max}^{\rm R}$: radio source size in arcsec measured from our new VLA data (8)  $\Delta t= -2.5 \times {\rm log}\frac{S^{\rm FIRST}_{\rm int}}{S^{\rm NVSS}}$. Sources with $\Delta t <$0.35 and $\ge$0.35 are classified as ``simple'' and  ``complex'' respectively in column (9).   The Teacup was not part of our VLA sample. It is shown here  because its radio properties are discussed  in the text. $^*$The NVSS flux of J0948+25 may be contaminated by a nearby source. }
\label{rad-ext}
\begin{tabular}{cccccccccc}
\hline
Source &     $S^{\rm FIRST}_{\rm peak}$  &   $S^{\rm FIRST}_{\rm int}$  & $S^{\rm NVSS}$     & log$(\theta^2)$ &  Class &    Observed (VLA) & $\Delta t$ & Class   \\ 
    &       mJy beam$^{-1}$  & mJy & mJy & & & & & &\\ 
 (1)	& 	(2)  &   (3)   & (4)	& 	(5)  &   (6)   & (7) & (8) & (9)   \\ \hline
J0802+25 &  29.37$\pm$0.15  &  30.61 & 30.3$\pm$1.0 &    0.018 &  Unres. & Unres & -0.011 & Simple   \\ 
J0841+01 &     2.69$\pm$0.15 & 3.99  & 6.8$\pm$0.5 &   0.172 &  Resolved  &   20.2$\arcsec$ & 0.579 & Complex \\
J0853+38 &  	16.55$\pm$0.11 & 17.93  &  42.0$\pm$1.6 &      0.035 & Unres.  & 210$\arcsec$ & 0.924 & Complex \\ 
J0907+46 &  29.17$\pm$0.13    & 35.10   & 47.1$\pm$1.5 &    0.080 &  Resolved & 27$\arcsec$ & 0.319 & Simple  \\
J0945+17 &	38.68$\pm$0.14    & 44.46   & 45.6$\pm$1.4 &      0.061 & Resolved  & 4.9$\arcsec$ &  0.027 & Simple \\
J0948+25 &  1.54$\pm$0.13 &  1.62 &  3.0$\pm$0.3$^*$   &   0.022 & Unres. & 1.3$\arcsec$ & 0.669 & Complex   \\
J1000+12 &   25.70$\pm$0.13 &  31.75 &  34.8$\pm$1.1 &    0.092 & Resolved & 17$\arcsec$ & 0.100 & Simple \\
J1108+06 &      9.30$\pm$0.13 & 9.84 &  11.1$\pm$0.5  &   0.025 & Unres. &  3$\arcsec$ & 0.131 & Simple \\ 
J1316+44 &  4.23$\pm$0.14 &  4.53 &  5.9$\pm$0.4  &   0.030  & Unres.& Unres.  & 0.287 & Simple \\ 
J1356+10 &   57.90$\pm$0.13 & 59.58 &  62.9$\pm$1.9 &      0.012 & Unres.  & $\ge$2.3 &  0.059 & Simple \\
J1437+30 &  63.91$\pm$0.14  &  67.08  & 68.3$\pm$2.5 &      0.021 & Unres. & 1.5$\arcsec$  &  0.020 & Simple \\ 
J1517+33 &       106.70$\pm$0.14  & 120.39 &  120.9$\pm$3.6 &       0.052 & Resolved & 3.8$\arcsec$ & 0.005  & Simple  \\  \hline
Teacup & 13.49$\pm$0.15	& 26.41 	& 	26.5$\pm$0.9  &   0.292 &  Resolved &  12$\arcsec$ & 0.004 &  Simple\\ 
\hline
\end{tabular}
\end{table*}

\section{Observations}
\label{observations}

\begin{table*}
\centering
\footnotesize
\caption{Log of GTC observations. $\lambda^{\rm TF}$ is the  wavelength used to tune the TF, taking into account the  shift to the blue of the central wavelength seen by the TF as the distance to the optical centre  increases (see Sect. 3.1). The superscripts $^{\rm TF}$ and $^{\rm cont}$ refer to the narrow band (TF) and continuum images respectively. The seeing sizes correspond to the FWHM.}
\label{log-optical}
\begin{tabular}{lcccccccccccc}
\hline
Source &   Date &  $\lambda^{\rm TF}$ & Filter$^{\rm cont}$  & $\Delta\lambda$& t$_{\rm exp}^{\rm TF}$ &    t$_{\rm exp}^{\rm cont}$ & Seeing$^{\rm TF}$ & Seeing$^{\rm cont}$  \\ 
		&   	&  \AA &  &  \AA&	sec  &   sec &	  $\arcsec$ &  $\arcsec$ \\  \hline
J0802+25 & 17/Feb/2020 & 7102 & f680/43    & 6098-6498 & 3$\times$900 &  3$\times$200 &  1.40$\pm$0.10 & 1.42$\pm$0.15   \\ 
J0841+01 &   16/Feb/2020   & 7296 &  f694/44 & 6057-6454 & 3$\times$900 &  3$\times$200 & 0.94$\pm$0.02 & 1.02$\pm$0.02   	 \\
J0853+38 &   16/Feb/2020   & 7408 &  f754/50 & 6090-6489 & 3$\times$900 &  3$\times$200 & 1.20$\pm$0.04 & 0.97$\pm$0.03  \\
J0907+46 &  16/Feb/2020 &  7665 &   f738/46  & 6126-6521 &   3$\times$900 &  3$\times$200 & 0.99$\pm$0.09 & 0.89$\pm$0.08    \\
J1000+12 &  16/Feb/2010 &  7543 & f709/45 & 5979-6370 &   3$\times$900 &  3$\times$200 & 1.07$\pm$0.02 & 1.09$\pm$0.03  \\
J1316+44  & 17/Feb/2020   &  7168 &   f680/43 & 6037-6433 &  3$\times$900 &  3$\times$200 & 1.15$\pm$0.15 & 1.24$\pm$0.16   \\ 
J1356+10 &  23/Jun/2019 &   7379 & f694/44 	& 5987-6379 &   3$\times$900 &  3$\times$200 & 0.90$\pm$0.04 & 0.87$\pm$0.03  \\
Teacup &  	16/Feb/2010  &   7128 &   f666/36 & 5947-6345 & 7$\times$900 &  3$\times$120 & 1.16$\pm$0.06 & 1.33$\pm$0.05 \\
J1437+30 &    16/Feb/2010 &  7175 &  f680/43& 6031-6427 & 3$\times$900 &  3$\times$200 & 0.95$\pm$0.04 & 1.07$\pm$0.05 &  \\ \hline
\end{tabular} 
\end{table*}

\subsection{10.4m Gran Telescopio CANARIAS}
\label{sec:GTC}

The Gran Telescopio Canarias (GTC) observations were performed on  23/06/2019 (program GTC28-19A)  and 16-17/02/2020 (GTC114-19B) for 10 QSO2 in the sample (Table \ref{log-optical}). 

H$\alpha$ Tunable filter (TF) images were obtained with the OSIRIS
instrument\footnote{http://www.gtc.iac.es/en/pages/instrumentation/osiris.php}
mounted on the 10.4m  GTC. This is an optical imager
and spectrograph that offers broad band photometry, tunable filter
imaging and both multi-object and long slit spectroscopy. OSIRIS consists of a mosaic of two 2048
$\times$ 4096 Marconi CCD42-82 (with a 9.4'' gap between them) and covers the
wavelength range from 0.365 to 1.05 $\mu$m with a field of view of 7.8$\arcmin \times$ 7.8$\arcmin$ and a pixel size of 0.127$\arcsec$. However, the OSIRIS standard
observation modes use 2$\times$2 binning, hence the effective pixel size during
our observations was 0.254$\arcsec$.

When using OSIRIS TF imaging mode the wavelength observed changes relative to the optical centre following the formula given by:

\begin{equation} 
{\lambda} = {\lambda_{0}}-5.04*r^{2}
\label{ec:TFlambda}
\end{equation}

 where $r$ is the distance in arcmin to the optical centre. 

For this reason, for each QSO2 a filter FWHM of 20 \AA\ was used at the QSO2s $z$ centred on the redshifted H$\alpha$ and taking into account the dependence of the wavelength observed with the red TF with distance relative to the optical centre described above. Also, to sample the continuum near the H$\alpha$+[NII] complex in each QSO we took one continuum image using an appropriate OSIRIS Red Order Sorter (OS) filters. These are medium-band filters (17 nm wide) initially used to isolate different interference orders within the TF, but that can also be used alone to produced direct imaging if needed. Both in direct TF and continuum (OS) imaging a 3-dither pattern was used by moving the telescope $\pm$15$\arcsec$ in RA and Dec  to correct for ghost images and cosmic rays. Total exposure times and corresponding order filters used are described in Table \ref{log-optical}, where the details on the observations are summarised.  The seeing size during the observations was in the range $\sim$0.9-1.4$\arcsec$ depending on the object. 

A detailed description on the TF reduction process can be found in \cite{Villar2017}. Briefly, the TF and the OS filter OSIRIS images were bias and
flat-field corrected as usual, using a set of bias frames and sky
flats. Then, images are sky subtracted by fitting a 1d polynomial both in x/y dimensions in order to remove the 
effect of the sky lines in the images (that appear as rings of emission over the CCDs due to the wavelength change produced by the TF described by Equation (\ref{ec:TFlambda})).  Longer exposures (typically 3$\times$900 sec and 3$\times$200 sec for the TF and continuum images respectively) where obtained to map the low surface brightness ($SB$) extended structures. Because the nucleus saturates in these frames, shorter exposures (90 and 60 sec for the TF and continuum images respectively) were also obtained to ensure the availability of unsaturated images for all objects. These will be mentioned only when relevant.

 The  TF H$\alpha$ images were calibrated in flux applying the method described in \cite{Cabrera2014}. Galactic extinction correction was not applied as it is  negligible. The short exposure images (and thus, not saturated) were used to compare the nuclear fluxes with those measured from the SDSS spectra. A  circular aperture of 3$\arcsec$ diameter, as   the SDSS fibre, was used for this. For each object, the SDSS flux was measured within a spectral window with the same central $\lambda$ and width (20\AA) as the TF image. The  TF nuclear fluxes are within $\sim$90\%  of the SDSS fluxes.

The H$\alpha$ and continuum images were then properly aligned  and scaled.
Based on the centroid of the
stars in the aligned images and the position of some important
features in the galaxies, we estimate that the alignment
accuracy is better than $\sim$0.5 pixels. The scaled 
 continuum images were then subtracted from the emission line images in order to produce the ``pure'' H$\alpha$ images. 

 Detection limits of the final H$\alpha$ images are in the range $3\sigma\sim$(2.5-5.4)$\times$10$^{-18}$ erg s$^{-1}$ cm$^{-2}$ arcsec$^{-2}$ depending on the object.

\subsection{Karl G. Jansky Very Large Array}
\label{sec:VLA}

\begin{table*}
\centering
\footnotesize
\caption{VLA  information about the  A and B-configuration observations.}
\label{tab:results}
\begin{tabular}{lccclcccccccccccc}
Source &  t$_{\rm exp}^A$ &  t$_{\rm exp}^B$ & \multicolumn{2}{c}{Beam$_{A}$} & \multicolumn{2}{c}{Beam$_{B}$} & S$^{A}_{\rm total}$ & S$^{B}_{\rm total}$ &  S$^{\rm 1.4\,GHz}_{\rm core}$ $^{\dagger}$ &  S$^{\rm 1.4\,GHz}_{\rm extended}$ $^{\ddagger}$ \\
    & min &  min & arcsec$^{2}$ & PA & arcsec$^{2}$ & PA & mJy & mJy & mJy bm$^{-1}$ & mJy \\
\hline
J0802+25 &  31 &  88 & 1.21$\times$1.10          & -84 & 5.52$\times$4.60 & -80 & 28    & 29    & 28         & $-^{*}$ \\ 
J0841+01 & 40  & 68  & 1.35$\times$1.18          & -16  & 5.27$\times$4.13 & -15  & 4.5   & 6.3   & 0.5$^{**}$ & 5.8      \\ 
J0853+38 &  36 & 72  & 1.29$\times$1.10          & -82 & 4.72$\times$4.47 & 64 & 33    & 68    & 17         & 51    \\ 
J0907+46 & 36  &  89 & 1.23$\times$1.04          & -84 & 4.63$\times$4.35 & 67 & 44    & 45    & 18         & 27    \\ 
J0945+17 & 30  & 60  & 1.21$\times$1.11          & -58 & 6.16$\times$4.19 & 58 & 44    & 43    & 26         & 17     \\ 
J0948+25 &  15 & 45  & 1.26$\times$1.13          & -72 & 6.03$\times$4.34 & 65 & 1.8   & 2.0   & 1.1        & 0.9      \\ 
J1000+12 & 30  &  77 & 1.23$\times$1.13          & -43 & 6.19$\times$4.34 & 54 & 32    & 34    & 19         & 15   \\ 
J1108+06 & 28  &  42 & 1.22$\times$1.03          & -22 & 6.40$\times$4.23 & -48 & 8.1   & 8.1   & 3.8        & 4.3     \\ 
J1316+44 &  27 &  64 & 1.25$\times$0.97          & 78  & 4.57$\times$4.32 & -76 & 3.3   & 3.8    & 2.8       & (1.0)$^{***}$ \\
J1356+10 &  28 & 170  & 1.31$\times$1.24          & 69  & 5.21$\times$4.44 & 44 & 58    & 59 &  52   & 7$^{\S}$ \\ 
J1437+30 & 30  &  60 & 0.78$\times$0.72$^{\S\S}$& -81 & 4.97$\times$4.50 & 67    & 55    & 58    & 40         & (18)$^{\S\S\S}$ & \\ 
J1517+33 &  30 & 60  & 1.26$\times$1.08          & -71 & 4.91$\times$4.30 & 84 & 116   & 115   & 69         & 46 & \\ 
\hline
\end{tabular} 
\flushleft 
$^{\dagger}$ Peak flux density at the location of the core in the A-configuration data.\\
$^{\ddagger}$ Total intensity of the source in B-configuration, excluding the peak flux density in A-configuration. \\ 
$^{*}$ Unresolved, with A- and B-configuration intensity consistent within the 10$\%$ uncertainty of the flux calibration.\\
$^{**}$ Assumed to be the weakest, most Eastern of the three central blobs in Fig. \ref{VLA0841}.\\
$^{***}$ Very uncertain, only marginally resolved in A-configuration.\\
$^{\S}$ Of which 6 mJy reflects the extension found in the A-configuration data, and 1.0 mJy is the integrated flux density of the emission found on large scales in the B-configuration data.\\
$^{\S\S}$ A-configuration data imaged with super-uniform weighting.\\
$^{\S\S\S}$ Only marginally resolved with super-uniform weighting.\\
All positions angles (PA) in the paper are quoted North to East.
\end{table*}

Observations with the VLA were performed during the period Feb$-$May 2019 in B-configuration and Aug$-$Oct 2019 in A-configuration (project VLA/19A-134), as well as 2h of Director's Discretionary Time (DDT) during the reconfiguration from B- to A-configuration on 14 Nov 2020 (project VLA/20B-428). We used the L-band system in standard continuum mode to cover the frequency range 1$-$2\,GHz. We performed snap-shot observations with standard phase, bandpass, and flux calibration, and visited each source at least three times to improve the (u,v)-coverage. 
The total on-source integration time per target was on average 60-90 min in B-configuration and 30 minutes in A-configuration (details are given in Table \ref{tab:results}). The DDT observations of J1356+10 where tapered as to only include baselines equivalent to those in the B-configuration, and added to the 90 min of data already taken for this source in B-configuration.

The data were calibrated using the VLA CASA pipeline. Afterwards, a round of phase-only self calibration was attempted, but for most sources this solution did not improve the imaging and was therefore not applied. The data were subsequently imaged in CASA using the task $tclean$ (\citealt{McMullin2007}; CASA Team et al. in prep).

For imaging, we used a natural weighting for the B-configuration data and Briggs weighting with robustness parameter 0.5 for the A-configuration data \citep{Briggs1995}. For the field of J1437+30, we also produced a map with super-uniform weighting in order to resolve bright radio structures on the smallest scales. To mitigate the effects from sidelobes of strong radio sources in the field, we imaged the full primary beam before doing the primary beam correction. However, because our target sources were located at the pointing centre, we did not need to use the wide-field imaging techniques to recover their structure and flux density.

Upper limits on the sizes of unresolved sources are given by the beam sizes in Table \ref{tab:results}.

\begin{table*}
\centering
\tiny
\caption{General properties of the radio structures.  (2) $d_{\rm max}^{\rm R}$: radio source size in kpc. Upper limits are provided for  spatially unresolved or marginally resolved (J1316+44) sources; (3) Are there radio structures related to the nuclear activity, at least partially?;  (4) morphological classification of prominent radio features; (5) Is there evidence for jet/gas interactions?; (6) Relevant references: R10: \cite{Rosario2010}; V14: \cite{Villar2014};  H15: \cite{Harrison2015};  B16: \cite{Bondi2016}; R17: \cite{Ramos2017}; J19: \cite{Jarvis2019}. The Teacup is shown separately because it was not part of our VLA sample. The radio information comes from H15 and J19. $^*$According to \cite{Bondi2016} the extended radio emission in J1108+06 is due to star formation and the radio core is  associated with one of the two AGN.}
\label{radprop}
\begin{tabular}{lccccc}
\hline
 Source   &   $d_{\rm max}^{\rm R}$  &    Related to AGN? & Nature of   & JGI?  & Ref.\\ 
      &  kpc &  &  radio features  &    \\ 
 (1)	& (2) & (3) &  (4)	 &  (5)\\ \hline
J0802+25 &  $<$2  &  Yes & Unknown  & Yes  & V14\\ 
J0841+01 &  40   &    Yes & Core, jet/lobe/hot spot   & Yes & \\
J0853+38 &  472 &  Yes & Core,  inner collimated jets ($\sim$4.5 and 9 kpc), large scale jet ($\sim$60 kpc) & Yes & \\ 
 &   &   &  radio hot spot, large and distant radio lobes ($\sim$472 kpc) &  \\ 
J0907+46 & 76  & Yes &  Core, inner collimated jets  & Yes  & \\
 &   &    & large scale bent jet ($\sim$56 kpc)   &  \\ 
J0945+17 &	11 &  Yes & Small ($\sim$2.3 kpc) jet, large scale ($\sim$11 kpc) jet or lobe  & Yes & J19 \\
J0948+25 &  $\sim$4    &   Unknown & Unknown &   No &  \\ 
J1000+12 & 43  & Yes & Core, jet-like structure ($\sim$2.6 kpc) & Yes  &  J19 \\
 &   &    &  hot spots, large scale distorted lobes (43 kpc) &  \\ 
J1108+06 & 9    &  Yes$^*$  &  Multiple components  & No   & B16\\ 
 &   &    & AGN radio core, extended star formation  &  \\ 
J1316+44 &  $\le$2  & Unknown   &  Consistent with SF & No &   \\ 
J1356+10 &  160  & Yes	&	 Jet ($\sim$5 kpc) & Yes &   J19  \\
 &   &   & large scale (160 kpc) radio lobes	 & &    \\
J1437+30 & $\sim$2.5 & Yes &   Unknown & Yes? &   \\ 
J1517+33 & 9  &  Yes  &  Core, bipolar bent jet ($\sim$9 kpc)  & Yes & R10  \\ \hline
Teacup &   19 &  Yes & Small scale ($\sim$1 kpc) radio jet & Yes  & H15,R17,J19 \\ 
 &   &    &  radio bubbles ($\sim$12 kpc) (lobes/AGN wind?)    &  \\  
\hline
\end{tabular} 
\centering
\tiny
\caption{General properties of the ionised nebulae.  (2) $d_{\rm max}$: maximum distance in kpc between detected ionised gas features; (3) $r_{\rm max}$: distance in kpc  from the QSO2 nucleus to the most distant ionised gas feature detected; (4)   morphological classification of prominent ionised features; (5) Is there gas outside the putative QSO2 ionisation cones?; (6)  Relevant references: G12: \citealt{Greene2012}; K12: \cite{Keel2012}; H15: \cite{Jarvis2019}; S18: \cite{Storchi2018}; V18: \cite{Villar2018}; F18:  \cite{Fischer2018}; J19: \cite{Jarvis2019}. J0948+25, J1108+06 and J1517+33 are not in this table because there are no emission line images available.}
\label{optprop}
\begin{tabular}{lccccccc}
\hline
 Source   &   $d_{\rm max}$  &  $r_{\rm max}$ &   Nature of prominent &   Outside  &  Ref. \\ 
    &   kpc  &  kpc &  ionised features  &   cones?  \\ 
 (1)	& (2) & (3) 	& (4) & (5) &  (6)    \\ \hline
J0802+25 & 38 & 21 &  Low surface brightness large ($\sim$38 kpc) asymmetric nebula  & Yes &   F18 \\ 
J0841+01 &  42  &  21 &   ionisation cone or giant  ($\sim$8 kpc) bubble?  & Yes &   S18 \\
&   &  &     bent arm tracing eastern radio source across 12 kpc  &      \\
&   &  & tidal features, TDGC &       &      \\
J0853+38 & 100  & 70  &    Bent arms tracing the radio source across $>$60 kpc &  Yes     \\ 
&   &  &     SF in spiral arms; distant patches   &       \\
J0907+46 &    68 & 46   &   Nebula ($\sim$13 kpc) tracing the inner radio source  & No?  &    \\
&   &  & tidal tail, TDGC        &      \\
J0945+17 &	 25 & 15 &   Giant bubble  candidate ($\sim$10 kpc), filaments  & No? & S18\\
J1000+12 & 112 & 90 &  Blob, jet-like feature ($\sim$9kpc),  giant bubble  candidate  ($\sim$10 kpc) &  No  &  J19 \\
 &   &  &  extension tracing the NW radio source,  tidal tail, TDGC &        \\
J1316+44 & 92	& 53	& 	Tidal bridge, distant patches   & Yes   \\ 
&   &  & widely spread SF in host galaxy &          \\
J1356+10 &119 & 92 &  Giant bubbles and  shells ($\sim$12-30 kpc)   &   Yes &  G12,J19  \\
&  & & large asymmetric nebula ($\sim$70 kpc), tidal brige/filament, TDGC       &     \\  
Teacup & 132 &  90  & Giant bubble ($\sim$10 kpc),  giant ($\sim$115 kpc$\times$87 kpc) nebula &  Yes      & K12,VM18    \\
&   &  &    arc (edge brightened cavity?), filament ($\sim$30 kpc),  distant patches &      \\
J1437+30 &  22 & 12 &  SF knot, low surface brightness asymmetric $\sim$12 kpc nebula & Yes &  F18 \\  \hline
\end{tabular} 
\end{table*}

\begin{figure}
\centering
\includegraphics[width=0.40\textwidth]{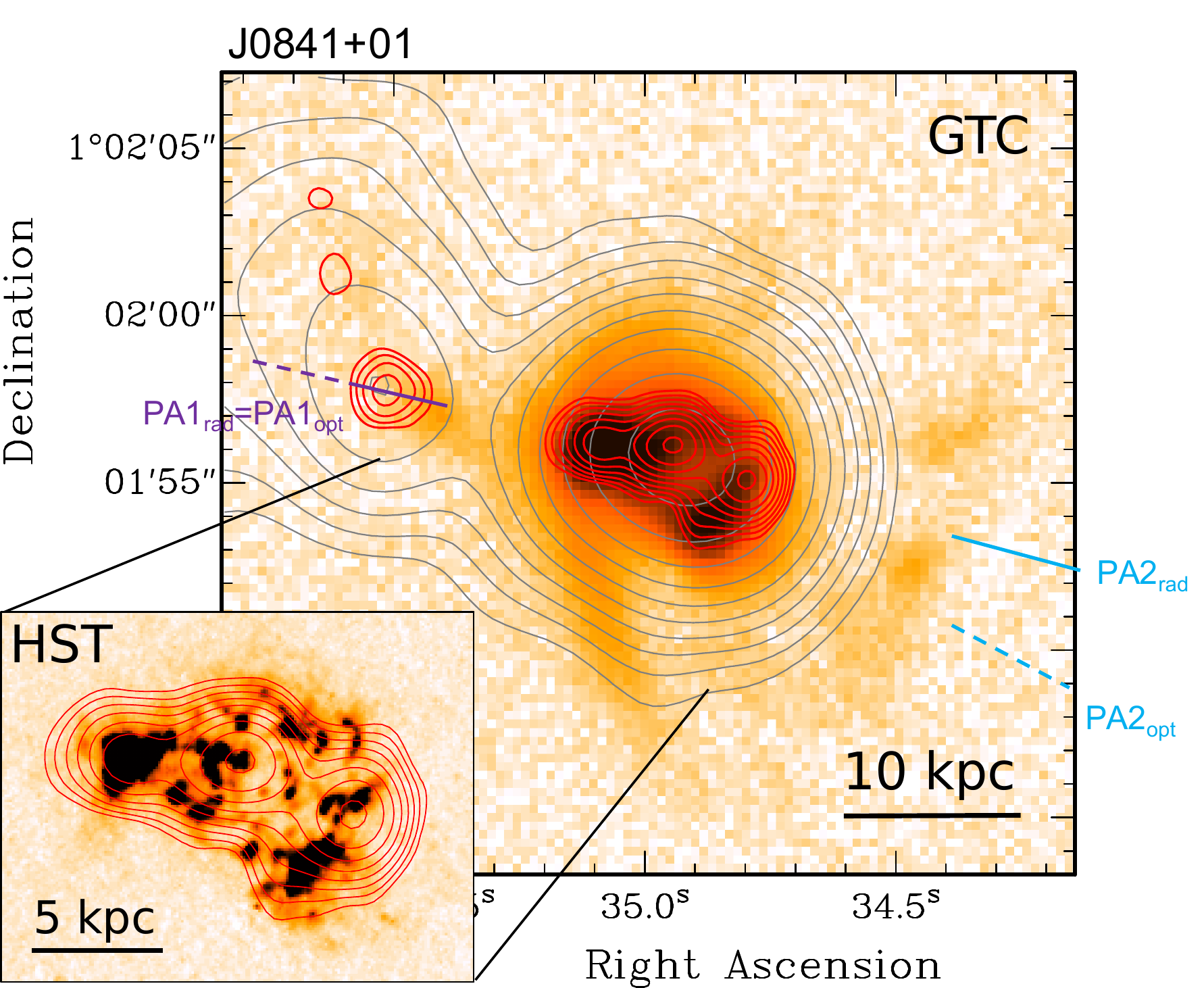}
\caption{GTC H$\alpha$ image of J0841+01 on a log scale, with overlaid contours of the VLA A-configuration (red) and B-configuration (grey) data. Contour levels are the same as in Fig.\,\ref{VLA0841}. The inset shows the contours of the VLA A-configuration data overlaid onto the smoothed HST [OIII]$\lambda$5007 continuum subtracted  image from \cite{Storchi2018}.   In these and other radio-H$\alpha$ overlays the  directions of the axes used to measure the degree of alignment between the radio (solid lines) and the ionised gas (dashed lines) structures are indicated. The same colour is used for axes   on the same side of the AGN (see Sect. \ref{sec:alignment}).}
\label{overlay0841}
\centering
\includegraphics[width=0.40\textwidth]{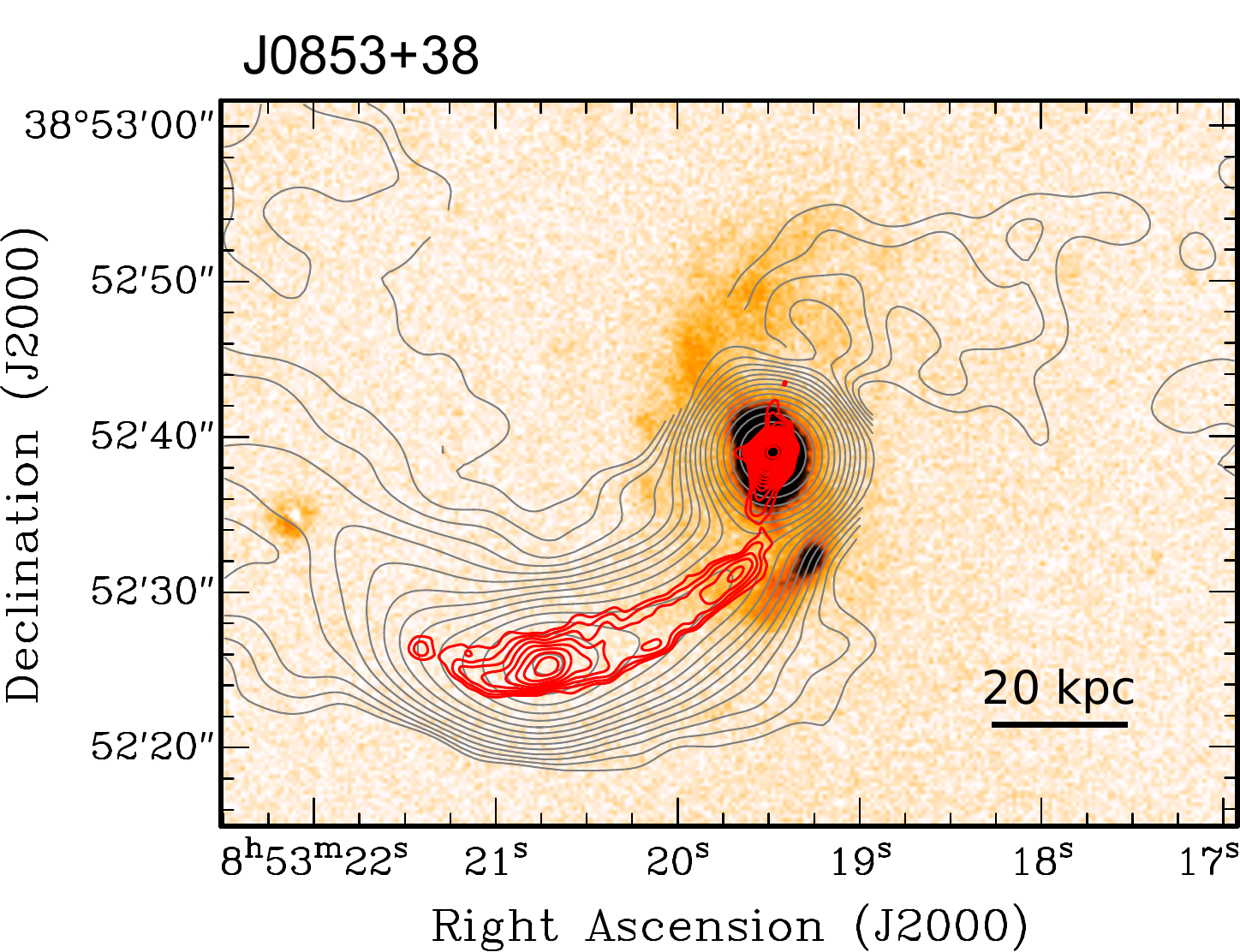}
\caption{GTC H$\alpha$ image of J0853+38, with overlaid contours of the VLA A-configuration (red) and B-configuration (grey) data. Contour levels are the same as in Fig.\,\ref{VLA0853}, and features indicated as image artefacts in the B-configuration data of Fig.\,\ref{VLA0853} have been omitted.}
\label{overlay0853}
\includegraphics[width=0.40\textwidth]{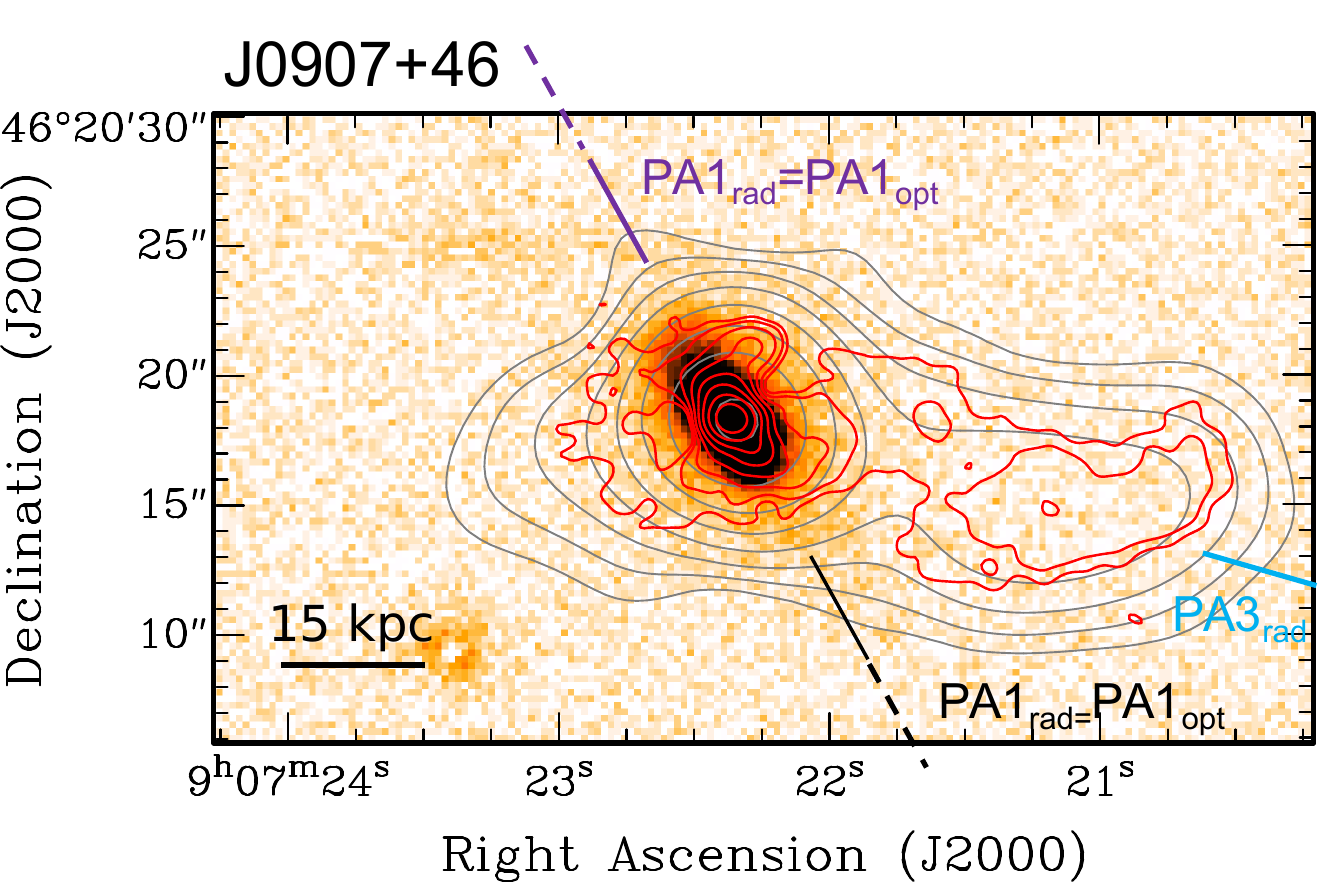}
\caption{GTC H$\alpha$ image of J0907+46, with overlaid contours of the VLA A-configuration (red) and B-configuration (grey) data. Contour levels start at 0.08 (0.2) mJy\,bm$^{-1}$ for the A (B) configuration data, and increase by factor 2.  Radio and H$\alpha$ axes as in Fig. \ref{overlay0841}.}
\label{overlay0907}
\end{figure}
\begin{figure}
\centering
\includegraphics[width=0.40\textwidth]{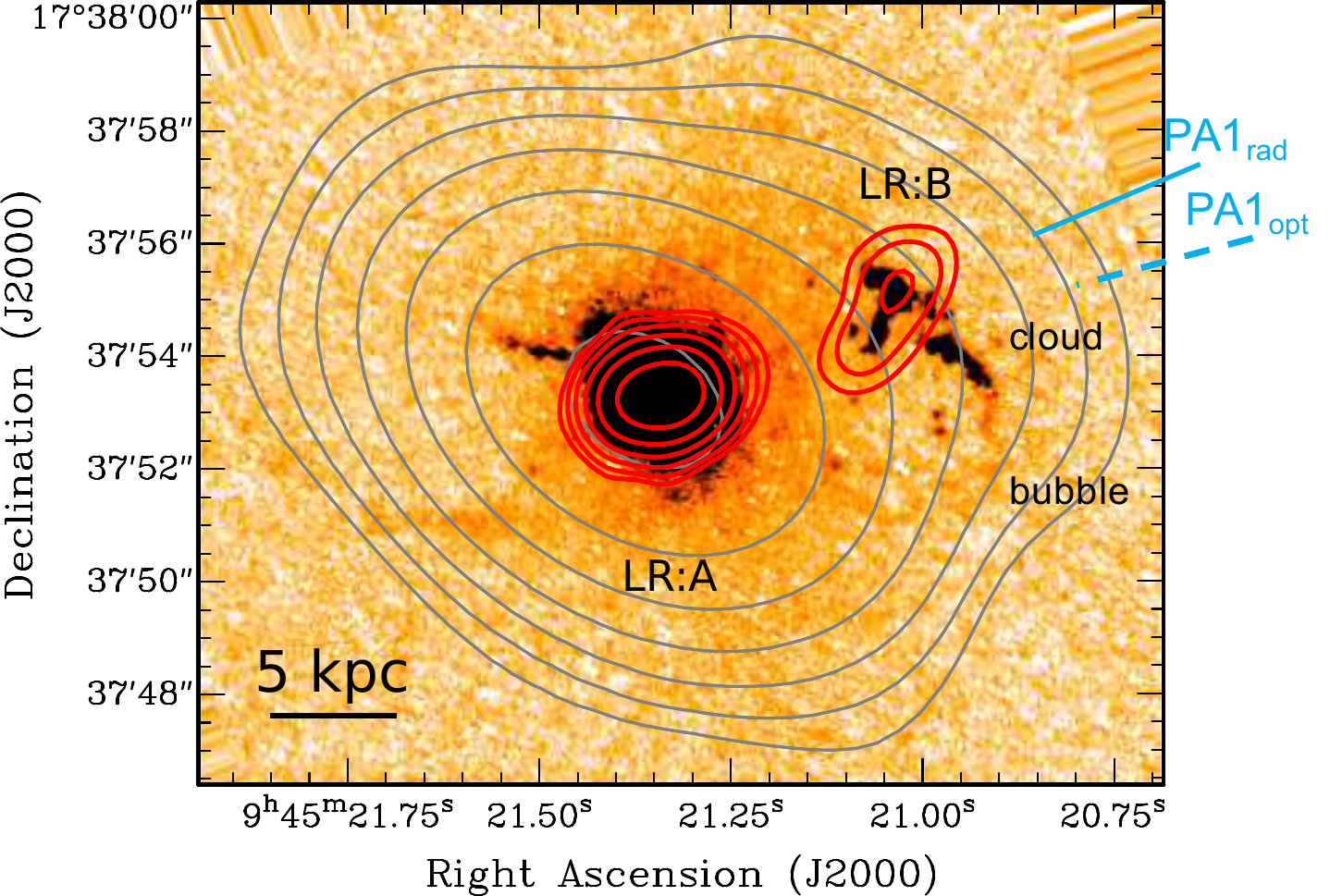}
\caption{HST/WFC FR551N (containing [OIII]) image of J0945+17, with overlaid contours of the VLA A  (red) and B-configuration (grey) data. Contour levels start at 0.45 (0.5) mJy\,bm$^{-1}$ for the A (B) configuration data, and increase by factor 2. LR:A and LR:B are the radio components identified  by \cite{Jarvis2019}. Radio and H$\alpha$ axes as in Fig. \ref{overlay0841}.}
\label{overlay0945}
\includegraphics[width=0.33\textwidth]{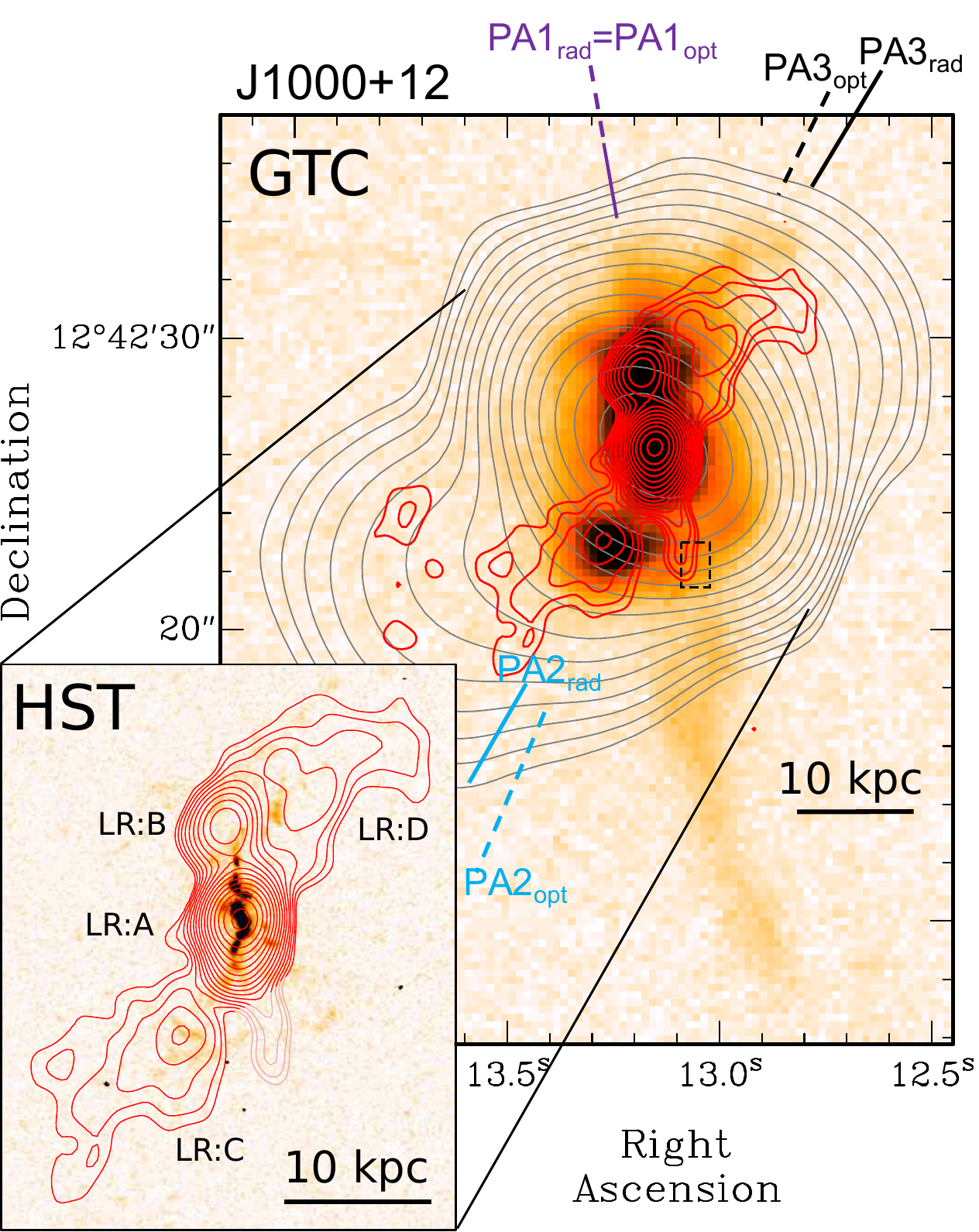}
\caption{GTC H$\alpha$ image of J1000+12, with overlaid contours of the VLA A (red) and B-configuration (grey) data. Contour levels are the same as in Fig.\,\ref{VLA1000}. The inset shows the contours of the VLA A-configuration data overlaid onto the HST WFC3/FQ508N narrow-band filter that contains [OIII] emission. The dashed black rectangle marks the approximate location of the giant [OIII] bubble proposed by \cite{Jarvis2019}. The narrow extension of radio emission at this location, which we coloured light-red in the small panel, is most likely dominated by artefacts of the snap-shot imaging.  LR:A to LR:D indicate the main  radio components with the same nomenclature as in \cite{Jarvis2019}.  Radio and H$\alpha$ axes as in Fig. \ref{overlay0841}.}
\label{overlay1000}
\end{figure}

\begin{figure}
\centering
\includegraphics[width=0.40\textwidth]{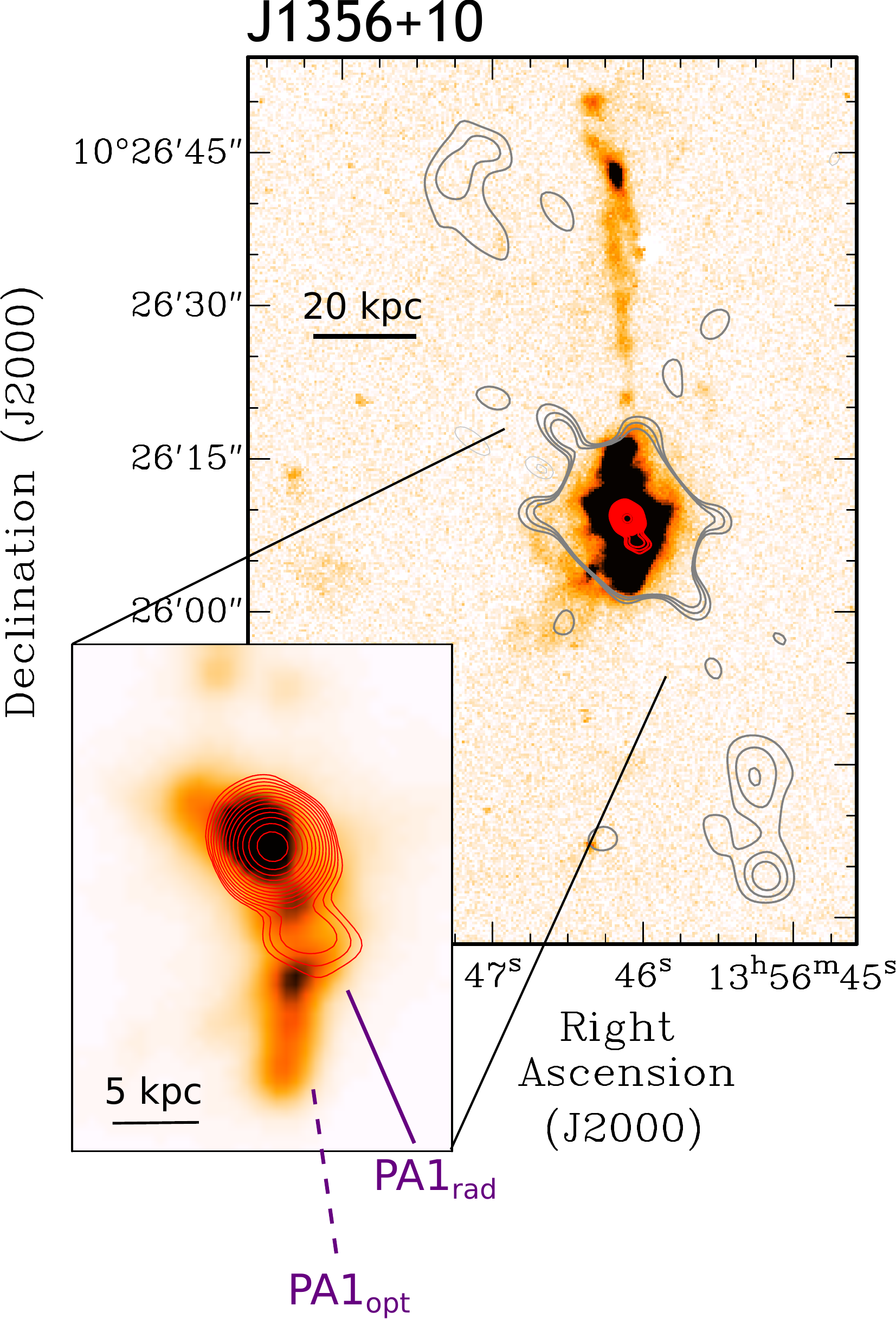}
\caption{J1356+10. GTC H$\alpha$ image of J1356+10, with overlaid contours of the VLA A-configuration in red, and the extended emission from the B-configuration data in grey (see Fig.\,\ref{VLA1356} for details). The inset highlights the central region of the galaxy, with the GTC image shown at a higher contrast.  Radio and H$\alpha$ axes as in Fig. \ref{overlay0841}.}
\label{overlay1356}
\includegraphics[width=0.32\textwidth]{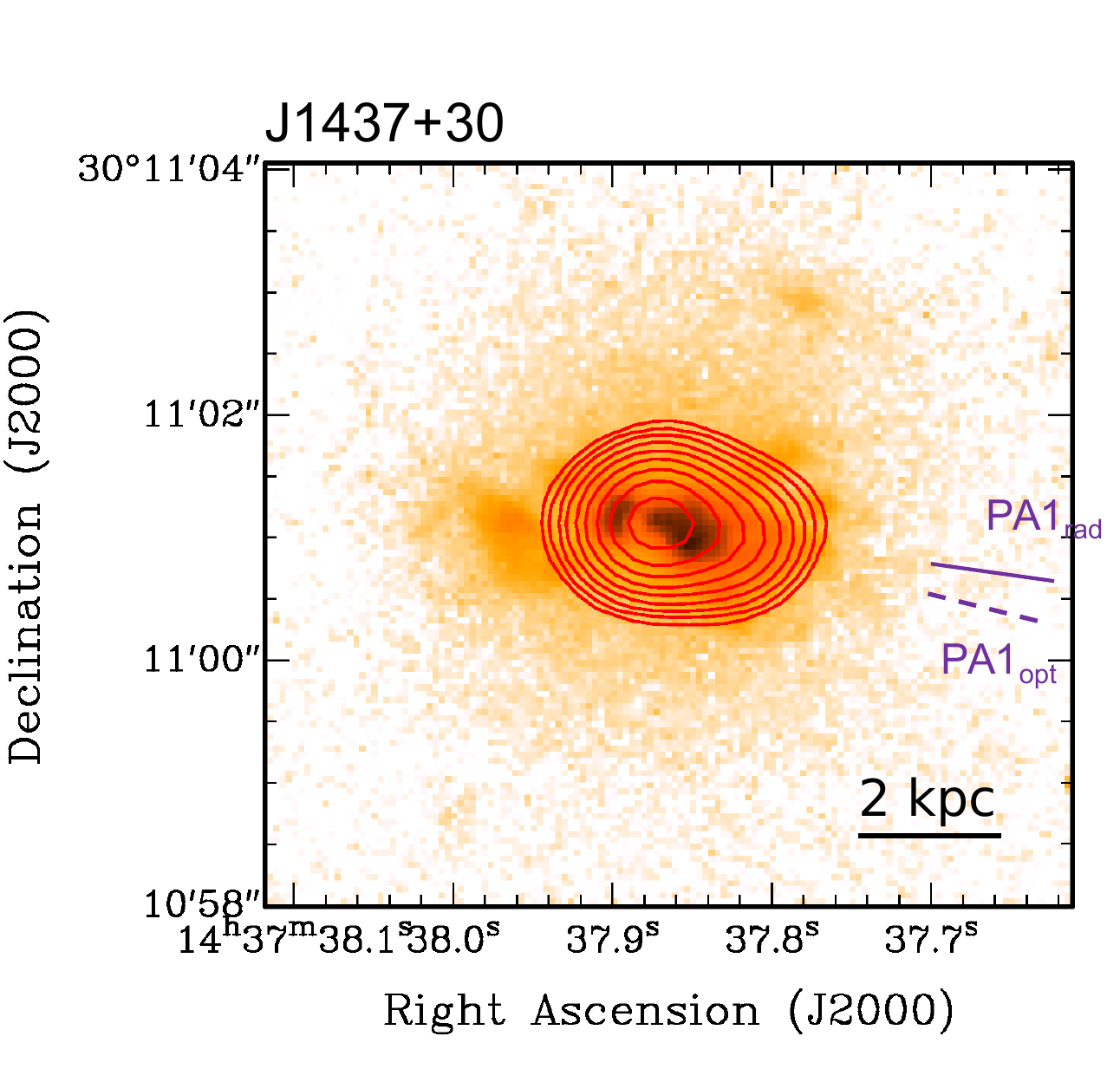}
\caption{J1437+30. HST/ACS image taken with the FR551N filter (covering [OIII]$\lambda$5007)  with overlaid contours of the VLA A-configuration data with super-uniform weighting from Fig.\,\ref{VLA1437}. The HST image      is contaminated by continuum, but the high $SB$ features trace the ionised gas morphology  accurately  (\citealt{Fischer2018}). Radio and H$\alpha$ axes as in Fig. \ref{overlay0841}.}
\label{overlay1437}
\end{figure}

\section{Results}
\label{results-global}

We will present  in this section the general properties of the radio and ionised gas in our sample. We also present in Figs. \ref{overlay0841} to \ref{overlay1437}, the overlays between the radio and H$\alpha$ maps. We refer the reader to the Appendix \ref{results-objects2} for  detailed descriptions and results of the  individual QSO2.  The information on the sizes and nature of the radio structures and the large scale ionised gas for all objects are summarised in Tables \ref{radprop} and \ref{optprop} and Figs. \ref{hist-rad} and \ref{hist-ion}.

\subsection{Size, morphology  and origin of the radio structures}
\label{results4p1}

The contribution of an AGN component to the total radio luminosity is confirmed in  11/13 QSO2, all except J0948+25 and J1316+44. For these two objects, the available information does not allow to discern  the origin (AGN and/or SF). 

\begin{figure}
\centering
\includegraphics[width=0.4\textwidth]{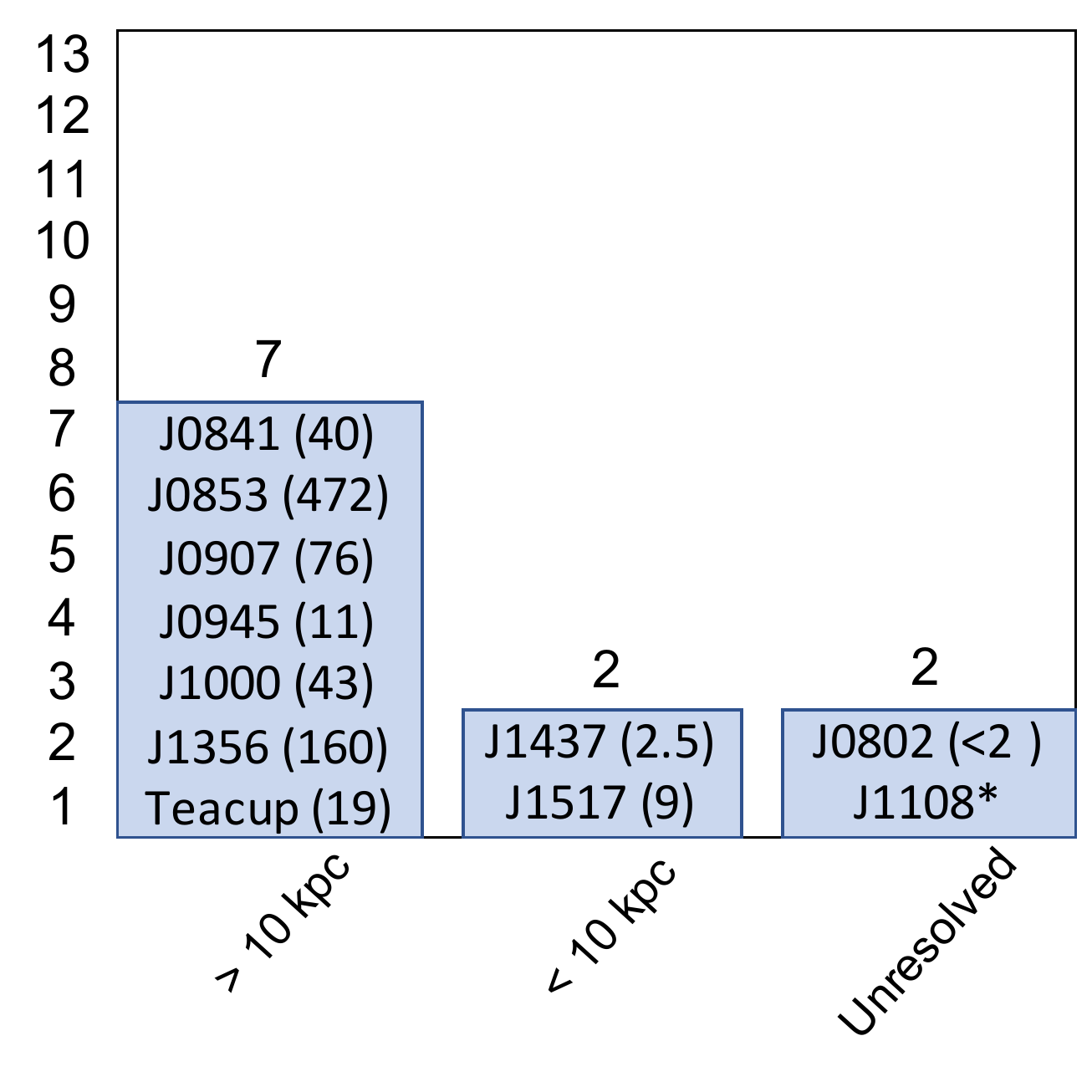}
\caption{Classification of the 11 QSO2 in our 13 QSO2 sample with confirmed AGN related radio emission according to the maximum size  of the AGN component (in brackets).  This coincides with $d_{\rm max}^{\rm R}$ (Table \ref{radprop}) except for J1108+06. It is highlighted with * because, although the radio source is extended for $d_{\rm max}^{\rm R}\sim$9 kpc, the AGN component is a compact core (\citealt{Bondi2016}).}
\label{hist-rad}
\end{figure}

  Out of these 11 QSO2,  the AGN radio component  is spatially unresolved in  J0802+25 and J1108+06 (see Appendix \ref{results-objects2}). The remaining 9  show extended  AGN related radio structures with sizes that range between $d_{\rm max}\sim$few kpc at the spatial resolution of our data and up to almost 500 kpc (Table \ref{radprop}, Fig. \ref{hist-rad}).   The relative contribution of these extended radio structures to the total radio luminosity at 1.4 GHz is in most cases in the range 30\% to 90\% (Table \ref{tab:results}).

 Structures reminiscent of, or confirmed to be jets are identified in all but J1437+33 (see also \citealt{Jarvis2019}).   The jet sizes are in the range 1-few kpc (J0945+17, J1000+12, J1356+10, the Teacup), $\sim$10 kpc (J1517+33), several tens  of kpc (J0841+01, J0853+38, J0907+46). Additional   extended  AGN related radio structures  (e.g.  lobes or radio bubbles, plumes, hot spots) are detected in J0853+38 (472 kpc), J0841+01 (40 kpc),  J1000+12 (43 kpc), J1356+10 (160 kpc), the Teacup (19 kpc). The radio component  at $\sim$11 kpc from the AGN in J0945+17 may be a jet or a lobe. 
The lobe emission from the 160 kpc radio source in J1356+10 is extremely faint, and additional observations are  necessary to unambiguously confirm and accurately image this emission.
The structure of the extended radio source in J1437+33 ($d_{\rm max}\sim$2.5 kpc) is unknown, but, as we argue in   Appendix  \ref{results-objects2}, it is most likely related to the nuclear activity.

Therefore, this  sample shows  a high fraction of AGN related radio sources that extend for at least typical effective radii of the spheroidal component of galaxies (\citealt{Urbano2019}) and often  much more.

\subsection{Size, line luminosity and nature of the large  scale ionised gas}
\label{results4p2}

\begin{table}[ht]
\centering
\footnotesize
\caption{Luminosities of the extended H$\alpha$ structures}
\label{halum}
\begin{tabular}{lccc}
\hline
Object &	$L_{\rm H\alpha}^{\rm ext}$  \\ 
		&	$\times$10$^{40}$ erg s$^{-1}$ \\ \hline
J0802+25	&	8.1$\pm$0.5 \\
J0841+01	&	43.5$\pm$4.2 \\
J0853+38	&	28.4$\pm$1.3 \\
J0907+46	&		17.4$\pm$0.9	\\
J0945+26 & 	207$\pm$49\\
J1000+12	&	203$\pm$4	\\ 
J1316+44	&	17.8$\pm$3.6 \\
J1356+10		&	235$\pm$9 \\
Teacup	&	139$\pm$9$^{\dagger}$\\
J1437+30	& 2.5$\pm$0.3\\  \hline
\end{tabular} 
\flushleft
$^{\dagger}$This value includes  also the contribution of the ionised bubbles. The nebula alone has  $L_{\rm H\alpha}\sim$(4.1$\pm$0.3)$\times$10$^{41}$ erg s$^{-1}$.\\
\end{table}

\begin{figure*}
\centering
\includegraphics[width=0.6\textwidth]{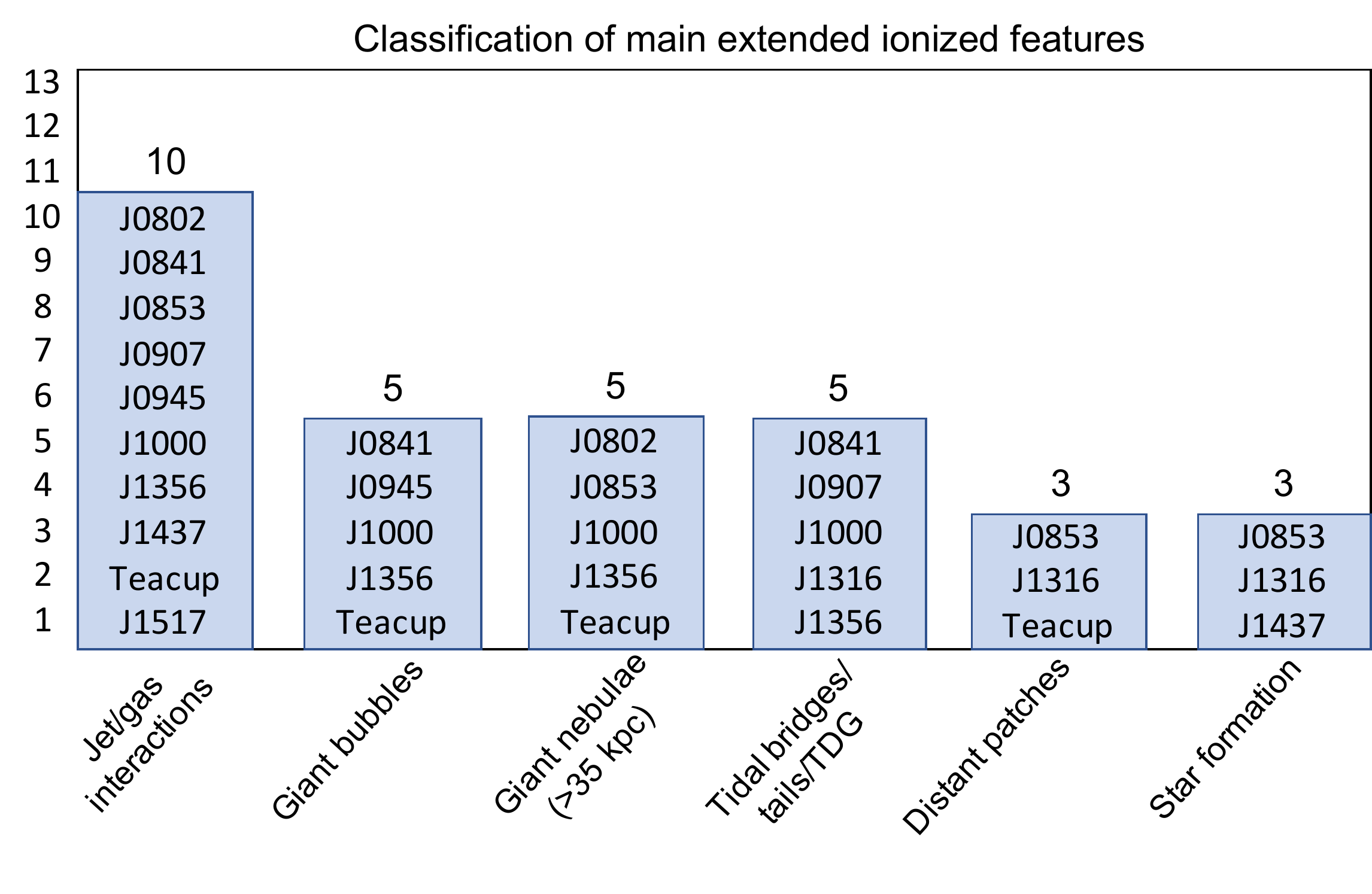}
\caption{The most interesting extended ionised features identified in the 13 QSO2 of our sample are classified in this histogram. 
"Giant nebulae" are large continuous reservoirs of gas with sizes $>$35 kpc not identified with bubbles, tidal features (bridges or tails), with star formation (e.g. spiral arms) or with detached features (e.g. distant patches). "Jet-gas interactions" refer to  emission line features that are closely  correlated with the radio morphology. This correlation indicates an interaction between the gas and the radio source. "Star formation" encompasses detached compact knots and SF in spiral arms.
"Distant patches" refer to detached patches of ionised gas located at large distances from the AGN (tens  of kpc).}
\label{hist-ion}
\end{figure*}

 We have found that QSO2 undergoing interaction/merger events appear to be invariably associated with ionised gas spread over large spatial scales, often  well outside the main body of the galaxies, with maximum distances from the AGN in the range $r_{\rm max}\sim$12-92  kpc, with a median value of 46 kpc.   The most spectacular case is the Teacup, which is associated with a $\sim$115$\times$87 kpc nebula (Fig. \ref{figTeacup}). The H$\alpha$ luminosities of the extended structures are in the range   $L_{\rm H\alpha}^{\rm ext}=$(2.5-235)$\times$10$^{40}$ erg s$^{-1}$ (median 3.5$\times$10$^{41}$ erg s$^{-1}$)  (Table \ref{halum})\footnote{Some contamination by [NII]$\lambda\lambda$6548,6583 is expected, although $<$30\%}. 
We show in Fig. \ref{hist-ion} the classification of the most interesting extended ionised features and the  objects where they have been identified.  For comparison, \cite{Balmaverde2021} measured   $L_{\rm H\alpha}^{\rm ext}$ for a sample of 37 3C low z radio galaxies. Six have nuclear [OIII] luminosities in the same range as our QSO2 sample. These have $L_{\rm H\alpha}^{\rm ext}=$(15-389)$\times$10$^{40}$ erg s$^{-1}$ (median 1.0$\times$10$^{42}$ erg s$^{-1}$). Although this comparison sample is small, the H$\alpha$ luminosities of the extended nebulae appear to be higher in the radio-loud sample.

 All QSO2  in our sample except J0907+46 show unambiguous evidence for mergers/interactions based on the distorted  optical continuum morphologies and the presence of prominent tidal continuum features. The H$\alpha$ image confirms this  is also the case for J0907+46 (Fig. \ref{fig0907a}). Many of the large scale ($\ga$10 kpc) ionised features identified in our sample are remnants of the interaction/process. In some cases, they trace entire or partial tidal tails, bridges or shells  (e.g.  J0841+01, J0907+46, J1000+12, J1316+44 and J1356+10;  see also \citealt{Villar2010,Villar2017}). 
 The extended ionised gas  frequently overlaps partially with wide large and amorphous scale continuum halos clearly associated with a merging or interaction event (e.g. J0802+25, J1437+30, J1356+10, the Teacup).

Five QSO2 in our sample present firm or tentative evidence for giant bubbles of ionised gas: J0841+01, J0945+17, J1000+06, J1356+10  and the Teacup. 
The sizes  $\sim$10$\pm$2 kpc are similar in all of them.   
 The Teacup (\citealt{Keel2012}), with clear morphological evidence, and J1356+10 and J1000+12, based on kinematic evidence, have been discussed  by other authors (\citealt{Greene2012,Jarvis2019}).

We have proposed that J0841+01 and J0945+17 are also associated with giant ionised bubbles.
We have identified a $\sim$10 kpc edge-brightened bubble on the W side of J0945+17 and, possibly, a counter-bubble to the E  (Appendix \ref{results-objects2},  Fig. \ref{fig0945}). The radio source is currently impacting  on the western bubble  and is enhancing the optical line emission at the location of the ``cloud'' identified by \cite{Storchi2018}. A past radio activity episode or a wide angle wind could have inflated the bubble(s).  J0841+01 is a dubious case. A striking set of ionised filaments is located between the two merging galaxies  (\citealt{Storchi2018}, Appendix \ref{results-objects2},  Fig. \ref{fig0841}).   Based on the morphology and filamentary appearance, we have proposed  that this is a giant ($\sim$8 kpc) ionised edge brightened bubble. The  apparent spatial overlap and similar size as   the radio source  (Fig. \ref{overlay0841}) suggests that it has been inflated by it. A wide angle wind cannot be discarded.   
Spectroscopy would be very useful to confirm or discard the nature of these  bubble candidates.

\begin{figure}
\centering
\includegraphics[width=0.45\textwidth]{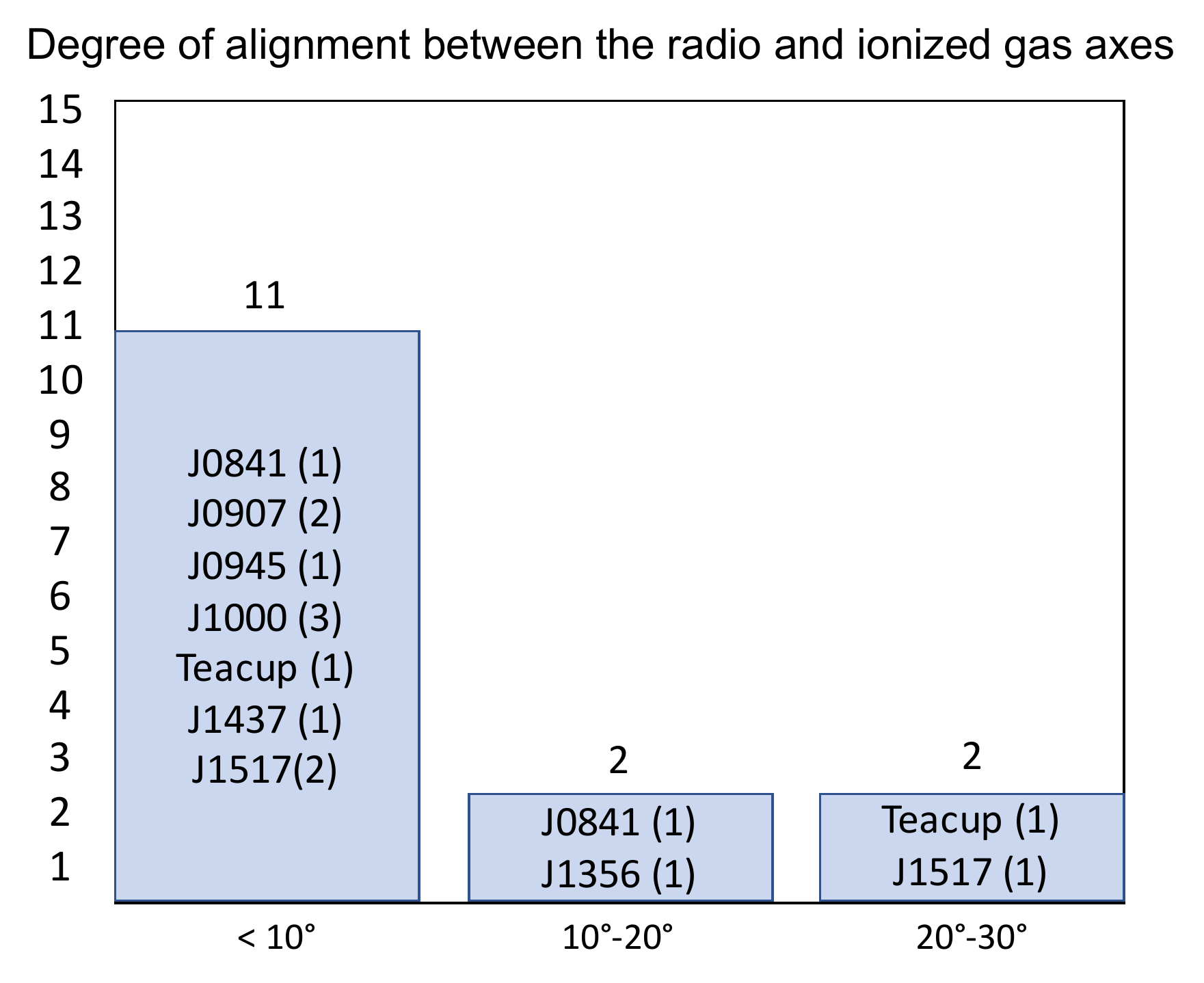}
\caption{Relative angles $\Delta$PA=$\lvert$PA$_{\rm rad}$-PA$_{\rm opt}\rvert$ between the radio and ionised gas axes  measured in the eight QSO2 indicated in this plot (see text). They have been grouped in three bins according to $\Delta\rm PA<10\degr$, $10\degr$$\le\Delta\rm PA<20\degr$ and $20\le\Delta\rm PA<30\degr$. For each object, the number of radio-H$\alpha$ angles in that specific bin is shown in brackets. The radio and gas axes are in general very closely aligned.  For seven objects (first bin), the axes are  identical or almost identical ($\Delta\rm PA<10\degr$) at least on one side of the AGN.}
\label{hist-alig}
\end{figure}

 A scenario in which a  radio jet ($\sim$few kpc) has inflated the bubbles has been also proposed for the Teacup (but see also \citealt{Zakamska2014}), J1000+06 and the southern bubble of J1356+10 (\citealt{Jarvis2019}).  

In addition to the ionised bubble candidates already reported in J1356+44 (\citealt{Greene2012,Jarvis2019}),  we have found  morphological   tentative evidence for  one more giant ($\sim$20 kpc) H$\alpha$ bubble in this system  (see Appendix \ref{results-objects2}). It may be part of the giant ($\sim$40 kpc) expanding structure  discussed by \cite{Greene2012}. The possibility that this feature is an ionised gas shell, remnant of the galactic interactions, cannot be discarded.

Tidal dwarf galaxy (TDG) candidates have been identified  as continuum and emission line compact objects at the tip of tidal tails in J0907+46, J1000+12,  J0841+01 and J1356+10. The formation of these recycled objects in the debris of merging/interacting systems has been observed frequently  and is predicted by simulations of colliding galaxies (e.g. \citealt{Bornaud2004,Lelli2015}).

\subsection{Radio-optical alignment}
\label{sec:alignment}

\begin{table}
\centering
\caption{Degree of alignment $\Delta$PA=$\lvert$PA$_{\rm rad}$-PA$_{\rm opt}\rvert$ between the H$\alpha$ and radio axes. Positive and negative PA values are measured N to E and N to W respectively.}
\label{tab-alig}
\begin{tabular}{lcccccc}
\hline
 Axis 1 &      \\ 
Source &  PA1$_{\rm rad}$	&	PA1$_{\rm opt}$	& $\Delta$PA1  \\
&	 ($\degr$)	& ($\degr$)	& ($\degr$) \\ \hline
J0841+01   & 77 	& 	77    &  0\\
J0907+46 &   29 &   29   & 0 \\
J0945+17 & -67  &   -75 & 8	 \\
J1000+12 &  10 &     10  &   0 \\ 
J1356+10  & -157  & -172  & 15  \\
 The Tecup &  65  & 65   & 0  \\
J1437+30 &  -98  &      -104   &   5        \\ 
J1517+33	&  117 & 109 & 8 \\
&  &      \\ 
  Axis 2 &      \\ 
Source &  PA2$_{\rm rad}$	&	PA2$_{\rm opt}$	&  $\Delta$PA2 \\ \hline
J0841+01 &    	-105 &		 	-118 &  13 \\
J0907+46 &   -151 &   -151  & 0 \\
J1000+12 &   150  & 157   & 7 \\ 
 The Tecup &  -99    &   -126  & 27 \\
J1517+33	&  -77 & -70 & 7 \\
 &  &      \\ 
 Axis 3 &      \\  
Source &  PA3$_{\rm rad}$	&	PA3$_{\rm opt}$	&  $\Delta$PA3 \\ \hline
J0907+46$^{\otimes}$ & -74    & N/A  &	N/A\\
J1000+12 &   -31   & -26  & 5 \\  
J1517+33	&  -92 & -112 & 30  \\ \hline
 \end{tabular} 
\flushleft {$^{\otimes}$J0907+46. PA3$_{\rm rad}$ refers to the axis that joins the radio core and the centroid of the distant, strongly bent jet/lobe (Fig. \ref{overlay0907}). H$\alpha$ emission has not been detected at similar distances and/or elongated at similar PA. Thus,  $\Delta$PA3 cannot be measured. PA3$_{\rm rad}$ is quoted for completeness.}
\end{table}

We have measured the degree of alignment between the ionised gas and the  radio structures. To do so, we determine the relative angle between    the radio and H$\alpha$ axes.  This study has been applied to the   QSO2 in our sample with spatially resolved AGN related radio structures and available emission line GTC and/or HST images (Table \ref{tab-alig}), except J0853+38. This object has been excluded due to the complexity of both the optical  and specially the radio morphologies, which results in the  identification of multiple axes. The Teacup radio axes  were measured in  \cite{Harrison2015}  published radio maps and the radio and ionised gas axes for J1517+33 were measured in Fig. 9 of \cite{Rosario2010}.

  To define the H$\alpha$ axes, we only consider the gas at $\la$30 kpc from the AGN.  The gas beyond  is in general associated with faint and distant  tidal features (tails, bridges) and  detached ionised patches, all located far from the  radio structures and likely unrelated to them. The companion  nucleus in J0841+01 has also  been excluded in the H$\alpha$  axis definition. 

 Both the ionised gas and radio morphologies are in general  asymmetric and complex and several axes can be identified in both cases. The main H$\alpha$ axes for objects with  well defined high surface brightness  features (e.g. bright   clouds, collimated nebulae)   were determined at both sides of the AGN by  lines that start at  the optical nucleus (expected approximate location of the AGN)  and run across such  features. The optical axes of objects with wide area, diffuse ionised gaseous structures (e.g. giant bubbles, amorphous halos, cones) were defined as the axes that start at the AGN location and bisect these regions.    One  axis only is identified if gas is detected on just one side of the optical nucleus.

To define the radio axes we followed a similar method.  Axes were identified at  one or both sides of the radio core, depending on the one or two sided radio morphology. All axes start at  the core  and extend up to  bright  compact radio features (e.g.  hot spots) or along structures clearly elongated in a preferential direction.  In the most complex cases, the radio source shows a preferential direction closer to the AGN and bends sharply further out (e.g. J0907+46 (Fig. \ref{overlay0907})  and  J1000+12 (Fig. \ref{overlay1000})).  This means that more than one  radio axis can be identified  at one side of the radio core.  

The  different radio and H$\alpha$ axes are shown in Figs. \ref{overlay0841} to \ref{overlay1437} and their PA  are in Table \ref{tab-alig}.  The relative angles $\lvert$PA$_{\rm rad}$-PA$_{\rm opt}\rvert$  at each side of the AGN are also shown in this table.

We have found that the radio and H$\alpha$ axes are very closely aligned  at both sides of the AGN in all objects (Table \ref{tab-alig} and Fig. \ref{hist-alig}).  Out of the 15 relative radio/optical angles measured,  11 form an angle $\lvert$PA$_{\rm rad}$-PA$_{\rm opt}\rvert<10\degr$. In the remaining  4 cases,  13$^{\degr}$$\le\lvert$PA$_{\rm rad}$-PA$_{\rm opt}\rvert\le$30$^{\degr}$, which also   imply   close alignments. 

Such close alignments show that the  H$\alpha$ morphologies at $<$30 kpc from the AGN and the radio morphologies are strongly shaped by  AGN related processes.

\subsection{Interactions between the radio source and the ambient gas}
\label{resultsInteractions}

9/13 QSO2 show evidence for interactions between the radio structures and the ambient gas on scales between $\sim$ few kpc and 10s of kpc depending on the object.  This conclusion is based both on our study and other authors work (see Appendix \ref{results-objects2} for all relevant references).

These 9 QSO2 show close morphological associations between emission line and radio features  which suggest  an interaction between the radio structures and the ionised gas on  scales between several kpc and 10s of kpc from the nucleus (see Appendix \ref{results-objects2} and Fig. \ref{overlay0841} to \ref{overlay1437}): J0841+01, J0853+38, J0907+46, J0945+17, J1000+12, the Teacup, J1356+10, J1437+30, J1517+33.   In these objects,  line emission  is enhanced near and/or at the location of radio features and sometimes tracing each other very closely. Spatially coincident signs of kinematic disturbance have been moreover identified   in J0945+17, J1000+12, J1356+10, the Teacup and J1517+33 (\citealt{Rosario2010,Ramos2017,Jarvis2019}).
 
Six of these objects show sharp radio bends at distances from several kpc up to 10s of kpc  from the nucleus: J0853+38, J0907+46, J0945+15, J1010+12, J1356+17 and J1517+33. This on its own is probably an indication of an interaction with the gaseous environment, which changes abruptly de direction of the radio source. This is supported by the fact that in all cases but J0907+45,  the bend is narrowly traced by the ionised gas morphology and/or occurs at locations where the gas presents  distinct kinematics that  deviates from the surrounding gaseous environment (\citealt{Rosario2010,Jarvis2019}). 

 Evidence for radio-gas interactions on scales $\la$2 kpc (equivalent to the radial size of the SDSS fibre) has been also reported in J0802+25, whose  radio source is unresolved in our maps. Such evidence was based on the extreme kinematics and relatively strong line emission of the nuclear ionised outflow (\citealt{Villar2014}).

 Evidence for radio-gas interactions in   J0948+25,  J1108+06 and J1316+44 cannot be confirmed or discarded.

\section{Discussion}
\label{discussion}

\subsection{Origin, properties and frequency of the radio sources}
\label{discussion5p1}

As explained in Sect. \ref{intro} the frequency of jets and related structures (e.g. hot spots, lobes) in QSO2 and, more generally, non radio loud quasars (RQ and RI)  is unknown. 

The $z<$0.2 QSO2 sample studied here shows a high frequency of objects with spatially extended radio structures which are reminiscent of or, confirmed to be, AGN related. The  sizes    vary between 1-few kpc and up to almost 500 kpc (Sect. \ref{results4p1}). Most extend well beyond the typical sizes of the spheroidal component of galaxies, sometimes well within the circumgalactic medium (CGM)\footnote{Following  \cite{Tumlinson2017}, we consider the CGM as the gas surrounding galaxies outside their disks or ISM and inside their virial radii}.

A high frequency for such structures was also found by  \cite{Jarvis2019} in their sample of 10 SDSS QSO2 at $z<$0.2 (3 overlap with our sample).  Most  exhibit extended radio structures on 1–25 kpc scales that are consistent with being radio jets. They span log($L_{\rm [OIII]}$/erg s$^{-1}$)=42.25-43.13 (log of the median 42.63)  and log($P_{\rm 1.4}$)=30.3-31.4 (median 31.1). The objects were selected to host nuclear ionised outflows (FWHM$_{\rm [OIII]}\sim$800-1800 km s$^{-1}$). This is not likely to introduce a severe bias, given the high frequency of ionised outflows in QSO2 (Sect. \ref{intro})  which moreover often show FWHM in this range   (e.g. \citealt{Villar2011,Mullaney2013,Harrison2014,Villar2016}). On the other hand, we have found that the detection of large scale radio sources appears to be  independent of the evidence and properties of  nuclear ionised outflows (e.g. J0841+01 and J0853+38).

The detection of large scale ($\ga$10 kpc) AGN radio structures in RQQ   is not frequent   (\citealt{Kellermann1994,Lal2010,Zakamska2004,Villar2017}),  but observations have been often limited in sensitivity and/or resolution. The Beetle is an interesting object on this regard. While prior FIRST and NVSS radio information suggested that the radio emission is unresolved and consistent with star formation, deep VLA observations showed that   it is associated with a $\sim$4 kpc jet and a  large scale ($\sim$46  kpc) radio source (\citealt{Villar2017}) which is interacting with the galactic and circumgalactic environments across tens of kpc.   Another interesting case is J0853+38.   Our new observations have revealed a large  and  highly asymmetric radio source of total extension $\sim$3.5' or $\sim$472 kpc  (Sect. \ref{results-objects2}).

It is not clear how these results fit in the general population of QSO2 since  our sample and that of \cite{Jarvis2019}  are affected by different sources of bias. They have relatively high radio luminosities, which   favours the detection of AGN radio sources, compact or extended (see Sect. \ref{intro} and \ref{sample}).  In more than half of our sample there was prior solid or tentative  evidence for the existence of extended radio sources. It is necessary to expand these studies to QSO2 samples of lower radio luminosities (\citealt{Kellermann2016}).
 We  also selected objects with a rich and wide spread CGM which may increase the probability of detecting large scale radio sources since the interaction with the ambient gas can enhance the radio emission, although, given the high incidence of merger/interaction signatures in QSO2 (70-75\%; \citealt{Bessiere2012,Urbano2019}) this criterion may not introduce a very restrictive bias.

\subsection{Jet gas interactions}

There is evidence  for radio-gas interactions in 10 of the 13 $z<$0.2 QSO2. In general, this is suggested by the close correlation between the radio and ionised gas morphologies  (Sect. \ref{resultsInteractions}) and is supported in some cases by the associated turbulent gas kinematics described by other authors.  Higher resolutions radio maps (e.g. J0802+25, J1437+30) and, most importantly, 2-dim spectroscopy would be very valuable to  search for unambiguous signatures in some cases (e.g. correlations between  the kinematic, physical and ionisation properties of the ionised gas and the radio morphology; \citealt{Clark1998,Villar2017}).

 The radio-gas interactions have been identified across different spatial scales, from the inner NLR close to the AGN ($\la$1-2 kpc) up to tens  of kpc, well outside the main body of the galaxies. The most extreme object on this regard is J0853+38, where the effect of the interaction is noticed across $\sim$58 kpc and up to at least 36 kpc from the AGN (Appendix \ref{results-objects2}).  Large scale radio-gas interactions occur in objects with and without prominent nuclear outflows (e.g. J0841+01, J0853+38 do not host  strong nuclear outflows).  They can occur also in objects where the relative contribution of the radio to IR luminosity (i.e. $q$) is  consistent with star formation (e.g. the Beetle; see Sect. \ref{radprop}). 

 Therefore, these radio sources provide a feedback mechanism that can act across different spatial scales from the nuclear region, and up to tens, even hundreds of kpc, across the ISM and well within the CGM (see also \citealt{Villar2017,Jarvis2019}). 

The effects of the  interactions are obvious both in the ambient gas and sometimes  in the radio source, in the form of strong morphological distortions. Sharp deflections have  been identified in  J0853+39, J0907+46 and J1000+12 and possibly also J0841+01. These are the four largest   radio sources ($\ge$40 kpc) in our sample.  Therefore,  when the radio sources succeed at escaping out to large distances from the AGN, they appear deflected. In our sample, this can be a consequence of two  things:  on one hand, the   widely spread gas is an obstacle for the advance of the radio source.   
   On the other, due to their modest radio power in comparison with powerful radio galaxies and radio loud quasars the radio sources can be more easily deflected. An implication is that  they can affect a larger volume and remain trapped for a longer time  (\citealt{Mukherjee2016}).

Five QSO2 in our sample present firm or tentative evidence for giant bubbles of ionised gas (Sect. \ref{results4p2}): J0841+01, J0945+17, J1356+10, J1000+06 and the Teacup. The sizes are  similar $\sim$10$\pm$2 kpc  in all of them. 
In addition, a $\sim$20 kpc bubble candidate has been identified in J1356+10.
One of the proposed mechanisms to inflate giant bubbles in galaxies are relativistic jets, both in active and non active galaxies, including the Milky Way (\citealt{Zhang2020}). This mechanism has been proposed by \cite{Jarvis2019} for  J1356+10, J1000+06 and the Teacup (see also \citealt{Lansbury2018}).  Based on the morphological correlation with the radio emission, this may also the case in J0841+01 and J0945+17 although this cannot be confirmed with the available data  (Appendix \ref{results-objects2}).

\subsection{Origin of the large scale ionised gas}
\label{discussion5p2}

 We have found that QSO2 in interaction/merger systems are in general   associated with ionised gas spread over tens  of kpc   well outside the main body of the galaxies (Sect. \ref{results4p2}). A high fraction of these large scale ionised features   are a consequence of  such interaction processes, since they are morphologically related to tidal remnants.  Given the high incidence of merger/interaction signatures in QSO2 ($\sim$70-75\%; \citealt{Bessiere2012,Humphrey2015,Urbano2019}),  this kind of features are expected to be very frequent in QSO2 in general.

These results  are consistent with   \cite{Stockton1987}, who found a correspondence between the incidence of strong extended ionised nebulae among  QSO1 and the presence of overt signs of strong galactic interactions, such as close companion galaxies and continuum tails or bridges. This is also the case of powerful strong line radio galaxies.
Recent results based on wide field integral field spectroscopy of 3C radio galaxies at $z<$0.3 have  revealed the widespread presence of filamentary ionised structures extending for tens of kpc (\citealt{Balmaverde2018}). The authors propose that they are likely the remnants of the gas rich mergers which triggered the AGN. Different works show that indeed these type of systems tend to show strong evidence for mergers/interactions (\citealt{Ramos2017,Tadhunter2016}).

Galactic interactions and mergers can spread the gaseous content of galaxies across huge volumes in the CGM. These remnants, which  are in general  detected in HI,  can be rendered visible as a consequence of  the nuclear activity.  The intense and hard radiation from the QSO can ionise the gas up to  many tens  of kpc.   Shocks driven during the interaction with large scale radio structures can also contribute (\citealt{Dopita1995,Villar2017}).  

 The close alignment between the axes of the ionised gas at $<$30 kpc from the AGN and of the extended radio structures  shows that the  H$\alpha$ morphology on these spatial scales and the radio morphology are strongly shaped by  AGN related processes (Sect. \ref{sec:alignment}).  
Other ionisation mechanisms unrelated to the nuclear activity must also be at work, since some of the large scale ionised features lie well outside  the QSO2 ionisation cones  (e.g. J0802+25, J0841+01, J0853+38, J1316+44, J1356+10, J1437+30, the Teacup; see also \citealt{Balmaverde2018}).  The cone  apertures can be approximately guessed by assuming the  axis is roughly aligned with the radio axis and a half opening angle of $\sim$45$\degr$. Star formation, shocks induced by ram pressure, heat conduction or magneto hydrodynamic waves are different possibilities (e.g. \citealt{Boselli2016}).

\subsection{The limiting size of the NLR}
\label{discussion5p3}

\cite{Bennert2002} found that the size of the NLR in Seyfert galaxies and RQ QSO1  is proportional to the square root of $L_{\rm [OIII]}$. This in turn is a proxy of AGN luminosity (\citealt{Heckman2004}).   Subsequent studies of SDSS QSO2   confirm this dependence  and suggest that there is a maximal radial size of the NLR of $R_{\rm NLR}\sim$6-8 kpc beyond which  the relation flattens (\citealt{Hainline2014}). The authors estimate this size  using a cosmology-independent measurement, which is the radius where the $SB$ corrected for cosmological dimming falls to $SB_{\rm max}=$10$^{-15}$ erg s$^{-1}$ cm$^{-2}$ arcsec$^{-2}$.  They propose that beyond this distance there is either not enough gas or the gas is over-ionised and does not produce detectable [O III]$\lambda$5007 emission.   Other works suggest instead a  smooth  continuation  of  the  size-luminosity relation out to large radii or a much larger break radius as previously proposed (e.g. \citealt{Fischer2018,Husemann2019}).

Different works, including our current study   demonstrate that much larger (tens  of kpc) reservoirs of ionised gas are frequently associated with  AGN  in the quasar regime.  This includes radio loud and radio quiet, type 1 and type 2 quasars and strong line powerful  radiogalaxies (Sect. \ref{discussion5p2}; see also \citealt{Villar2018}). These features often have  very low $SB$ $\sim$few$\times$10$^{-17}$ erg s$^{-1}$ cm$^{-2}$ arcsec$^{-2}$. At the median redshift of the sample studied by \cite{Hainline2014} ($z=$0.56),    such features would have SB$\sim$10$^{-17}$ erg s$^{-1}$ cm$^{-2}$ arcsec$^{-2}$ taking into account cosmological dimming, which is $\ll SB_{\rm max}$. 

 The high $SB$ gas within $R_{\rm NLR}$ most likely traces gas that has been photoionised by the AGN. Thus, in order to put our results in the context of these works, it would be necessary to determine the ionising mechanism of the large scale gas and measure the maximum distance at which  photoionisation by the QSO dominates the gas excitation. What is clear  is that large scale gas well beyond $R_{\rm NLR}$ is abundant   for many QSO2 (quasars in general), often dispersed in the CGM by merger/interactions (see also \citealt{Storchi2018}). 
On the other hand, gas photoionised by the quasar has been detected at tens  of kpc from the AGN in several objects (e.g. \citealt{Shopbell1999,Villar2010,Villar2018}). 

 It is  clear that mapping gas with $SB\ll$10$^{-15}$ erg s$^{-1}$ cm$^{-2}$ arcsec$^{-2}$ is essential to trace its true spatial distribution.

\section{Summary and conclusions}
\label{conclusions}

 We have investigated the incidence of radio induced feedback  in a sample of 13 optically selected SDSS QSO2 at $z<$0.2. None are radio-loud. For this, we have searched for signs of radio-gas interactions by characterising  and comparing the morphologies of the ionised gas and the radio structures.  The first is traced by narrow band H$\alpha$ images obtained with the GTC 10.4m telescope and the Osiris instrument. The second is traced by VLA radio maps obtained with A and B configurations to achieve both high resolution to study the galaxy environment, as well as brightness sensitivity for imaging radio sources on tens to hundreds of kpc scales.

The main conclusions are:

$\bullet$  The radio luminosity has an AGN component in 11/13 QSO2. Such component is spatially extended in   9 of them  (jets/lobes/bubbles/other). Their relative contribution  to the total radio luminosity at 1.4 GHz is in most cases in the range 30\% to 90\%. The maximum sizes are in the range $d_{\rm max}^{\rm R}\sim$few kpc to  $\sim$500 kpc.   Of special interest is the  tentative detection of a  160 kpc radio source associated with the well known radio-quiet  J1356+10 QSO2. In most cases, the radio source extends beyond the typical size of the spheroidal component of  the galaxy and often well into the circumgalactic medium.

$\bullet$  QSO2 undergoing  galaxy interaction/merger events appear to be invariably associated with ionised gas spread over large spatial scales, often  well outside the main body of the galaxies.  Given the high incidence of merger/interaction signatures in QSO2,  this is expected to be frequent in QSO2 in general.  The maximum distances from the AGN are  in the range $r_{\rm max}\sim$12-90  kpc, with a median value of 46 kpc.  The line luminosities are in the range $L_{\rm H\alpha}=(2.5-235)\times 10^{40}$ erg s$^{-1}$. 
 The nature is diverse including giant ionised bubbles, tidal features, giant ($\ga$35 kpc) nebulae, tidal dwarf galaxy candidates and distant patches at tens  of kpc from the AGN. 

$\bullet$ The axis of the ionised gas at $<$30 kpc from the AGN and the axis of the extended radio structures are in general closely aligned at both sides of the AGN in the 8 objects with available  measurements.  Seven objects show radio/H$\alpha$ relative angles   $<$10$^{\degr}$ at least on one side of the AGN.
  This shows that the  H$\alpha$ morphology on these spatial scales and the radio morphology are strongly shaped by  AGN related processes.

$\bullet$ The detection of ionised gas at tens  of kpc from the AGN in most of our sources  questions the existence of a maximum size for the NLR $\sim$6-8 kpc. Mapping the ionised gas at surface brightness $\ll$10$^{-15}$ erg s$^{-1}$  cm$^{-2}$ arcsec$^{-2}$ is essential to characterise the true ionised gas distribution.

$\bullet$ Evidence for radio-gas interactions has been found in 10/13  QSO2, that is, all but one of the QSO2 with confirmed AGN radio components. This conclusion is based both in our study   and other authors. Therefore, when AGN radio sources are present, they tend to  interact with the gaseous  environment. The interactions occur across different spatial scales, from the scale of the nuclear NLR   up to tens  of kpc, well outside the main body of the galaxies.   Thus, these radio sources provide a feedback mechanism that can act across different spatial scales from the nuclear region, and up to tens, even hundreds of kpc, across the ISM and well within the CGM (see also \citealt{Villar2017,Jarvis2019}). 

$\bullet$ Large scale ($>$10  kpc) radio-gas interactions occur in objects with and without prominent nuclear ionised outflows.

$\bullet$ When the radio sources succeed at escaping out to large distances from the AGN ($\ga$several kpc), they are frequently deflected. Therefore, the effects of the  interactions are sometimes also obvious  in the radio sources.    The  obstacle posed by the widely spread gas and the modest radio power in comparison with powerful radio galaxies and radio loud quasars favour the deflection of  the radio sources.

Although this sample cannot be considered representative of the general population of QSO2 it reinforces the idea that low/modest power large spatial scale radio sources can exist in radio-quiet QSO2, and provide a source of feedback  on scales   of the spheroidal component of  galaxies and much beyond well into the circumgalactic medium, in  systems where radiative mode feedback is expected to dominate.

High resolution, high sensitivity radio observations
 and 2-dim wide-field integral field spectroscopy of larger QSO2 samples, specially with lower radio luminosities, will be very valuable to determine
how frequent jets and related structures
are in the general population of non-radio loud quasars,  to constrain
their sizes and morphologies and evaluate their impact on the  surrounding environment.  Ultimately, this will provide a deeper knowledge on the role of radio mode feedback in quasars in general.

\section*{Acknowledgments}

Based on observations made with the Gran Telescopio Canarias (GTC), installed in the Spanish Observatorio del Roque de los Muchachos of the Instituto de Astrofísica de Canarias, in the island of La Palma. Based also on data obtained with the Karl G. Jansky Very Large Array (VLA).  The National Radio Astronomy Observatory is a facility of the National Science Foundation operated under cooperative agreement by Associated Universities, Inc.  We have also used Hubble Space Telescope archive data.
 We  thank the GTC staff for their support with the optical observations and the VLA staff for executing the radio observations. Thanks to Pieter van Dokkum and Philip Cigan for their useful advice on the optimum removal of cosmic rays of the HST images using the iraf task lacos$\_$im.cl (\citealt{VanDokkum2001}). 

MVM, AC, AAH and EB acknowledge support from  grants AYA2016-79724-C4-1-P  and/or  PGC2018-094671-BI00 (MCIU/AEI/FEDER,UE). E. B. acknowledges the support from Comunidad de Madrid through the Atracci\'on de Talento grant 2017-T1/TIC-5213. MVM, EB and AAH  work was done under project No. MDM-2017-0737 Unidad de Excelencia “Mar\'\i a de Maeztu” Centro de Astrobiolog\'\i a (CSIC-INTA).  
AH was supported by Fundação para a Ciência e a Tecnologia (FCT) through the research grants UIDB/04434/2020 and UIDP/04434/2020, and an FCT-CAPES Transnational Cooperation Project.  The National Radio Astronomy Observatory is a facility of the National Science Foundation operated under cooperative agreement by Associated Universities, Inc.

 This research has made use of: 1) the VizieR catalogue access tool, CDS,
 Strasbourg, France. The original description of the VizieR service was
 published in \cite{Ochsenbein2000};   2) data from Sloan Digital Sky Survey. Funding for the SDSS and SDSS-II has been provided by the Alfred P. Sloan Foundation, the Participating Institutions, the National Science Foundation, the U.S. Department of Energy, the National Aeronautics and Space Administration, the Japanese Monbukagakusho, the Max Planck Society, and the Higher Education Funding Council for England. The SDSS Web Site is http://www.sdss.org/; 3) the Cosmology calculator by \cite{Wright2006};
4) the NASA/IPAC Extragalactic Database (NED), which is operated by the Jet Propulsion Laboratory, California Institute of Technology, under contract with the National Aeronautics and Space Administration.

\begin{appendix}

\section{Results on individual objects}
\label{results-objects2}

\begin{figure*}[ht]
\centering
\includegraphics[width=0.75\textwidth]{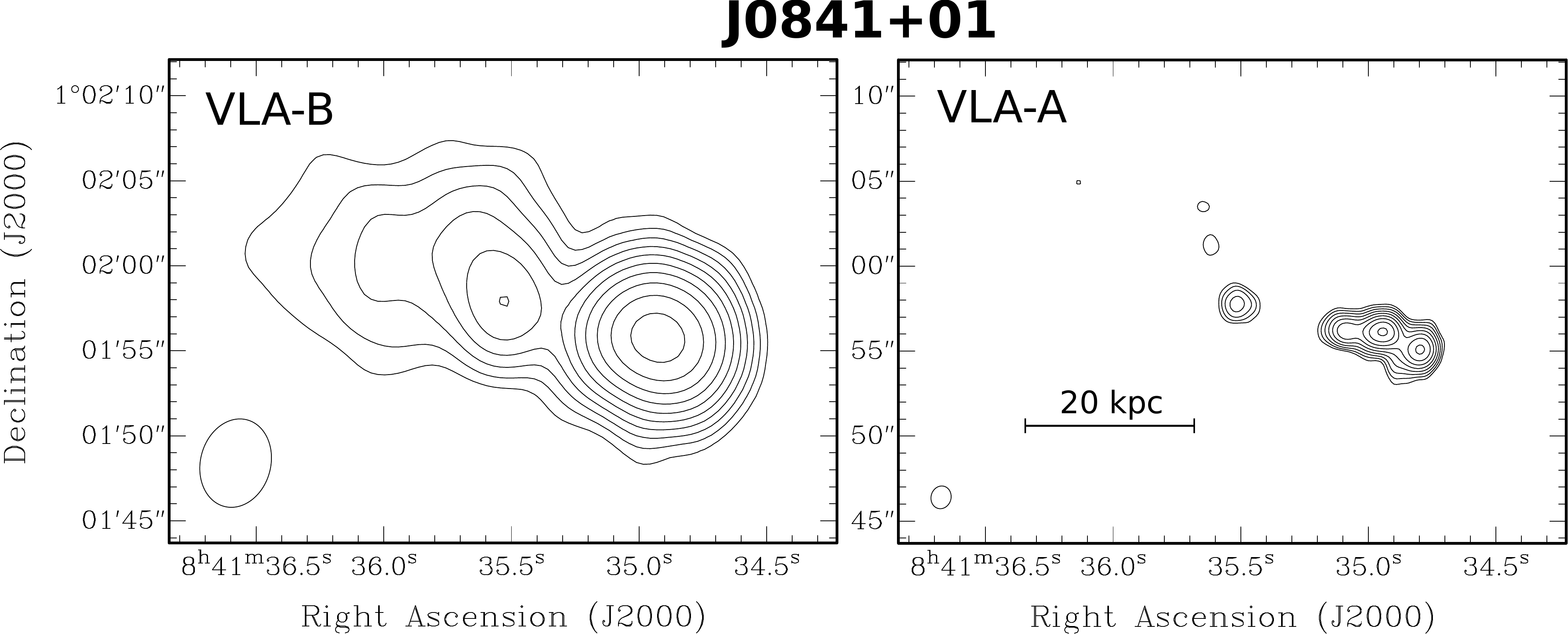}
\caption{VLA radio maps of J0841+01. Left: B-configuration map. Contour levels start at 100\, $\mu$Jy\,beam$^{-1}$ and increase by factor $\sqrt{2}$. Right: A-configuration map. Contour levels starts at 67 $\mu$\,Jy\,beam$^{-1}$ and increase by  factor $\sqrt{2}$. The synthesised beam is shown in the bottom-left corner of the maps.}
\label{VLA0841}
\centering
\includegraphics[width=0.8\textwidth]{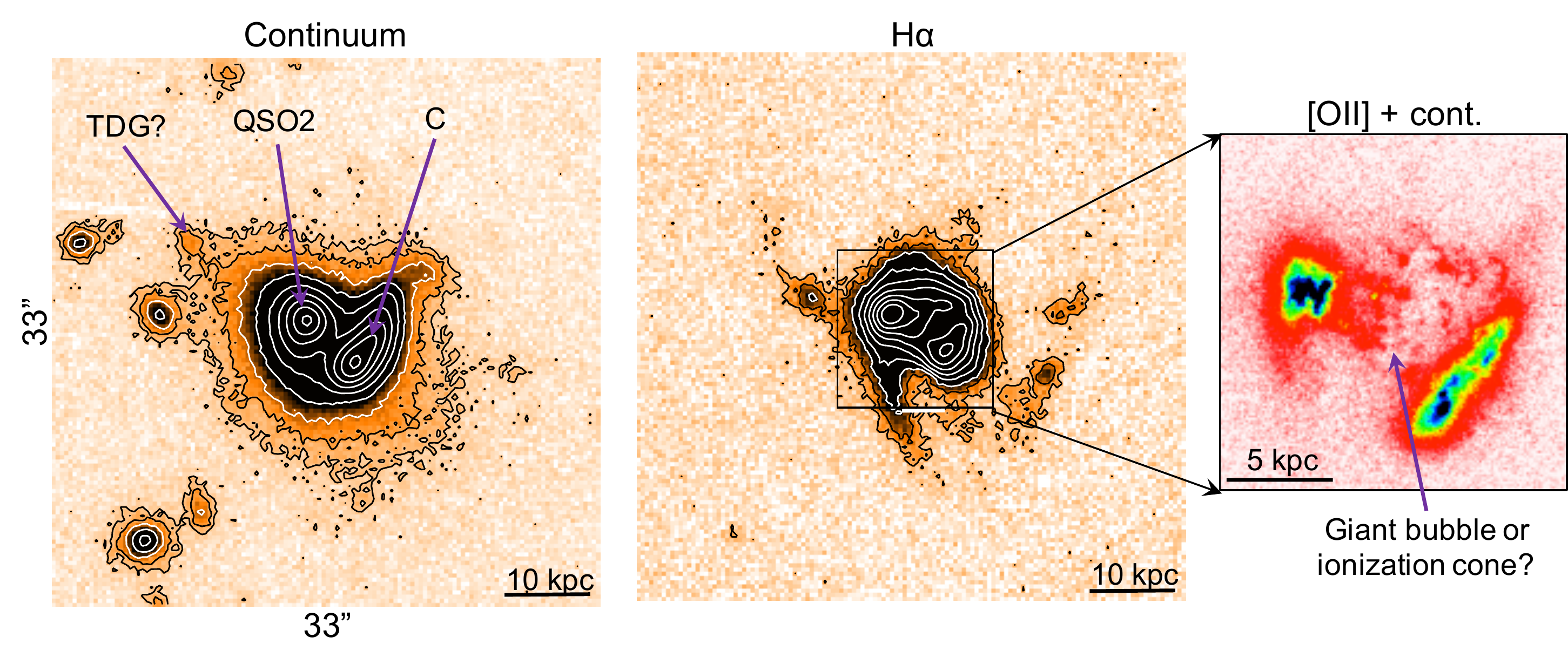}
\caption{J0841+01. GTC continuum (left) and H$\alpha$ images (middle). Contour values start at 3$\sigma$ and increase with factor $\times$2. For the H$\alpha$ image, $\sigma$=4.1$\times$10$^{-19}$ erg$^{-1}$ cm$^{-2}$ pixel$^{-1}$. QSO2 and C  indicate the QSO2 and the companion galaxy. TDG is a tidal dwarf galaxy candidate (see text). The small right  panel zooms in the central region of the HST WFC3/UVIS F438W image, which  contains the [OII]$\lambda$3727 doublet.  The intricate filamentary structures between the two galaxies are dominated by gas emission.   They are reminiscent of a giant ionised bubble. In this and all figures the H$\alpha$ images have been continuum subtracted. N is always up and E is left.}
\label{fig0841}
\end{figure*}

We present here detailed descriptions  and results for the  QSO2 in our VLA sample  and the Teacup, grouped by radio source physical size.
The main results  regarding the sizes and nature of the radio structures and the large scale ionised gas for all objects are summarised in Tables \ref{radprop} and \ref{optprop} respectively. They are complemented  with results from other studies in the literature.

\subsection{Large scale radio sources ($>$10 kpc from the AGN)}
\label{appendix-large}

By ``large'' we mean extended $>$10 kpc from the AGN, which   is significantly larger than the typical effective radius of the spheroidal component of QSO2 hosts (\citealt{Urbano2019}). Thus, these sources   may affect volumes as large as galaxy spheroids and beyond.

\begin{figure*}
\centering
\includegraphics[width=0.9\textwidth]{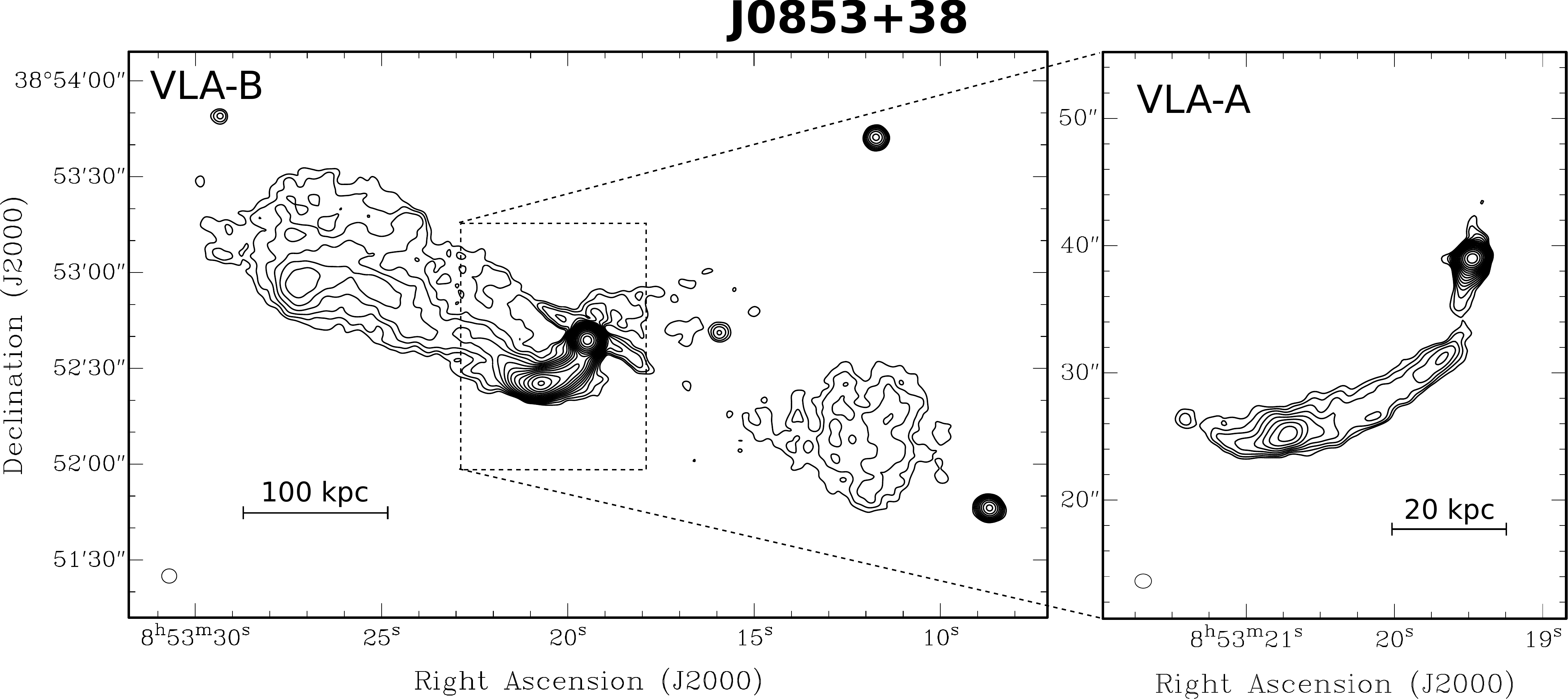}
\caption{VLA radio maps of J0853+38. Same as Fig.\,\ref{VLA0841}. Contour levels start at 60 $\mu$Jy\,beam$^{-1}$ (B-configuration) and 100 \,$\mu$Jy\,beam$^{-1}$ (A-configuration) and increase with factor $\sqrt{2}$. Note that in the B-configuration map, the prominent streak that crosses the core from NE to SW across $\sim$3$^{\prime\prime}$ aligns with a spike in the Point-Spread-Function (PSF), and is thus likely an artefact of the snap-shot observations.}
\label{VLA0853}
\centering
\includegraphics[width=0.75\textwidth]{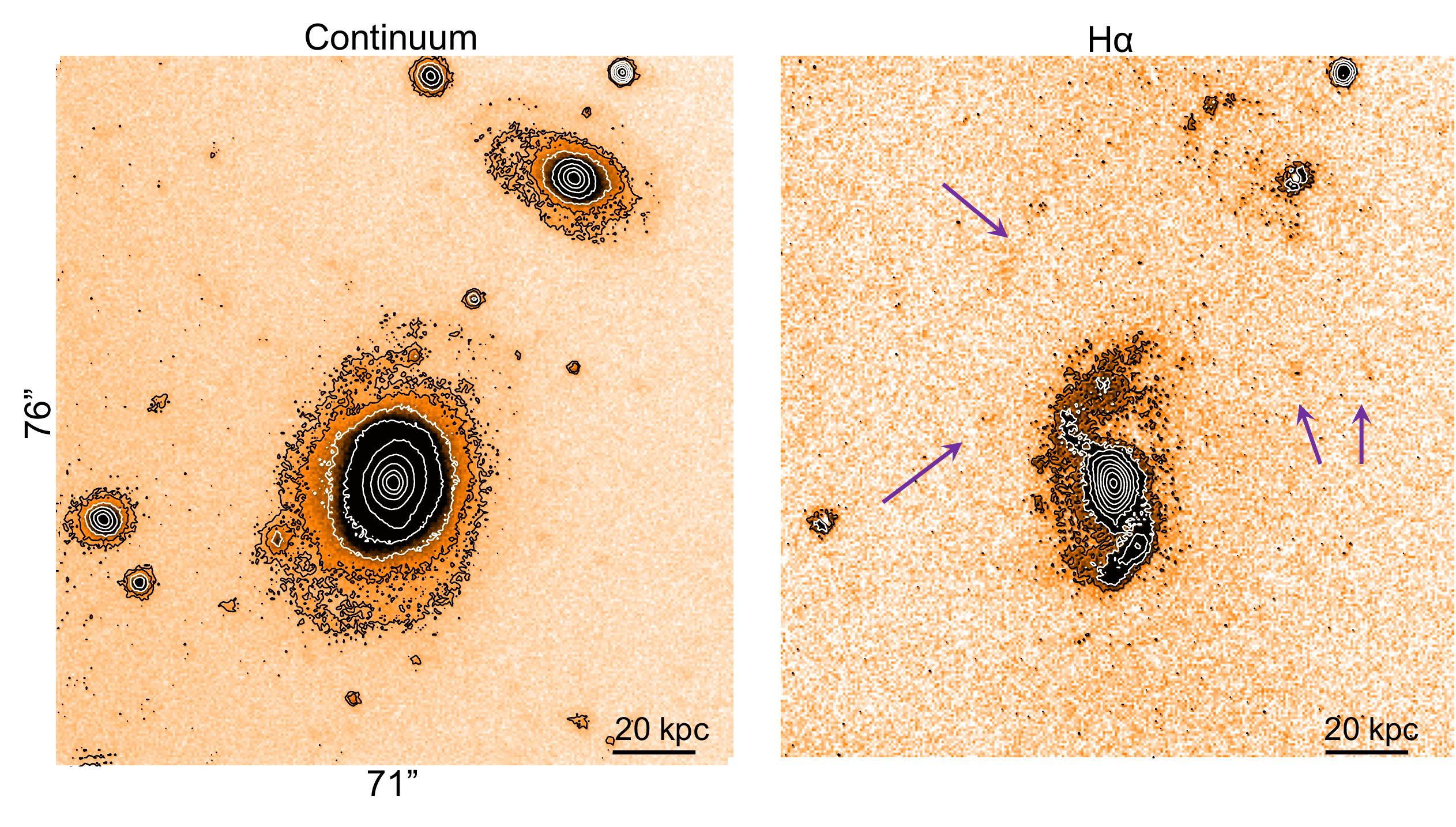}
\caption{J0853+38. GTC continuum and H$\alpha$ images. Contour values start at 3$\sigma$ and increase with factor $\times$2. For the H$\alpha$ image, $\sigma$=3.4$\times$10$^{-19}$ erg$^{-1}$ cm$^{-2}$ pixel$^{-1}$. The purple arrows  in the right panel mark distant patches of H$\alpha$ emission.}
\label{fig0853}
\end{figure*}

 $\bullet$ J0841+01 ($z$=0.110)  

This dual  or offset AGN  system  consists  of two  galaxies  with distorted morphologies undergoing a major merger. They are separated by $\sim$4$\arcsec$ or 8 kpc  (e.g. \citealt{Greene2011,Comerford2015}).

 The GTC continuum  and H$\alpha$  images are presented in Fig. \ref{fig0841}.  The  interacting galaxies are clearly seen in both. The HST image obtained with the WFC3/UVIS and the F438W filter is also shown (central $\lambda_0$=4320 \AA~  and FWHM=695 \AA. Program 12754; PI J. Comerford). It includes the redshifted [OII]$\lambda$3727, which dominates the emission from the filamentary structures between the galaxies.  The HST spatial resolution is essential to appreciate the shape and the morphological substructure  of this gas.  \cite{Storchi2018} proposed that this ionised gas delineates the ionisation cone  of the QSO2 hosted by the eastern galaxy.  Based on the edge-brightened appearance and the curved edges, we suggest   that this gas  traces a  giant ($\sim$8 kpc) ionised bubble inflated by an outflow driven by the  QSO2.

The radio source of the RQ J0841+01 (Fig. \ref{xu}) has a maximum extension of $d^{\rm rad}_{\rm max}\sim$20$\arcsec$ or $\sim$40 kpc  (Fig. \ref{VLA0841}). We show in Fig. \ref{overlay0841}  the overlay between the low (B configuration, black contours) and high (A configuration, red contours) VLA radio maps with the H$\alpha$ GTC and the  [OIII]$\lambda$5007 continuum subtracted  HST/ACS FR551N image from \cite{Storchi2018}  (program 13741; PI: T. Storchi-Bergmann).   Three compact radio components are identified: a radio core is associated with the QSO2 nucleus, a brighter radio knot is located between the two galaxies,  overlapping with the intricate filamentary gas, and  a third, outer  knot  overlaps with the northern side of the companion galaxy and is spatially shifted $\sim$2$\arcsec$ relative to its optical nucleus.   This suggests that this third radio feature  represents a hot-spot or lobe at the end of a  jet associated with the  QSO2.

The middle radio component  overlaps with the  giant  bubble candidate.  Soft X-ray extended emission spatially coincident with this warm ionised gas has been  detected  (\citealt{Comerford2015}).  This is     reminiscent of the ~10 kpc radio, X-ray and optical emission line bubble associated with the Teacup (\citealt{Keel2012,Harrison2015,Lansbury2018}). In this QSO2 it is not clear whether the bubble has been inflated by a wide angle AGN driven wind or by the relativistic radio source (see also \citealt{Jarvis2019}). The second scenario is favoured in J0841+01 by the similar size of the bubble and the triple radio source.
The bubble could be ionised by the QSO2 (\citealt{Storchi2018}) and/or  by shocks induced by the outflow. The hot shocked gas may also produce  the  soft X rays. 
 Although projection effects cannot be discarded, the bubble seems to  reach the western  companion galaxy which may thus be also affected by the giant outflow.  Spatially resolved spectroscopy would  be essential to investigate this scenario further.

The large GTC collecting area  allows to detect  additional faint  ionised gas spread over a large area  beyond the main bodies of both galaxies (Fig. \ref{fig0841}, middle panel).  The existence of  ionised gas all over the place, including locations well outside the putative AGN ionisation cones, suggests that local excitation mechanisms unrelated to the AGN are  at work at some positions.

H$\alpha$ emission is detected towards the NE up to a maximum distance   $r_{\rm max}\sim$21 kpc from the QSO2   and up to $\sim$14 kpc  to the SW of the companion nucleus.  The maximum distance  between the detected ionised gas features is $d_{\rm max}\sim$42 kpc. These  structures  are reminiscent of tidal features.        Most show no or very faint identifiable continuum counterpart at the sensitivity of the GTC image.   The H$\alpha$ knot at the tip of Eastern tail is an exception. The spatial location, the compact morphology and the relatively strong continuum emission suggest that it is a tidal dwarf galaxy (TDG; e.g. \citealt{Bornaud2004}).  

An alternative scenario is also possible. Several  radio knots seem to track the eastern H$\alpha$ tail (Fig. \ref{fig0841}).  The advancing radio jet may be interacting with the ambient gas along the tail possibly enhancing its H$\alpha$ emission.  At the same time, the jet is being deflected by the gas. 

No evidence for an ionised outflow has been identified in this object (see also \citealt{Greene2011}). The non detection of the outflow may be a consequence of the dilution due to the large size of the 3$\arcsec$ SDSS fibre.

\begin{figure}
\centering
\includegraphics[width=0.36\textwidth]{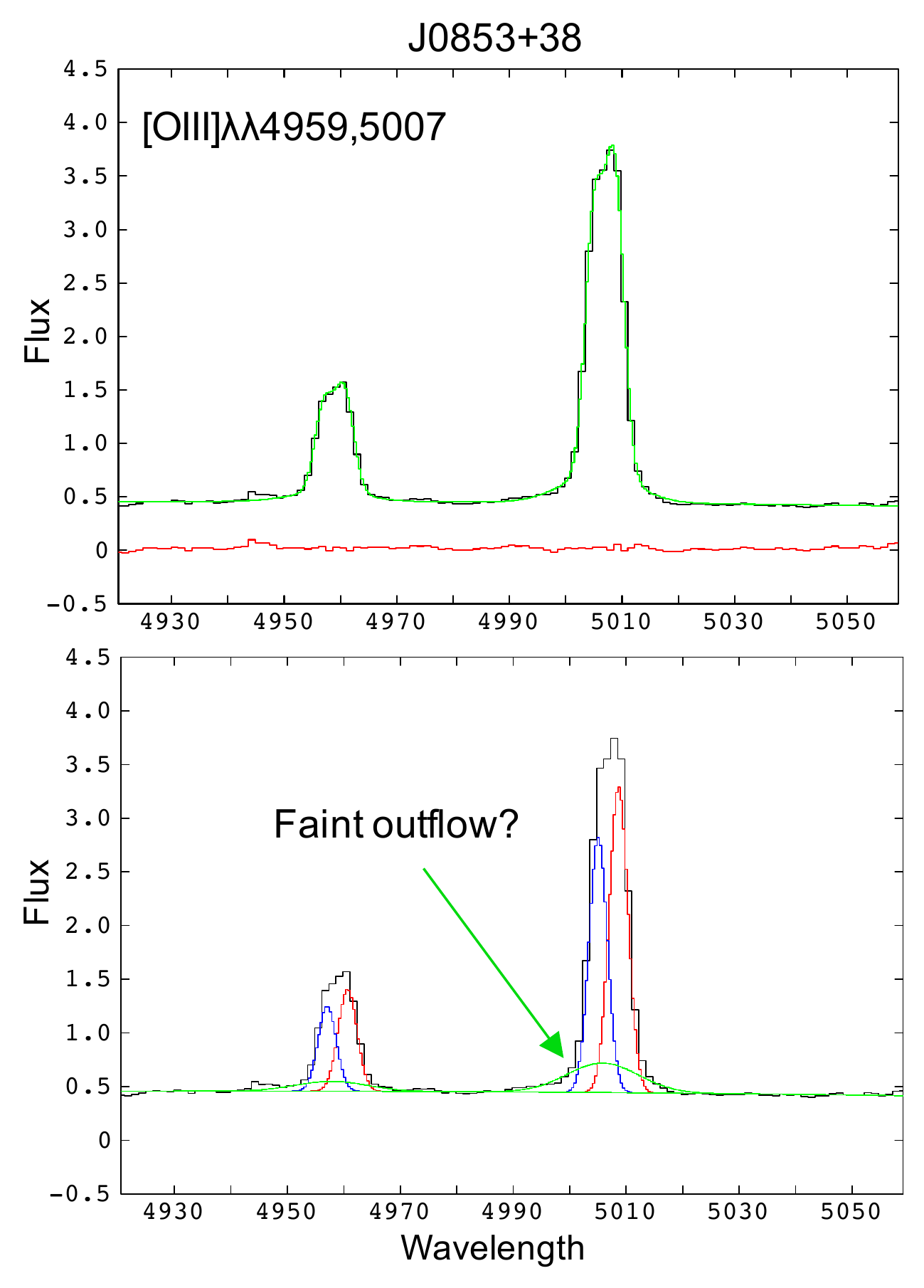}
\caption{[OIII] doublet SDSS  spectrum of J0853+38.  Wavelength in \AA.  A  fit (green) and residuals (red) are shown in the top  panel with the data (black). The three kinematic components isolated in the fit are shown in the bottom panel with different colours. The broadest component may trace an ionised outflow, but higher S/N data is needed to confirm this. Wavelength in \AA. Fluxes in units of 10$^{-15}$ erg s$^{-1}$ cm$^{-2}$ \AA$^{-1}$.}
\label{outflow0853}
\end{figure}

$\bullet$ J0853+38 ($z$=0.127)

 This RI   QSO2  (Fig. \ref{xu}) has a remarkable  radio morphology (Fig. \ref{VLA0853})\footnote{The map shown in Fig. \ref{VLA0853} contains artefacts imprinted by the PSF near the bright radio core as a result of the snap-shot observations. We will mention only the features that are unambiguously real}. The high resolution (VLA A-configuration) map  shows  a radio core and two narrow jets extending N ($\sim$4.5 kpc) and S ($\sim$9.0 kpc) of the AGN.  Beyond a gap in the radio emission, the high $SB$  jet bends sharply and extends towards the East for $\sim$60 kpc. This jet was already apparent in the FIRST data. The VLA B-configuration map reveals that it is part of a much larger  and  highly asymmetric radio source of total extension $\sim$3.5' or $\sim$472 kpc. The bent jet extends towards the NE with a wiggling morphology that ends in a  hot spot  at $\sim$1.5$\arcmin$ or $\sim$202 kpc  from the QSO2. This is embedded in a large  lobe, which shows diffuse emission mostly on the northern side along the jet and hot-spot.

Towards the SW another  radio lobe  with no identifiable hot spot is detected at similar distance from the QSO2. The northern $\sim$4.5 kpc inner jet detected in the A configuration map appears   to bend  $90\degr$ towards the West in the B configuration   image.  Faint radio emission detected between this structure and the SW lobe suggests that the radio lobe may be continuous, as on the eastern side, but Doppler de-boosted and thus receding.   The low $SB$ radio emission   fills an area of at least $\sim$20,000  kpc$^2$ in projection.  Therefore, the relativistic electrons  have the potential to  inject energy across a huge   volume and large distances from the AGN.

The GTC images of J0853+38 are shown in Fig. \ref{fig0853}.  The QSO2 host  is a spiral galaxy and so is SDSS J085317.82+385311.7 at $\sim$37.5$\arcsec$  (84 kpc) to the NW ($z_{\rm ph}=$0.123$\pm$0.010; SDSS database). 
The H$\alpha$ image shows the spiral arms of this companion  prominently. This  is not detected in the VLA maps. Both galaxies may be interacting, although this is not conclusive based on our data.

The H$\alpha$ image shows extended ionised gas  all around  the QSO2 nucleus. Some of it overlaps with the  spiral arms  and is naturally explained by star formation. This is further supported by the presence of some compact regions  on the eastern side.

  The most prominent ionised features extend up to $\sim$36 kpc  North and $\sim$22 kpc South of the AGN.  They are not tightly correlated with the spiral arms  and the main axis  is  misaligned relative to the continuum major galaxy axis (Fig.\ref{fig0853}, right). 

The overlay between the low (B configuration) and high (A configuration) VLA radio maps with the H$\alpha$ GTC image is shown in  Fig. \ref{overlay0853}.  Both  N and S of the QSO2, the ionised gas and the  radio source trace each other. The prominent bends  of the radio source suggest   that it  is propagating through  a  gas rich environment that deflects it (\citealt{Heesen2014}). Simultaneously  the radio source appears to be  affecting the H$\alpha$ morphology by redistributing and/or exciting the  gas it encounters as it advances.

There is a gap of H$\alpha$ emission at $\sim$10 kpc South of the AGN. The southern radio structure  bends at or near this location.  This morphological correlation suggest that the  radio source   has cleared  this region of gas. At the same time, this interaction  forces it to  change its trajectory.

Therefore, the J0853+38 radio source is interacting with the ambient gas at different locations up to at least $\sim$36 kpc from the AGN. In fact, the  radio source is so large that it has the potential to affect a huge volume in and around the QSO2 host.

Several patches of H$\alpha$ emission indicated with  purple arrows in the middle panel of Fig. \ref{fig0853} are seen at different locations at tens  of kpc  from the QSO2 and the companion ($r_{\rm max}\sim$31$\arcsec$ or 70 kpc and $d_{\rm max}\sim$100 kpc). 
They may be separate gas patches, but it appears more natural that they are fragments of a much larger  probably neutral gas reservoir made of tidal remnants from the interaction between both galaxies. Local ionisation by star formation or other mechanisms could produce H$\alpha$ at these particular locations.

As in J0841+01, there is no conclusive evidence for an  ionised outflow in the NLR of SDSS J0853+38 as traced by [OIII]. The lines in the SDSS spectrum are    best reproduced by two  narrow components   with FWHM=155$\pm$10 and 129$\pm$12 km s$^{-1}$ respectively, shifted by 213$\pm$10 km s$^{-1}$. A  faint blue excess asymmetry is detected  which may be the signature of a faint outflow, but the kinematic properties are  too uncertain. A possible fit is shown in Fig. \ref{outflow0853}. The broadest component contributes $\sim$16\% of the total line flux. It has FWHM=948$\pm$70 km s$^{-1}$ and similar velocity as the  faintest  narrow component.

Therefore, radio induced feedback can work across giant spatial scales also in objects with no or  faint  nuclear  outflows.

$\bullet$ J0907+46 ($z$=0.167)

\begin{figure*}[ht]
\centering
\includegraphics[width=0.8\textwidth]{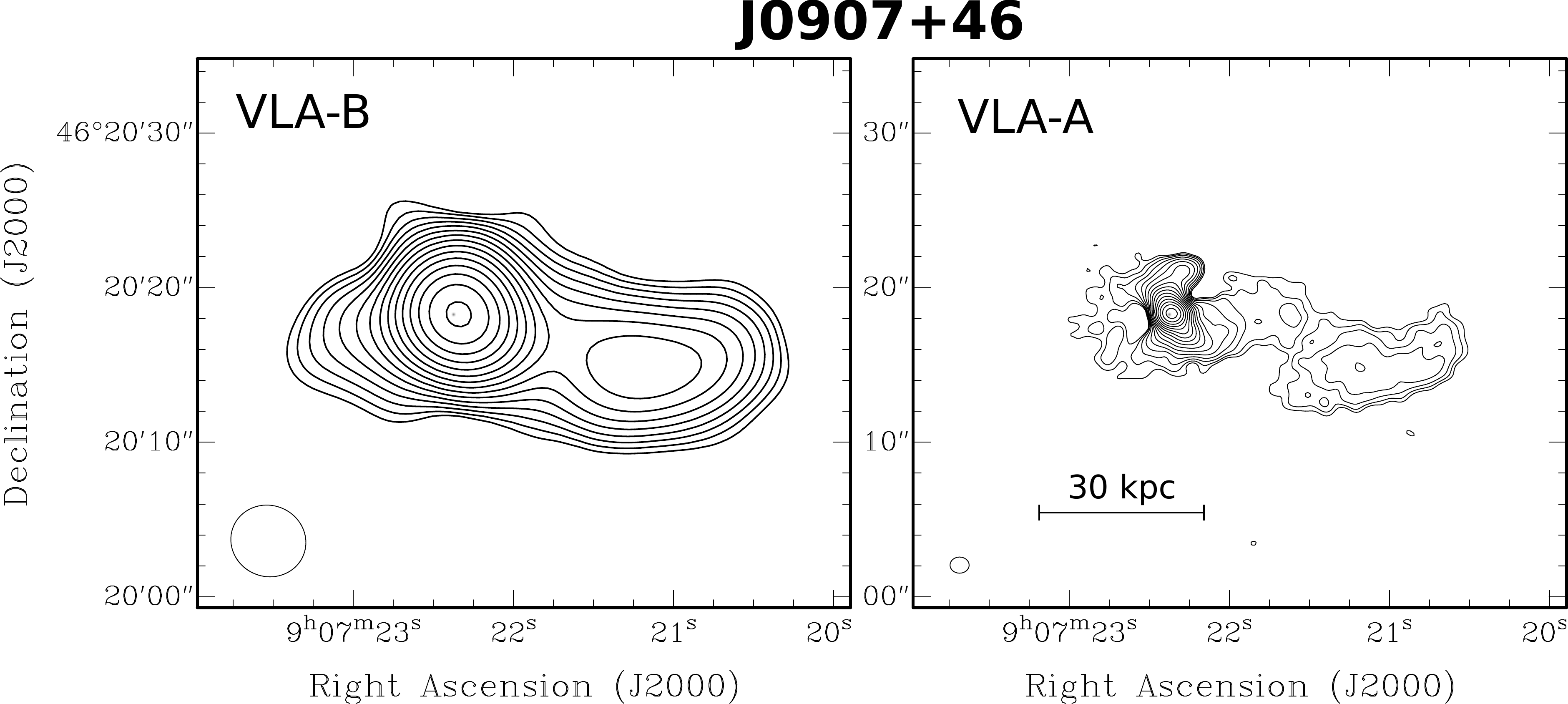}
\caption{VLA radio maps of J0907+46. Same as Fig.\,\ref{VLA0841}. Contour levels start at 200 $\mu$Jy\,beam$^{-1}$ (B-configuration) and 80 \,$\mu$Jy\,beam$^{-1}$ (A-configuration) and increase with factor $\sqrt{2}$.}
\label{VLA0907}
\centering
\includegraphics[width=0.8\textwidth]{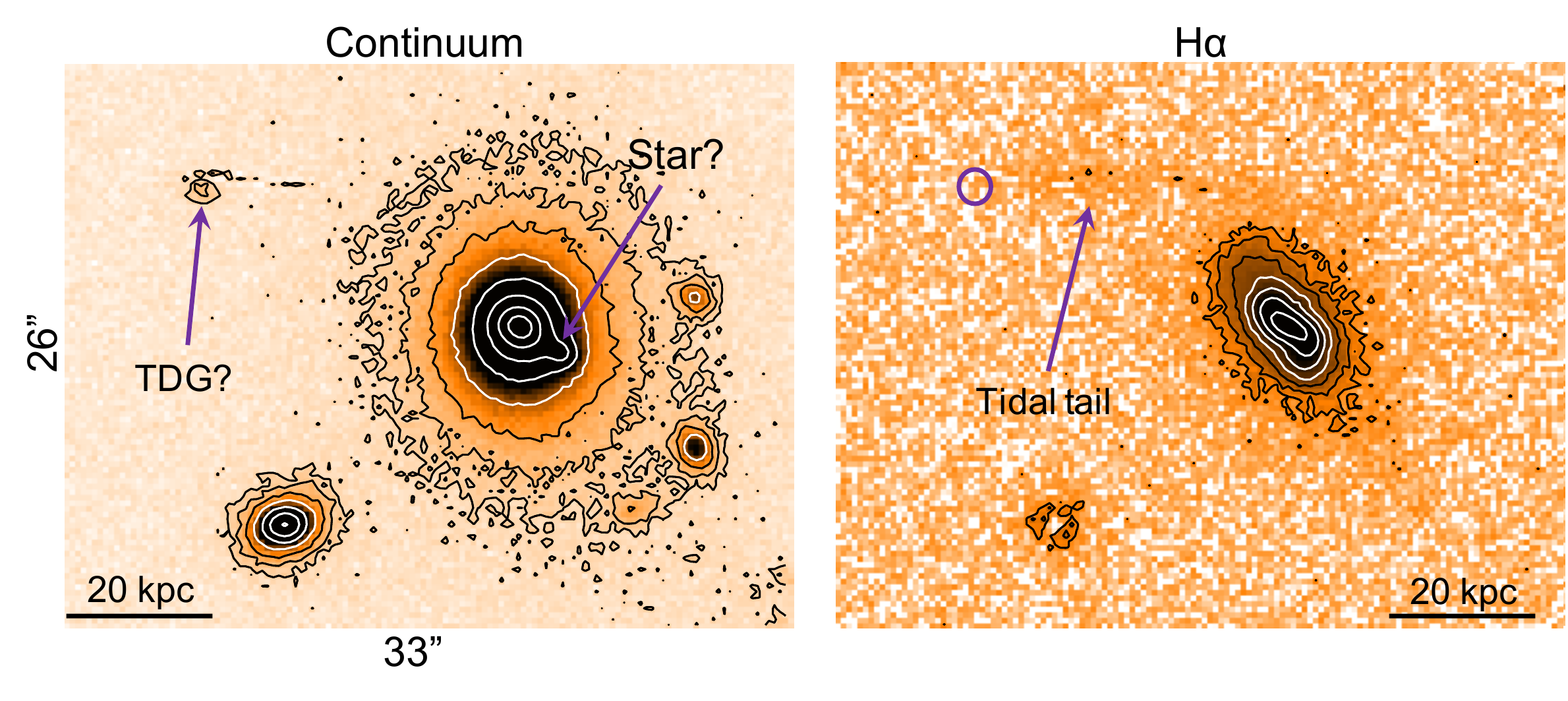}
\caption{J0907+46. GTC continuum and H$\alpha$ images.   GTC continuum and H$\alpha$ images. Contour values start at 3$\sigma$ and increase with factor $\times$2. For the H$\alpha$ image, $\sigma$=2.4$\times$10$^{-19}$ erg$^{-1}$ cm$^{-2}$ pixel$^{-1}$. The purple arrows  in the right panel mark distant patches of H$\alpha$ emission. The contours highlight the different shape of the continuum and H$\alpha$ isophotes. The small continuum source in the left panel  is a tidal dwarf galaxy (TDG) candidate. Its location in the H$\alpha$ image is shown with a small circle. }
\label{fig0907a}
\end{figure*}

\begin{figure}
\centering
\includegraphics[width=0.36\textwidth]{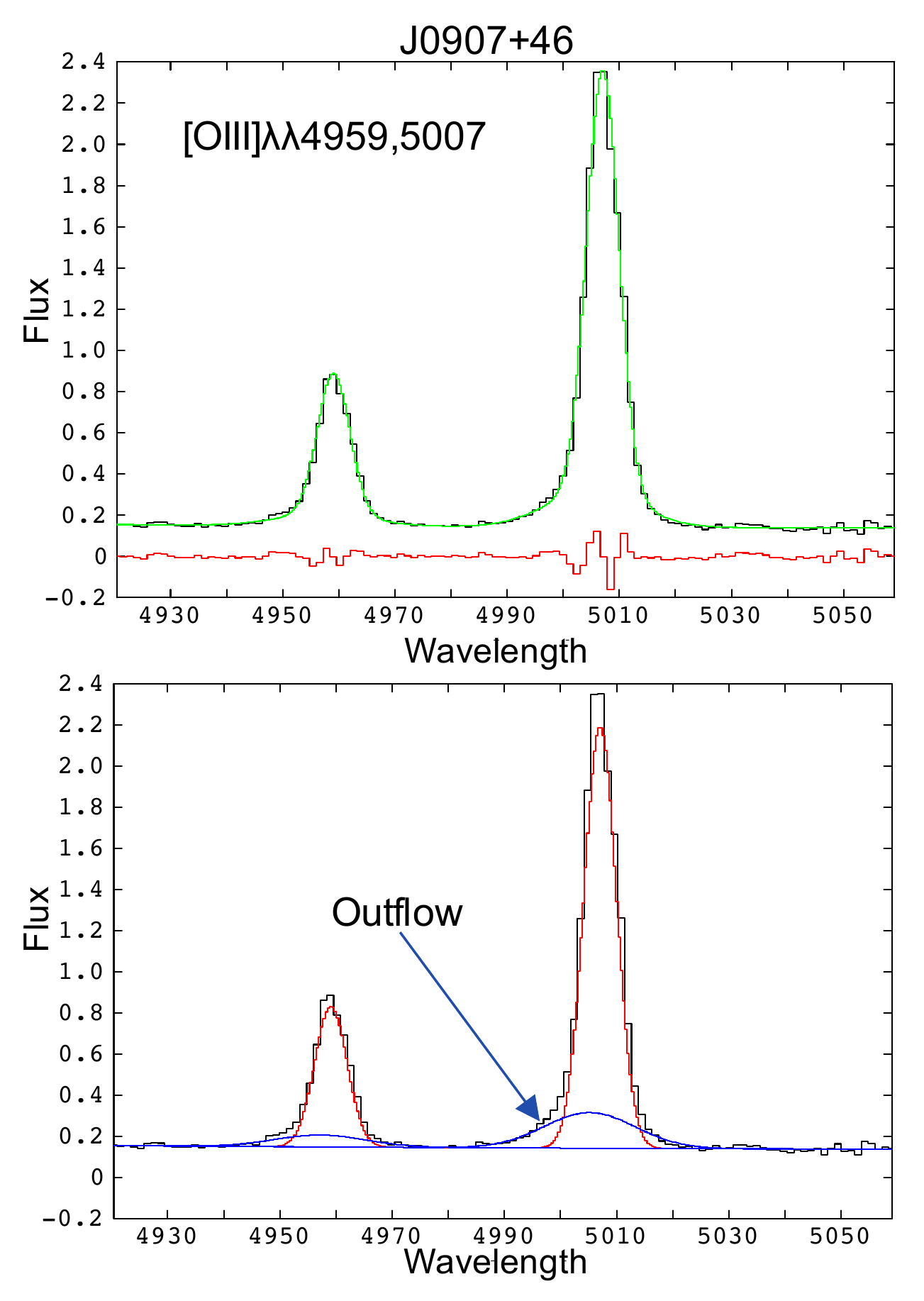}
\caption{[OIII] doublet SDSS  spectrum of J0907+46. Panels and line colour code as in Fig. \ref{outflow0853}. 
 The broadest of the two kinematic components isolated  in the fit traces an ionised outflow (blue, bottom panel). Wavelength in \AA. Fluxes in units of 10$^{-15}$ erg s$^{-1}$ cm$^{-2}$ \AA$^{-1}$.}
\label{outflow0907}
\end{figure}

Little is known about this RI QSO2 (Fig. \ref{xu}).   The FIRST radio image shows an extended source ($\sim$20$\arcsec$ or 57 kpc) which is appreciated with greater detail in 
our VLA data  (Fig. \ref{VLA0907}). It is a complex, highly asymmetric source of total size $d^R_{\rm max}\sim$76 kpc. The radio emission is extended near the radio core for $\sim$13 kpc along  PA $\sim$29$\degr$.   At both ends the inner  radio jet  bends towards the W. On the southern edge,  it bends  almost $\sim$90$\degr$. It  extends for $\sim$20 kpc more where it bends again and extends further up to $\sim$48 kpc. To the East, some low $SB$ emission is detected up to $\sim$14 kpc from the AGN.

We show in Fig. \ref{fig0907a} the GTC continuum and H$\alpha$ images. The QSO2 host is an almost circular galaxy with no apparent signs of mergers/interactions in the continuum image. The H$\alpha$ emission   in the host galaxy is elongated along PA $\sim$29$\degr$ (like the radio axis PA) with a total size of $\sim$ 31 kpc. It is strongly misaligned relative to the continuum morphology. Low $SB$ emission extends further out up to 24 kpc from the AGN to the SW. In addition, a long ($\sim$44 kpc) H$\alpha$ filament extends from  the northern side of the galaxy towards the East. It is reminiscent of a tidal tail,  although  no signs of morphological disturbance are  apparent in the continuum image. The filament ends up precisely at the location of a small continuum source which may be a TDG  at the tip of the tidal tail (e.g. \citealt{Bornaud2004}). It is located at $r_{\rm max}\sim$46 kpc from the galaxy nucleus.

We show in Fig. \ref{overlay0907}  the GTC H$\alpha$ image of J0907+46  with overlaid contours of the VLA A  (red) and B-configuration (grey) data. The H$\alpha$ morphology and the inner radio structures are aligned and have similar extensions ($\sim$13 kpc). At the edges of the brightest H$\alpha$  emission the radio source appears to bend sharply. This suggest that   the radio source and the ambient gas are interacting. This may cause the deflection of the radio source.

An  [OIII] outflow is isolated in the SDSS spectrum with FWHM=1095$\pm$97 km s$^{-1}$ and blueshifted by -110$\pm$40 km s$^{-1}$ relative to the narrow core of the line. It contributes $\sim$19\% of the total [OIII] flux (Fig. \ref{outflow0907}).

$\bullet$ J0945+17 ($z$=0.128)

\begin{figure*}
\centering
\includegraphics[width=0.72\textwidth]{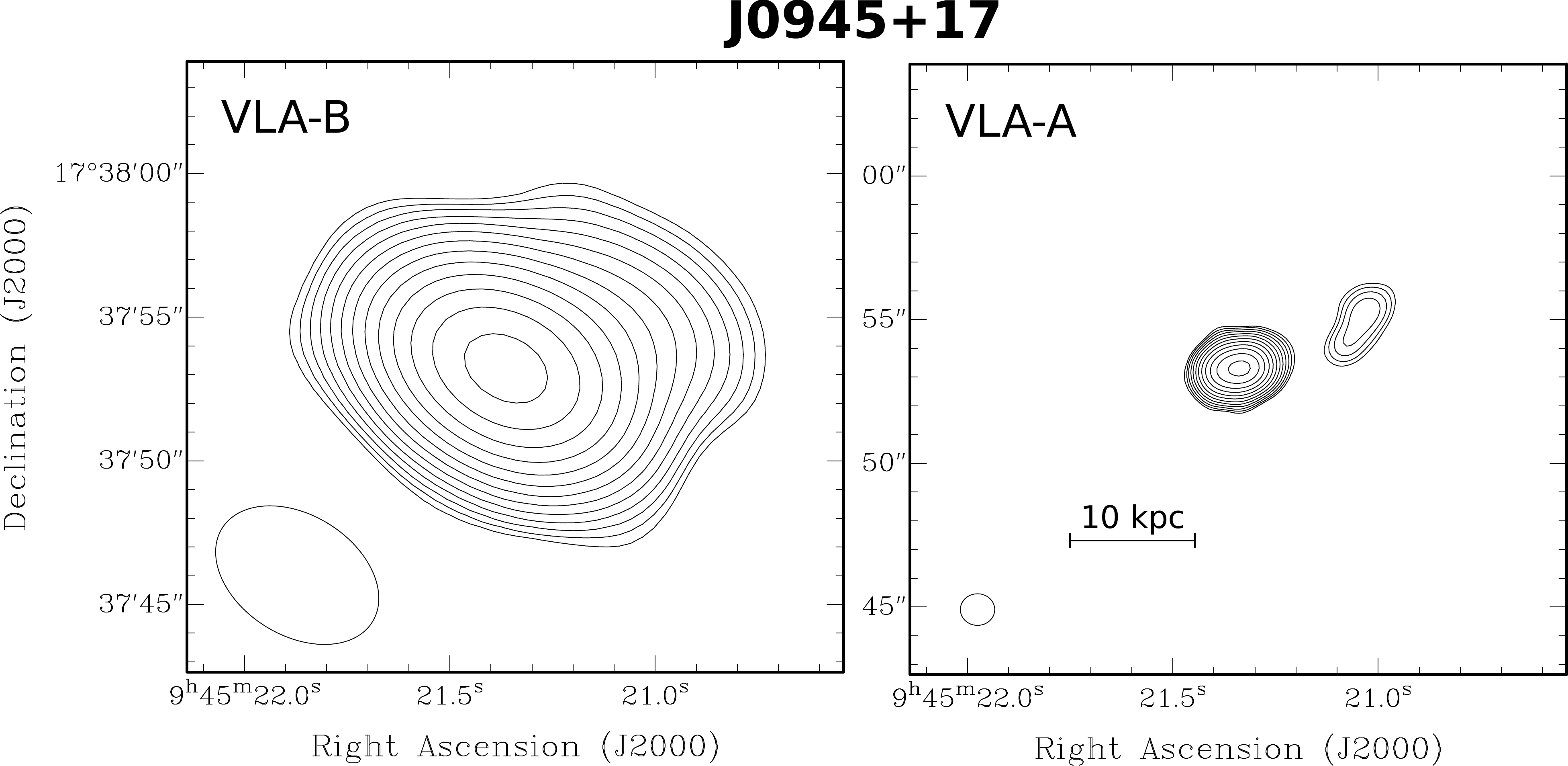}
\caption{VLA radio maps of J0945+17. Same as Fig.\,\ref{VLA0841}. Contour levels start at 500 $\mu$Jy\,beam$^{-1}$ and increase with factor $\sqrt{2}$ for both B- and A-configuration.}
\label{VLA0945}
\centering
\includegraphics[width=0.9\textwidth]{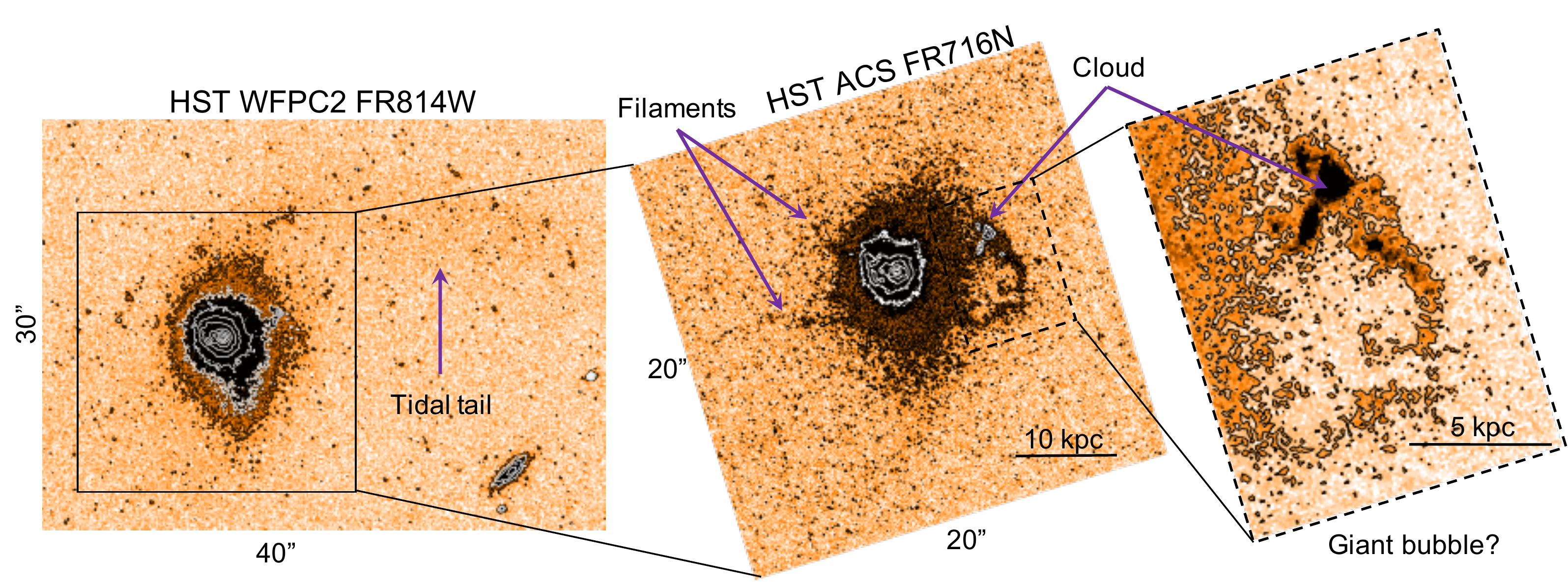}
\caption{J0945+17. Left: HST WFPC F814W continuum image. This filter covers the spectral range 7617-8921 \AA~(6752-7909 \AA~rest frame) and is thus dominated by continuum.     Middle panel: HST ACS /WFC FR716N image containing H$\alpha$+[NII] (the images are not flux calibrated). It has been rotated to show N up and E left. The contours start at 3$\sigma$ and increase with a factor $\times$2 (continuum image) and $\times$3 (narrow band image).  The region of the cloud and the giant bubble candidate is zoomed in the right panel.   The  extended filamentary structures  are dominated by line emission. The ``cloud'' at $\sim$11 kpc to the West was identified by \cite{Storchi2018}.  It is part of a spectacular ensemble of ionised filaments  whose morphology is reminiscent of a giant bubble.}
\label{fig0945}
\end{figure*}

This RQ/RI QSO2 (Fig. \ref{xu})  shows an excess of radio emission above that expected from star formation  ($q$=1.35$\pm$0.10; Sect. \ref{sample}).   The new VLA  maps are shown in Fig. \ref{VLA0945}.

No GTC images are available for this objet. The HST WFPC2 F814W  continuum image (program 6346; PI K.D. Borne) shows    signatures of galactic interactions (Fig. \ref{fig0945}). The host galaxy is  morphologically distorted and has a $\sim$56 kpc long tidal tail  stretching towards the W.  

\cite{Jarvis2019} presented 1-7 GHz high (0.22$\arcsec \times$0.21$\arcsec$ beam) and low-resolution ($\sim$1.2$\arcsec \times$0.9$\arcsec$ beam) radio imaging (VLA and e-MERLIN) of this QSO2.  The  VLA map shows 
two spatially distinct features. One overlaps with the QSO2 host (LR:A, using their nomenclature) and a separate component at $\sim$5$\arcsec$ or $\sim$11 kpc to the NW (LR:B), which they identify with a secondary jet or lobe.   LR:A is resolved with   e-MERLIN,  into a compact core associated with the QSO2 nucleus and a  $\sim$1$\arcsec$ ($\sim$2.3 kpc) curved feature, possibly a bent jet.  It  terminates at a region of  brightened blueshifted [OIII] emission. They propose that  the jet is hitting a cloud, both pushing the gas away and deflecting the jet.

\begin{figure}
\centering
\includegraphics[width=0.36\textwidth]{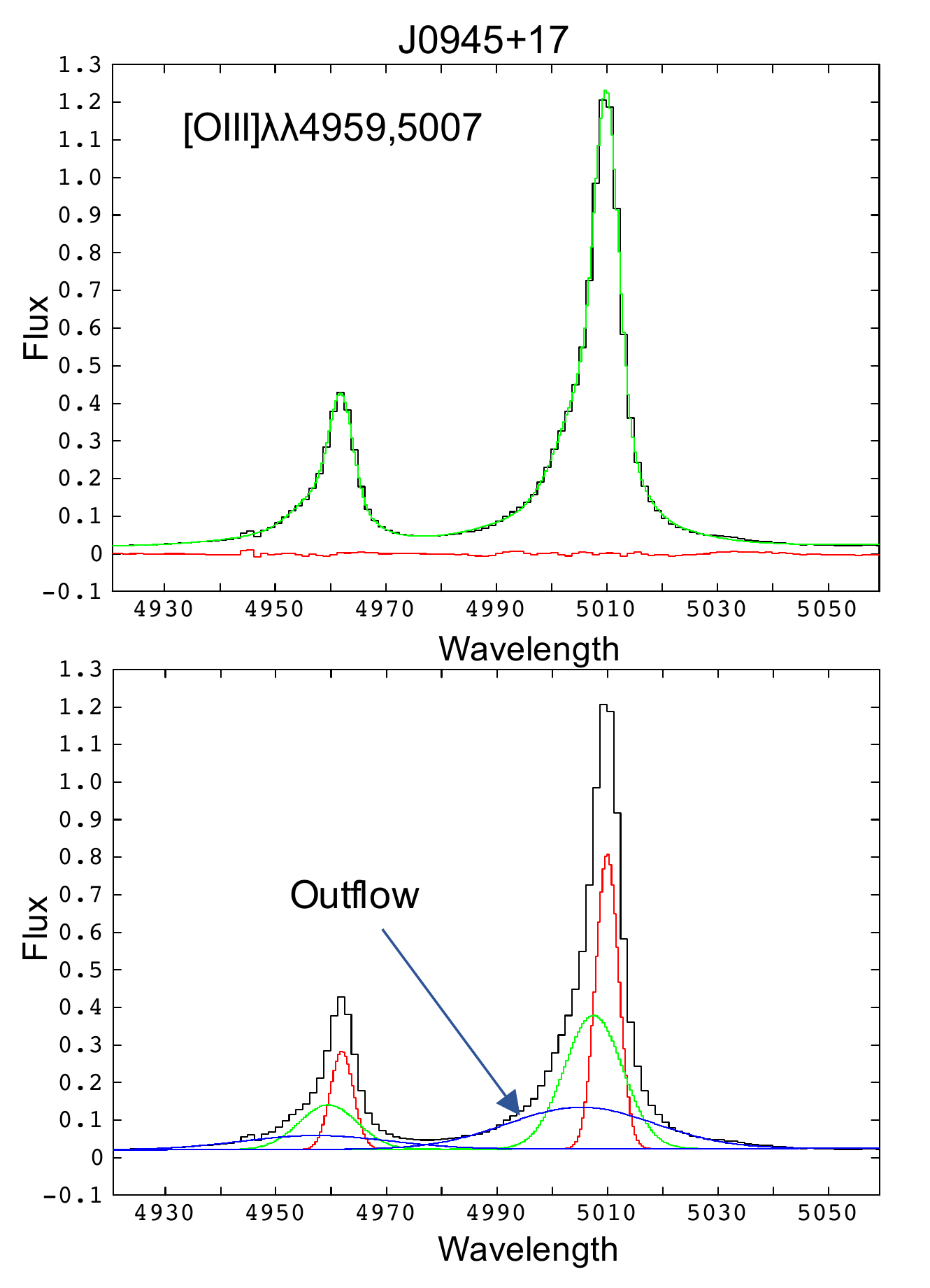}
\caption{[OIII] doublet SDSS  spectrum of J0945+17. Panels and line colour code as in Fig. \ref{outflow0853}. 
 The broadest of the three kinematic components isolated in the fit traces a prominent ionised outflow (blue, bottom panel). Wavelength in \AA. Fluxes in units of 10$^{-14}$ erg s$^{-1}$ cm$^{-2}$ \AA$^{-1}$.}
\label{outflow0945}
\end{figure}

Further indirect evidence that this small jet is triggering a nuclear ionised outflow is provided by the SDSS spectrum (Fig. \ref{outflow0945}). 
A strong outflow with FWHM$\sim$1935$\pm$35 km s$^{-1}$ and blueshifted by $V_s$=-252$\pm$18 km s$^{-1}$ is identified in the [OIII] lines. The large values  of the ratios   $R=\frac{\rm FWHM_{\rm [OIII]^{\rm outflow}}}{\rm FWHM_{\rm core}}$=5.9$\pm$0.3 (this is, the FWHM of the outflow component compared with the FWHM of the narrow line core) and $R2=\frac{F(H\beta)^{outflow}}{F(H\beta)^{total}}$=0.31$\pm$0.02 (ratio between the flux of the broad component and the total H$\beta$ flux) suggest that  this outflow is triggered by a radio source (\citealt{Villar2014}).

Our new  VLA A and B configuration maps (Fig. \ref{VLA0945})   were obtained with $t_{\rm exp}$= 30 and 60 min on source respectively  (prior radio  maps were obtained with 5-10 minute exposure).  The two radio components are LR:A and LR:B in \cite{Jarvis2019}.   LR:B overlaps with a region  of ionised gas that  was  studied  by \cite{Storchi2018} using narrow-band [O III] and  H$\alpha$+[N II] HST images. They describe a ``cloud''   of ionised gas that could be either reminiscent of a merger/interaction or due to a previous outflow from the AGN.  

 Although our radio maps do not  reveal additional structures, novel results are obtained from the  comparison with the HST images. The  ACS F716N image containing H$\alpha$+[N II]  (see  \citealt{Storchi2018}  for details. Program 13741; PI: T. Storchi-Bergmann) is shown in  Fig. \ref{fig0945}  with the continuum image.   We also show in Fig. \ref{overlay0945} the overlay of the VLA A and B configuration maps and the HST/WFC FR551 image containing [OIII]$\lambda$5007 (same HST program). 

The ``cloud'' is detected in both the line and the continuum images.  It may be  a small galaxy that has been hit by the jet.    The  elongated morphology of LR:B  is tightly correlated with a clear enhancement of the line emission in two knots that are part of the ``cloud''  (see also Fig. \ref{fig0945}). We discard that the line and radio emissions are  produced locally by star formation.  The 1.4 GHz luminosity of LR:B is log($P_{\rm1.4~GHz}$)=30.46 in erg s$^{-1}$. The implied star forming rate would be 109 M$_{\rm \odot}$ yr$^{-1}$ (\citealt{Condon1992,Hodge2008}) which in turn would result on  an unrealistically high $L_{\rm H\alpha}\sim$1.4$\times$10$^{43}$ erg s$^{-1}$. This is $\sim$125 times higher than that roughly  inferred from the H$\alpha$+[NII] HST image, accounting for the continuum and ignoring [NII] contamination. Given the morphology and high $SB$ of the ``cloud'', it appears unlikely that reddening could explain this discrepancy.

We propose a different scenario.   The ``cloud''   is part of a spectacular filamentary structure whose morphology is reminiscent of a giant $\sim$10 kpc bubble  (see  right panel in Fig. \ref{fig0945}).    It may be a giant outflow, which has been inflated by  a wide angle AGN driven wind or by the radio source. The emission is enhanced in the ``cloud'', because  the LR:B jet or lobe has hit the bubble at this location.  The ``cloud'' appears to be deflecting the radio source as well. The radio axis and the main bubble axis  are misaligned by $\sim$30$\degr$. This could be a consequence of jet precession and multiple episodes of jet activity (\citealt{Jarvis2019}).  There are two very narrow ionised filaments  of lengths $\sim$18 and 4 kpc  to the East.  It is unclear whether they trace the edge of a counter-bubble at the opposite side of the AGN.

\begin{figure*}
\centering
\includegraphics[width=0.72\textwidth]{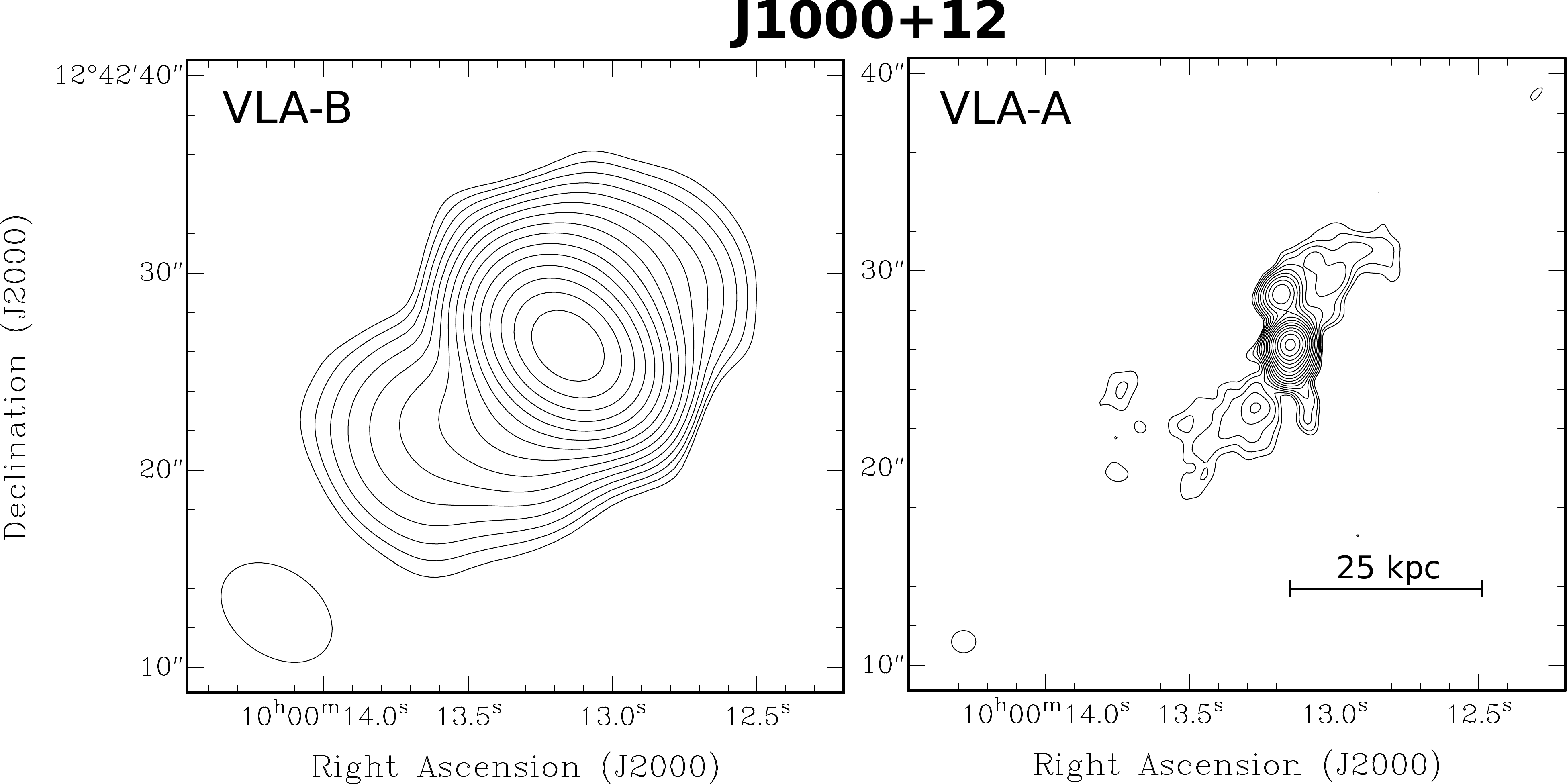}
\caption{VLA radio maps of J1000+12. Same as Fig.\,\ref{VLA0841}. Contour levels start at 150 $\mu$Jy\,beam$^{-1}$ (B-configuration) and 93 $\mu$Jy\,beam$^{-1}$ (A-configuration) and increase with factor $\sqrt{2}$. Note that the extension that stretches $\sim$2$^{\prime\prime}$ south of the core is likely dominated by artefacts of the snap-shot observations, because a spike in the prominent pattern of the Point-Spread-Function (PSF) crosses the two bright regions in the centre.}
\label{VLA1000}
\centering
\includegraphics[width=0.85\textwidth]{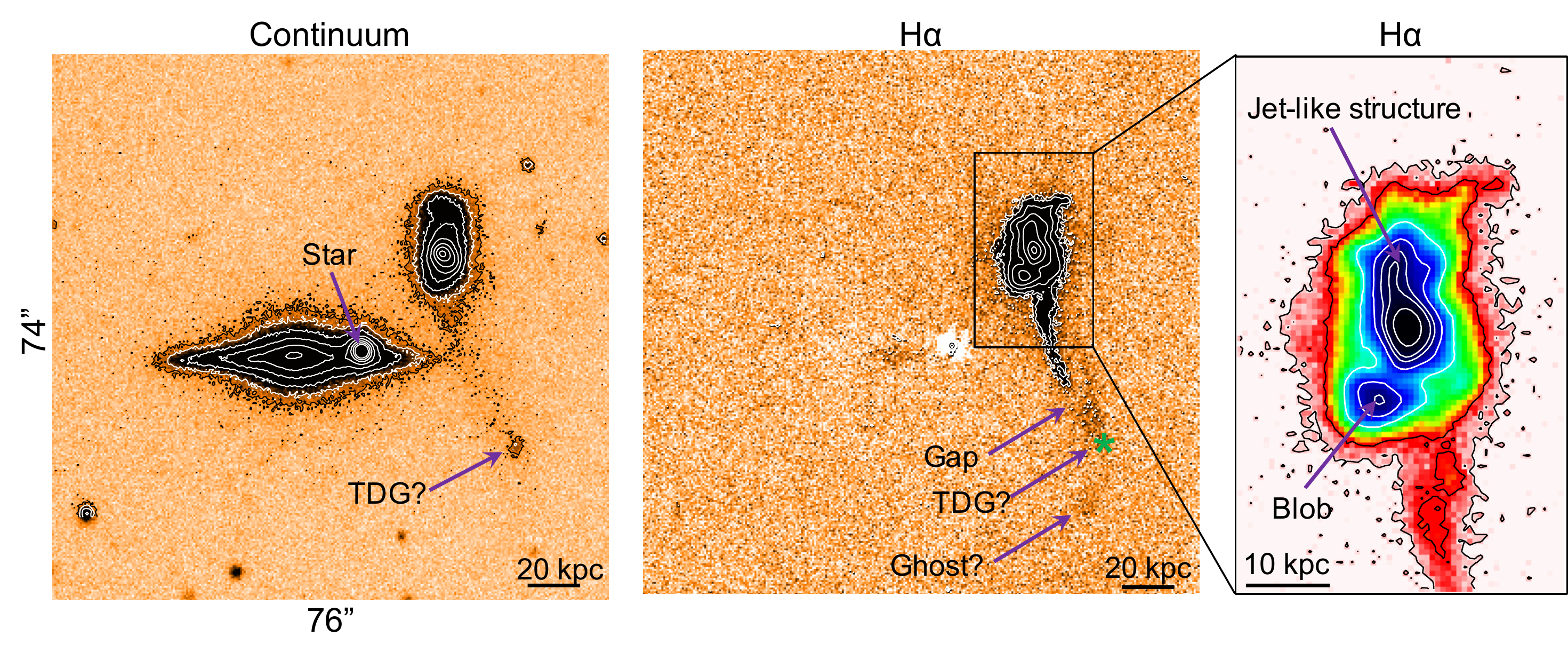}
\caption{J1000+12. GTC continuum and H$\alpha$ images.  Contour values start at 3$\sigma$ and increase with factor $\times$2 and $\times$3 for the continuum and H$\alpha$ images respectively. For the H$\alpha$ image, $\sigma$=3.8$\times$10$^{-19}$ erg$^{-1}$ cm$^{-2}$ pixel$^{-1}$. The right panel shows  the  H$\alpha$ emission in a smaller region centred on the QSO2 with a different colour palette. The small green * symbol in the middle panel shows the expected location of the TDG candidate identified in the continuum image. Several additional features mentioned in the text are indicated with purple arrows.}
\label{fig1000}
\end{figure*}

$\bullet$ J1000+12 ($z$=0.148).   This RQ/I QSO2 shows a clear excess of radio emission ($q=$1.08$\pm$0.09) relative to that expected from star formation (Fig. \ref{xu}).  A [OIII] outflow  is identified in the SDSS spectrum  with FWHM=1069$\pm$19 km s$^{-1}$ blueshifted by $V_{\rm S}=$-63$\pm$10 km s$^{-1}$  (see also \citealt{Sun2017}).  

  J1000+12 was also part of \cite{Jarvis2019} sample. They observed it with the VLA from 1$-$8 GHz with resolution ranging from 0.3$-$4.8$^{\prime\prime}$ and exposure time 5$-$7 minutes per target. In addition, they obtained deep e-MERLIN observations at 1.4 GHz. 
They  identified large scale radio emission over $\sim$25 kpc   divided in 4 components. Using their nomenclature these are LR:A (which encompasses core and jet)  and three lobes (LR:B, LR:C and LR:D) which trace a strongly bent morphology that is likely a sign of deflection. The high resolution map shows  a  $\sim$1$\arcsec$ ($\sim$2.6 kpc) jet like structure to  the S of the radio core  and a hot spot at  $\sim$6.4 kpc to the N.

The authors compare  the spatial distribution of the radio and [OIII] morphologies. Both   morphologies and the gas kinematics  are complex.  [OIII] is enhanced along the probable southern jet. They propose that the line splitting and blue asymmetry  at its base  is a signature of  a $\sim$10 kpc  outflowing bubble being launched by the jet. 

We show in Fig. \ref{VLA1000} our VLA maps of J1000+12, which have 6$-$11$\times$ longer integration times than the observations by \citet{Jarvis2019}.   The B-configuration map does not reveal radio structures  on significantly larger scales than previously identified.  Several faint knots  are detected in the A-configuration image that prolong the LR:C SE lobe in \cite{Jarvis2019}  towards the East. We measure a maximum extension of the radio source of $d^{\rm R}_{\rm max}\sim$43 kpc.
    
We show in Fig. \ref{fig1000} the GTC continuum and H$\alpha$ images of J1000+12. 
Both show a highly distorted  morphology due to a merger or interaction event. The large galaxy to the SE is most likely unrelated to the QSO2. The SDSS photometric redshift   $z_{ph}=$0.06$\pm$0.02 is significantly lower than that of the QSO2.  The non-detection   in the H$\alpha$ image further supports a different $z$.

A long tidal tail extends from the QSO2 host towards the South  for at least $\sim$72 kpc. It ends at the location of a small continuum source which may be a tidal dwarf galaxy (see also J0841+01, J0907+46 and J1356+10).  

The H$\alpha$ image shows a complex gaseous environment extending for tens  of kpc. The tidal tail is clearly seen.  It shows a gap at $\sim$50 kpc  which is not apparent in the continuum image (Fig. \ref{fig1000}). Beyond, this gap the line emission becomes more diffuse and ends at the expected location of the TDG  candidate identified in the continuum image (Fig. \ref{fig1000}, middle panel).  The patch of low $SB$ H$\alpha$ emission  detected  further out, at $\sim$90 kpc from the QSO2 nucleus is probably a ghost, since it is seen only in one of the tunable filter frames.

The H$\alpha$ and radio overlay is shown in Fig. \ref{overlay1000}. The overlay between the radio map and the HST/WFC3/UVIS2  FQ508N filter image containing [OIII]$\lambda$5007  (program 4730; PI  A. Goulding) is also shown in the small panel. We adopt \cite{Jarvis2019}  nomenclature to identify  the main radio structures. The ionised gas and radio 
morphologies are clearly correlated  across large spatial scales (see also \citealt{Jarvis2019}), which suggests that they are interacting. The brightest H$\alpha$ and radio features overlap spatially. The long  and narrow, jet-like H$\alpha$ feature extending  $\sim$9 kpc towards the North from the AGN  overlaps with  LR:B  (see also Fig. \ref{fig1000}, right panel).  The velocity of this gas is  blueshifted relative to the adjacent gas and the nuclear emission, which is additional evidence for the interaction  (\citealt{Jarvis2019}).
Further out, the ionised gas  traces the bend of the LR:D radio lobe. Moreover, the H$\alpha$ blob $\sim$10 kpc SW of the AGN, overlaps with  the location of the highest LR:C  radio lobe flux. 

Notice that none of the above ionised features  overlap with the $\sim$10 kpc giant, jet-driven outflowing bubble  hypothesised  by \cite{Jarvis2019} based on the [OIII] high positive asymmetry and  splitting of the line profile. 
 Its location is indicated in Fig. \ref{overlay1000}.  It overlaps partially with a narrow radio extension, but this is most likely an artefact  of our snap-shot observations (see Fig. \ref{VLA1000}).

$\bullet$ J1356+10 ($z$=0.123)

\begin{figure*}
\centering
\includegraphics[width=0.70\textwidth]{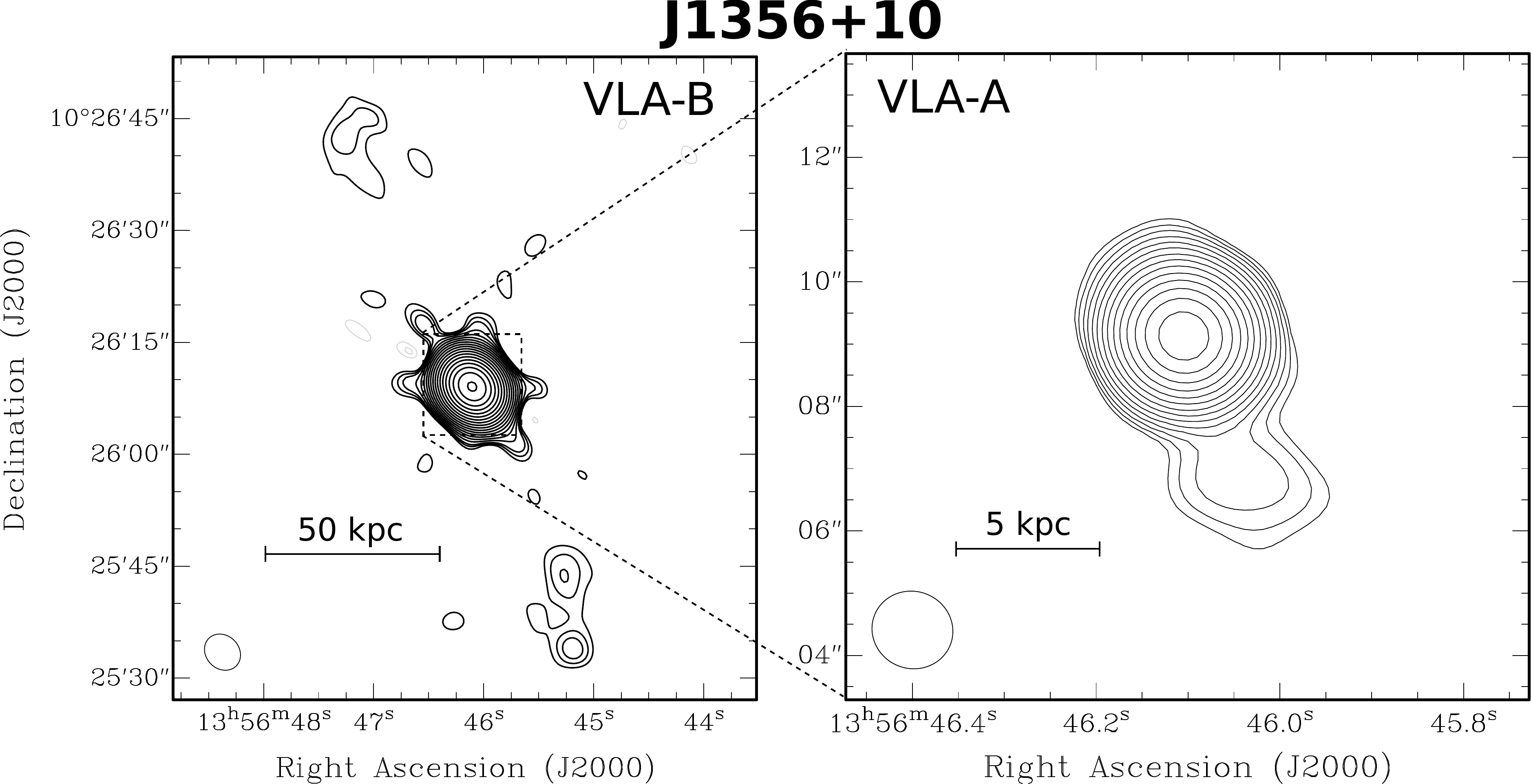}
\caption{VLA radio maps of J1356+10. Same as Fig.\,\ref{VLA0841}. Contour levels for the B-configuration data start at 3.5, 4.5, 5.5$\sigma$ and then increase in steps of $\sqrt{2}$, with $\sigma$\,$\sim$\,37 $\mu$Jy\,beam$^{-1}$ the local rms noise. At these low levels, image artefacts are visible around the strong core. Negative contours are shown at the same level in grey. Contours of the A-configuration data
start at 450 $\mu$Jy\,beam$^{-1}$ and increase with factor $\sqrt{2}$.}
\label{VLA1356}
\centering
\includegraphics[width=0.82\textwidth]{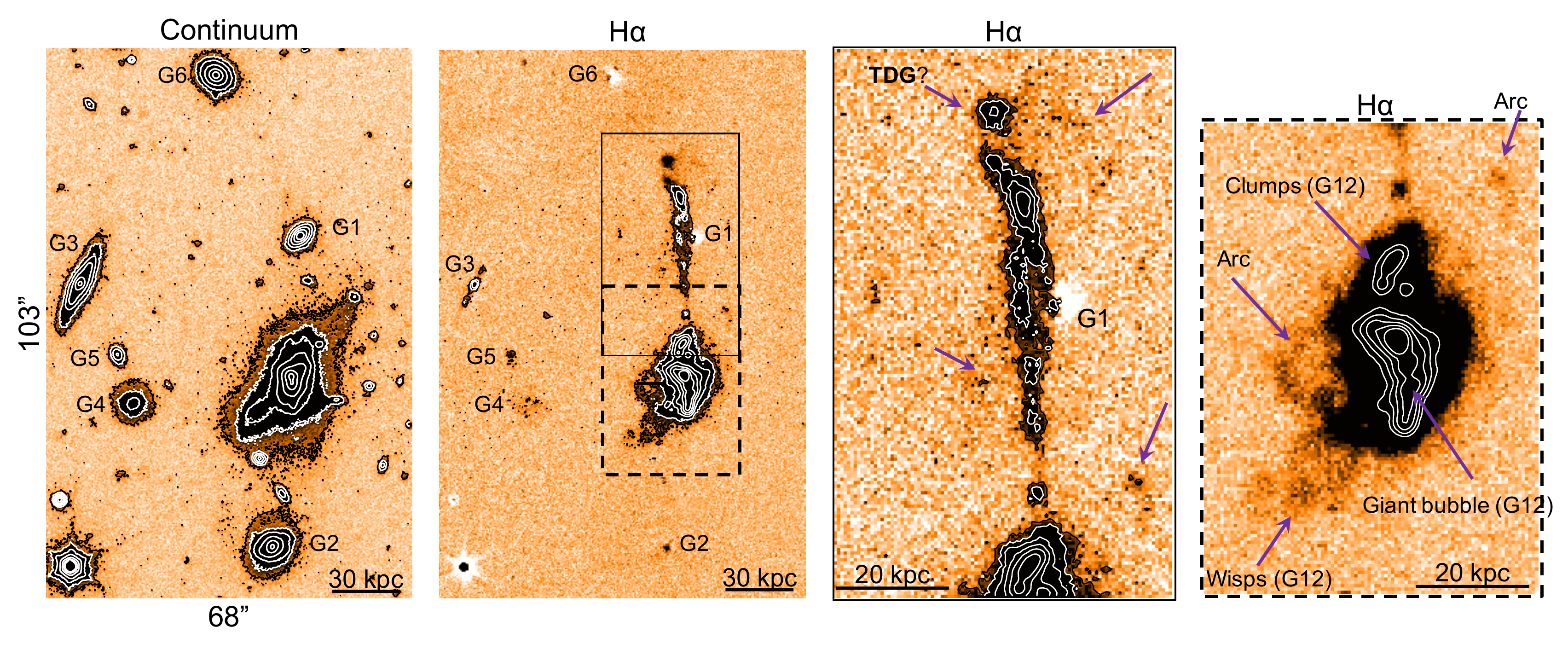}
\caption{J1356+10. GTC continuum and H$\alpha$ images.  Contour levels in the continuum (first panel) and  H$\alpha$ images (second panel) start at 3$\sigma$ and increase with factor $\times$2.5. For the H$\alpha$ image, $\sigma$=4.4$\times$10$^{-19}$ erg s$^{-1}$ cm$^{-2}$ pixel$^{-1}$. Galaxies G1, G3, G4 and possibly G2, G5 and G6 are members of  the same group as the QSO2 (see text). The rectangular areas in the 2nd panel are zoomed in the 3$^{th}$ (top rectangle) and 4$^{th}$ (bottom rectangle) panels. The contours in these trace the morphology within the high $SB$ regions. The purple arrows in the last panel show the location of the giant 12 kpc  bubble, the clumps  and wisps identified by Greene et al. (\citeyear{Greene2012}). Additional faint H$\alpha$ features  are also indicated. Ionised gas is detected up to $\sim$16 kpc E and W of the long tidal filament (3rd panel).}     
\label{fig1356}
\includegraphics[width=0.9\textwidth]{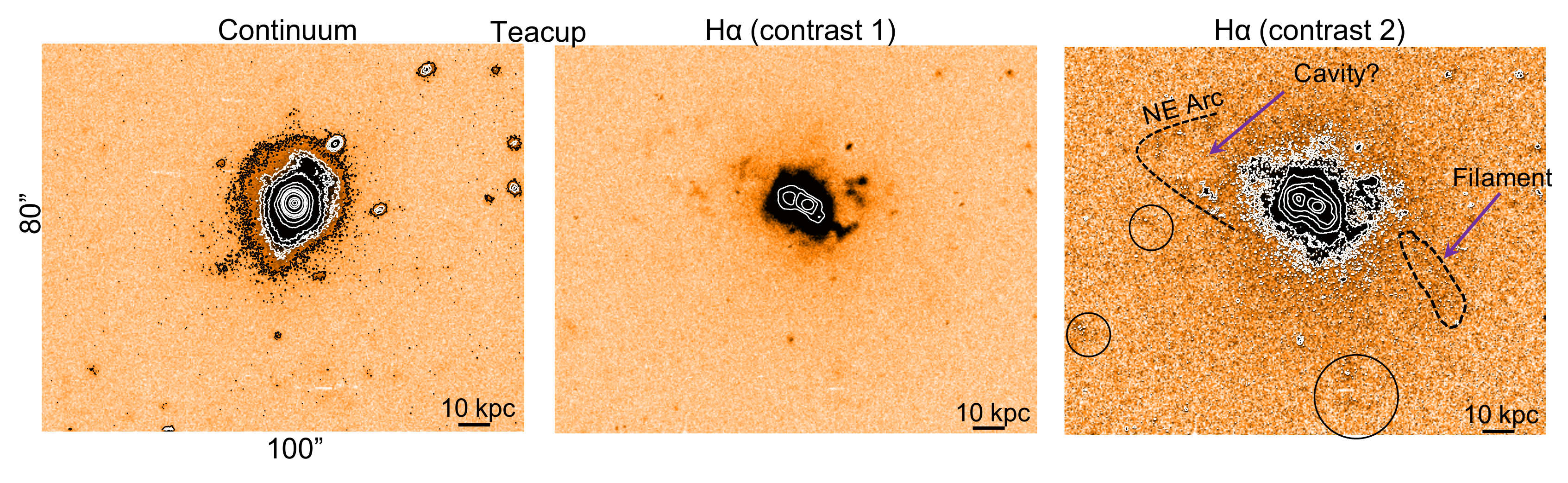}
\caption{The Teacup. Continuum  (left) and H$\alpha$ images with two different contrasts (middle and right panels) to highlight  individual features and the giant low surface brightness halo (right). Contour values start at 3$\sigma$ and increase with factor $\times$3 for the H$\alpha$ image ($\sigma$=3.2$\times$10$^{-19}$ erg$^{-1}$ cm$^{-2}$ pixel$^{-1}$)  and $\times$1.5 for the continuum image.  The contours in the middle  panel are shown to identify the well known ionised bubbles. The different features marked on the right panel are discussed in the text. The black circles  enclose distant H$\alpha$ patches.}              
\label{figTeacup}  
\end{figure*}

  This RQ/RI QSO2 (Fig. \ref{xu})  has been widely studied for two main reasons: it  is a dual AGN candidate and it hosts a giant ionised outflow candidate. \cite{Greene2011} identified  a pair of giant ($\sim$12 kpc) expanding shells or bubbles  centred on the northern nucleus, which hosts the QSO2. They propose that these are part of a larger   expanding structure of total size $\sim$40 kpc  (see also \citealt{Shen2011,Comerford2015}).
 
The origin of the outflow is unclear. It may have been driven by the quasar radiation (\citealt{Greene2011}) or by  a radio jet (\citealt{Jarvis2019}). These  authors identified  an elongated, bent radio source, possibly a jet, extending $\sim$5-6 kpc from the QSO2 nucleus that ends at the base of the southern  bubble. It overlaps  with a region of turbulent ionised gas, possibly  induced by its interaction with the radio source. They propose that  jet induced outflows have inflated the bubble.

We show in Fig. \ref{fig1356} the GTC continuum and H$\alpha$  images of J1356+10. There is  a high density of galaxies    at $\la$1$\arcmin$ from the QSO2 compared with neighbouring sky areas. G1, at 26$\arcsec$ (57 kpc) to the N, is an interacting companion  connected by a tidal feature to the central merging system  (\citealt{Greene2012}).  
G3 and G4, and maybe also G2, G5 and G6, belong to the same group. 
 G3 is a disk galaxy  with SDSS photometric $z_{\rm ph}$=0.143$\pm$0.026 (consistent with the QSO2). All attempts to subtract the continuum from the TF image leave similar H$\alpha$ residuals tracing the nucleus and the edge-on disk  (Fig. \ref{fig1356}, middle panel). 
 Although G4 and G5  have  higher $z_{\rm ph}$=0.176$\pm$0.034  and 0.379$\pm$0.148, the residual H$\alpha$ emission from both suggests a similar $z$ as J1356+10.
 According to the SDSS database, G2 and G6 have $z_{\rm ph}$ consistent  with  the QSO2. 
No reliable H$\alpha$ residuals are detected and, thus, the $z$ cannot be confirmed with our data.

The continuum image (Fig. \ref{fig1356}, left panel) shows the well known highly disturbed morphology of the system.  Continuum  is detected across large spatial scales centred on the QSO2 and the companion nucleus, with a maximum extension in the SE-NW direction of  $\sim$92 kpc.   It traces stellar material that has been spread during the merger over a  large volume.

The long H$\alpha$ filament  extending up to 42$\arcsec$ or $\sim$92 kpc North of the QSO2 was discovered by \cite{Greene2012} with long slit spectroscopy. It seems to connect the QSO2 host with G1,   and extends further out up to $\sim$32 beyond this small galaxy. 
Line emission from  G1 cannot be confirmed. There is a faint elongated continuum feature on its northern side which may be a tidal remnant of the interaction with the QSO2 host.

The tidal filament is invisible in the continuum image  and  ends on a  bright separate H$\alpha$ knot. The compact morphology  and the spatial location at the tip of the  filament suggest that it is  a TDG (see also and J0841+01, J0907+46 and J1000+12).  Very faint H$\alpha$ extensions  are detected  up to $\sim$16 kpc  to the E and W of the filament in the direction perpendicular to it (third panel of Fig. \ref{fig1356}).

 J1356+10  is also associated with a large, morphologically disturbed reservoir of ionised gas which is  widely spread around the central nuclei and overlaps partially  with the amorphous continuum halo. The deep GTC image shows that  it extends up to  $\sim$37 kpc from the QSO2 nucleus to the SE and across a maximum extension of  $\sim$70 kpc in the SE-NW direction.  Part of this reservoir was traced across  40 kpc along PA 45$\degr$ with long slit spectroscopy by \cite{Greene2012}.   The giant 12 kpc bubble candidate, wisps and northern clumps identified by these authors are indicated  in Fig. \ref{fig1356} (right panel)  (see also \citealt{Jarvis2019}).

We highlight in Fig. \ref{fig1356}  (right panel) a remarkable  feature with no  continuum counterpart   which is reminiscent of an edge brightened $\sim$20 kpc bubble.  It may be part of  the giant $\sim$40 kpc hypothetical expanding structure mentioned above. Considering also the southern $\sim$12 kpc jet-driven bubble candidate,  multiple outbursts driven by different mechanisms are suggested. 2-dim spectroscopy would be essential to investigate this further, by characterising the kinematics  and ionisation mechanism across the  entire  system. 

The messy H$\alpha$ morphology is reminiscent of  the triple SMBH merger system NGC6240 ($z=$0.024; \citealt{Yoshida2016,Muller2018}).  H$\alpha$ is detected across  $d_{\rm max}\sim$92 kpc and shows multiples loops, bubbles, knots and filaments.   The complex nebular morphology and kinematics  have been proposed to be a combination of a double outflow (one AGN driven and one starburst driven) and the merger  (\citealt{Nardini2013}). A similar scenario  of multiple outflows and a complex merger may apply to J1356+10.  

The VLA A and B-configuration maps of J1356+44 are shown in Fig. \ref{VLA1356} (see also Fig. \ref{overlay1356}).  The $\sim$5 kpc southern extension seen in the high resolution map was interpreted as a radio jet or lobe by \cite{Jarvis2019}. It coincides with a gap in  H$\alpha$  (small panel in Fig. \ref{overlay1356}).
  Our B-configuration map  (Fig. \ref{VLA1356}, left) reveals emission on scales of $\sim$160 kpc. It is reminiscent of the brightest regions of two diffuse  radio lobes with PA axis $\sim$22$\degr$, which coincides with the PA of the inner jet/lobe.    The emission in the outer lobes of J1356+10 is very faint, and the original data was plagued by image artefacts, therefore additional DDT time was granted to verify this large-scale emission (Sect. \ref{sec:VLA}). Our combined B-configuration data shows persistent large-scale radio emission at a level of 4$-$5$\sigma$ level. Nevertheless, future synthesis observations are needed to unambiguously confirm and accurately image this extended emission in J1356+10.

Assuming that the large scale radio source is real, it would mean that the radio structures have escaped the  QSO2 host and have advanced  a huge distance of $\sim$80  kpc, well within the CGM. This radio source would have  the potential to provide a feedback mechanism that may  affect the properties of the rich, complex environment across $\sim$160  kpc (\citealt{Villar2017}).

$\bullet$ The Teacup ($z=$0.085)

 This RQ QSO2 (Fig. \ref{xu}) has been widely studied.  The system hosts  a giant outflow whose effects are noticed up to $\sim$10-12 kpc from the AGN and is responsible for the  bubble-like morphology of the ionised gas and diffuse radio emission (Keel et al. \citeyear{Keel2012}, Gagne et al. \citeyear{Gagne2014}, Ramos Almeida et al. \citeyear{Ramos2017}, Harrison et al. \citeyear{Harrison2015}). The outflow may have been generated by a wide angle AGN driven wind or by the $\sim$1 kpc   radio jet.  The  almost coincident  direction of the central ($\sim$1 kpc) ionised outflow  and the radio axis suggests that the radio jet has triggered the nuclear outflow at least (\citealt{Ramos2017,Jarvis2019}).  

  Based on long slit GTC spectroscopy at PA 60$\degr$ and 90$\degr$, \cite{Villar2018} reported the discovery of a $\sim$100 kpc ionised nebula associated with this object. 
  We proposed that it is part of the CGM of the QSO2 host, which has been populated with tidal debris by galactic interactions and rendered visible thanks to the illumination by the AGN.   The optical/radio bubbles appear to be expanding across this medium.

We included the Teacup in our GTC (but not VLA) sample with the aim of mapping the morphology of the giant nebula. The optimum scaling factor for the continuum image that produces the best subtraction for most sources in the field is not adequate for the Teacup host. Strong negative residuals remain at the location of  large scale diffuse continuum features associated with the galaxy. Thus, the scaling factor underestimates the continuum level. Since  no prominent emission lines contaminate the continuum filter,  the spectrum in those areas  must be quite steep. This is supported by the SDSS spectrum, which reveals the bluest spectrum among all QSO2 in our sample. We thus scale the continuum and TF images to produce the best subtraction of all the Teacup continuum features.  The results are shown in Fig. \ref{figTeacup}.  

Faint H$\alpha$  emission is detected across a huge area of total extension 115 kpc $\times$ 87 kpc (major and minor axis respectively). The surface brightness is (0.5-few)$\times$10$^{-17}$ erg s$^{-1}$ cm$^{-2}$ arcsec$^{-2}$.
It is bordered towards the NE by an arc  with vertex at $\sim$56 kpc  from the QSO2 (see third panel in the figure). The NE arc and the nebula  have  a similar main axis as the bubble and the radio jet  (\citealt{Harrison2015}).  This suggests  that the nebular morphology is  influenced on very large scales by the nuclear activity. The total luminosity  of te nebula is $L_{\rm H\alpha}$=(4.1$\pm$0.3)$\times$10$^{41}$ erg s$^{-1}$. This does not include  the contribution of the $\sim$10 kpc ionised bubbles, which have  (4.0$\pm$0.2) and (2.8$\pm$0.5)$\times$10$^{41}$ erg s$^{-1}$ for the NE and SW bubbles respectively.

 Our long slit spectroscopy showed    line emission all the way from the QSO2 up to the NE arc and not beyond.  The projected area within the arc therefore appears to be filled with gas (\citealt{Villar2018}), but the  $SB$ is significantly lower   (Fig. \ref{figTeacup}). All together, it appears that the NE arc   encompasses   an edge brightened cavity.  

Several additional  diffuse H$\alpha$ patches   are detected  (Fig. \ref{figTeacup}, black circles in the bottom left panel). The most distant is at $r_{\rm max}\sim$90 kpc from the QSO2 nucleus. They have no or very faint continuum counterparts.  They have $L_{\rm H\alpha}\sim$(1-3)$\times$10$^{39}$ erg s$^{-1}$.
Towards the SW of the QSO2 an elongated, very low surface brightness $\sim$30 kpc long feature (``filament'' in the figure) is hinted.

In addition to AGN related processes,  additional ionisation mechanism(s)   must be at work, since line emission is detected all over the place, up to 10s of kpc from the QSO2. There is gas all around the AGN and thus, a large fraction is  outside the QSO2 ionisation cones.

\subsection{Spatially extended small  radio sources ($\sim$few kpc from the AGN).}
\label{appendix-small}

$\bullet$ J0948+25 ($z$=0.179)

\begin{figure}
\centering
\includegraphics[width=0.38\textwidth]{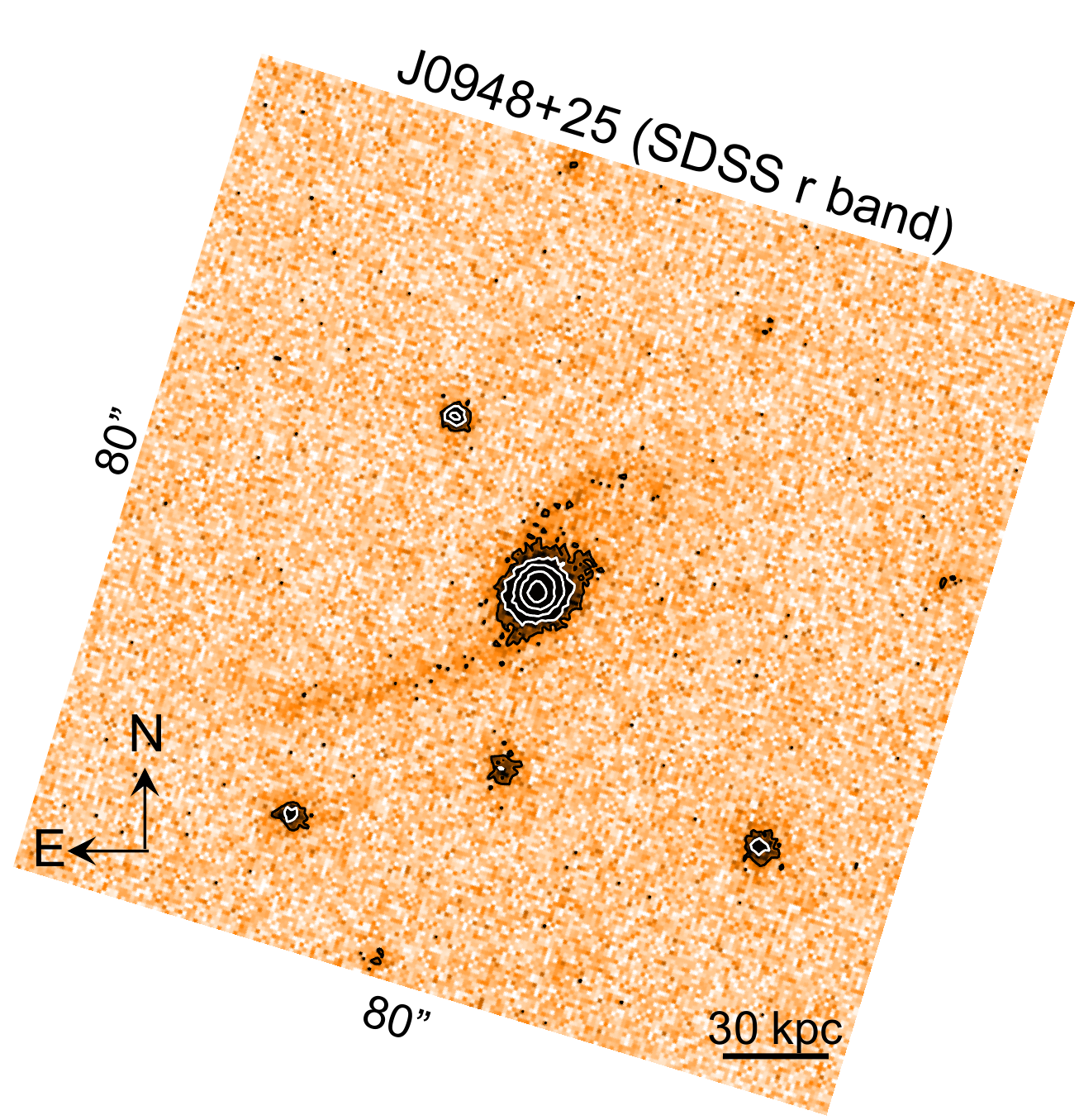}
\includegraphics[width=0.39\textwidth]{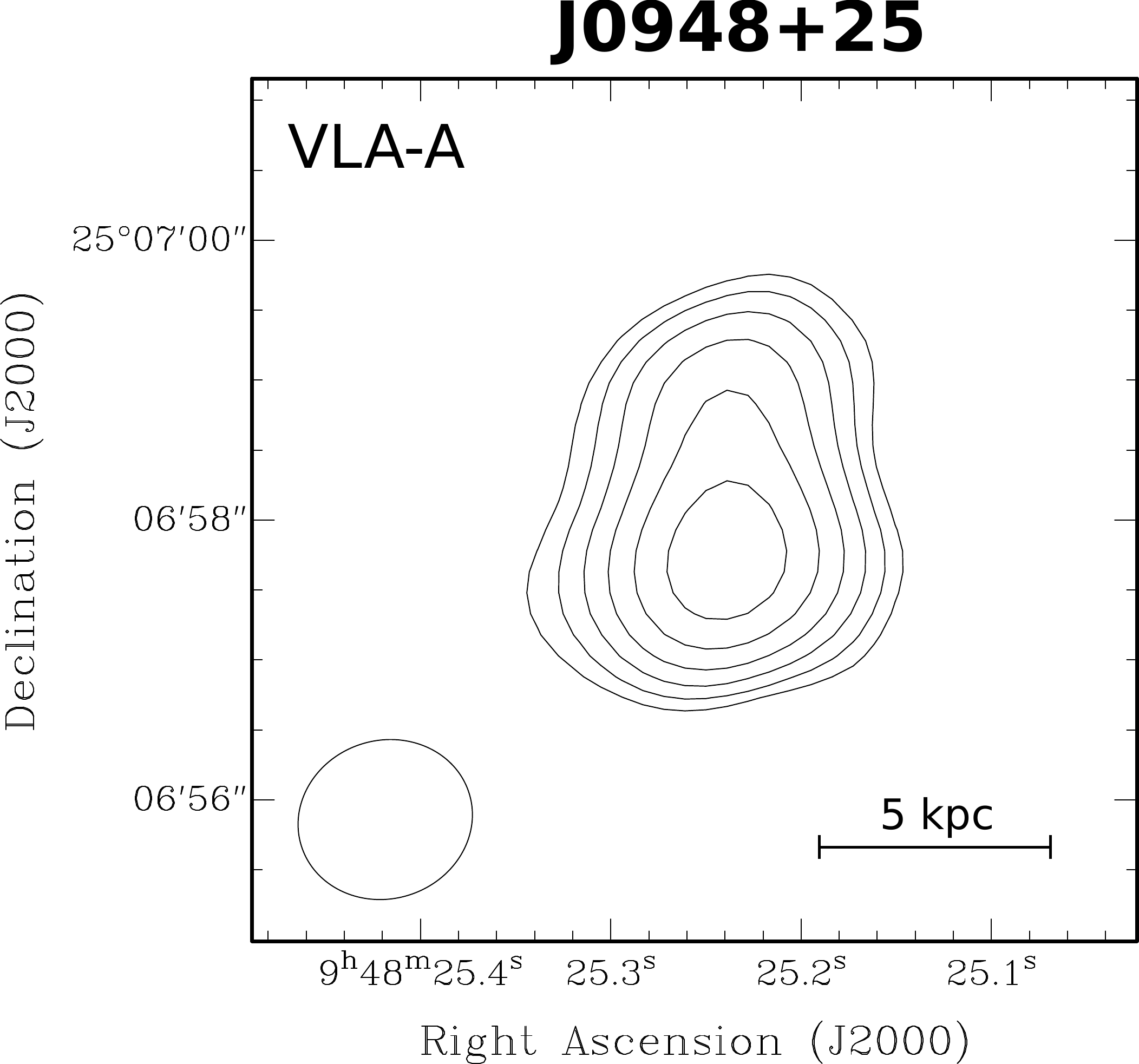}
\includegraphics[width=0.39\textwidth]{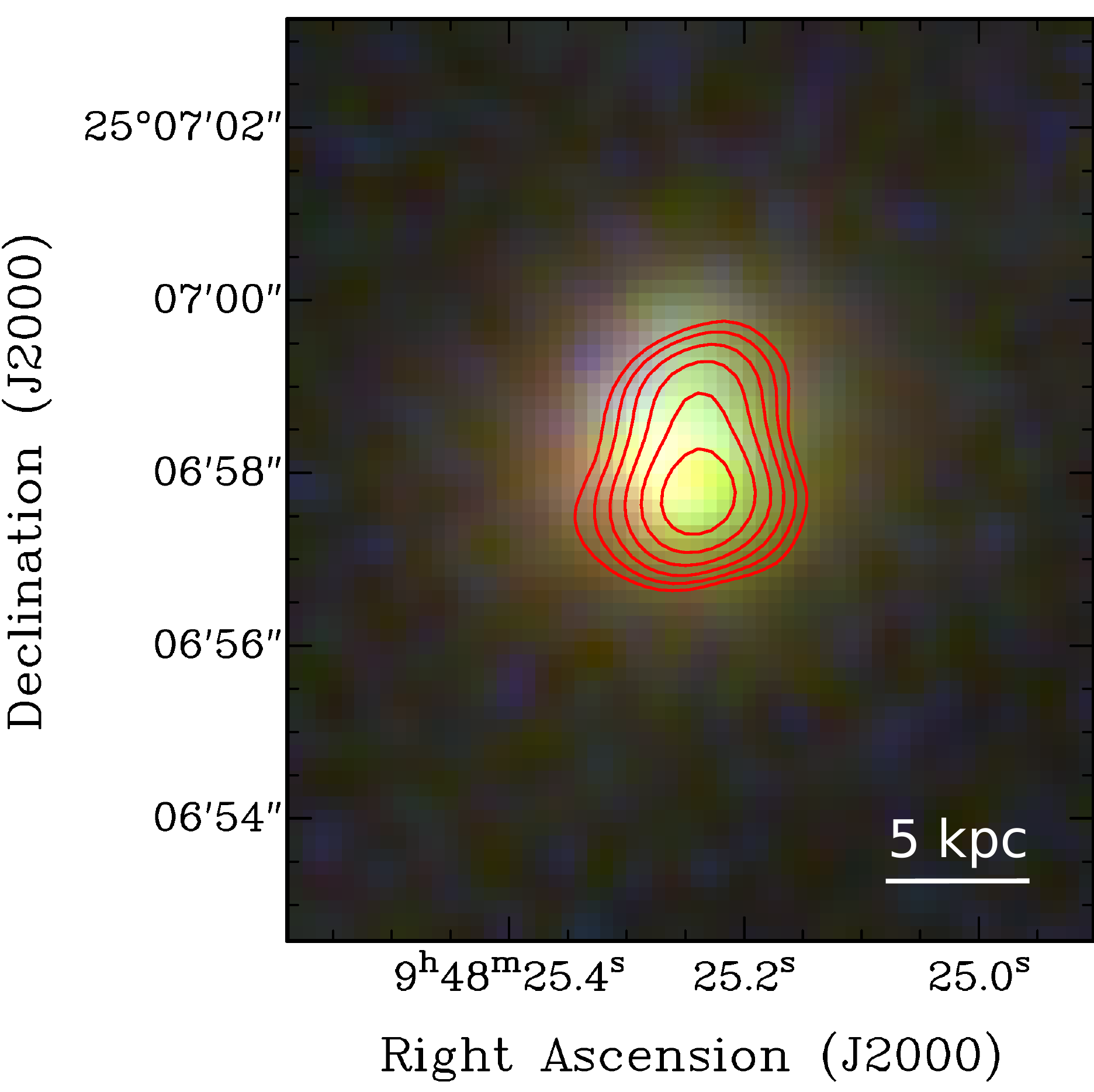}
\caption{J0948+25 ($z$=0.179). Top:  SDSS r band image.  Contour levels  start at 3$\sigma$ and increase with factor $\times$2. Middle: A-configuration map. Contour levels start at 150 $\mu$Jy\,beam$^{-1}$ and increase with factor $\sqrt{2}$. The radio source is confined within the galaxy size. This source is unresolved in B-configuration.  Bottom: The SDSS 3-colour image (u,g,r)  of the  host galaxy  is shown with overlaid contours of the VLA A-configuration data. This colour map shows more clearly the similar elongation angle and size of the host and the radio source.}
\label{fig0948}
\end{figure}

Only the SDSS images are available for this RQ QSO2 (Fig. \ref{xu}). Two long tidal tails are  clearly identified (Fig. \ref{fig0948}, top). The longest extends towards the SW for  $\sim$29$\arcsec$ or 87 kpc.

 The radio source is resolved only in  the  A-configuration VLA map, which shows a small $\sim$1.0-1.5$\arcsec$  extension towards the North from the core (Fig. \ref{fig0948}).  $L_{\rm 60 \mu m}$ and $q$ are not available   (Sect. \ref{sample}) so we cannot discern whether the radio emission has an AGN contribution (e.g. small jet?) or is instead dominated by star formation.  The  central region of the QSO2 host appears to show a similar elongation (Fig. \ref{fig0948}, bottom). 

\begin{figure}
\centering
\includegraphics[width=0.36\textwidth]{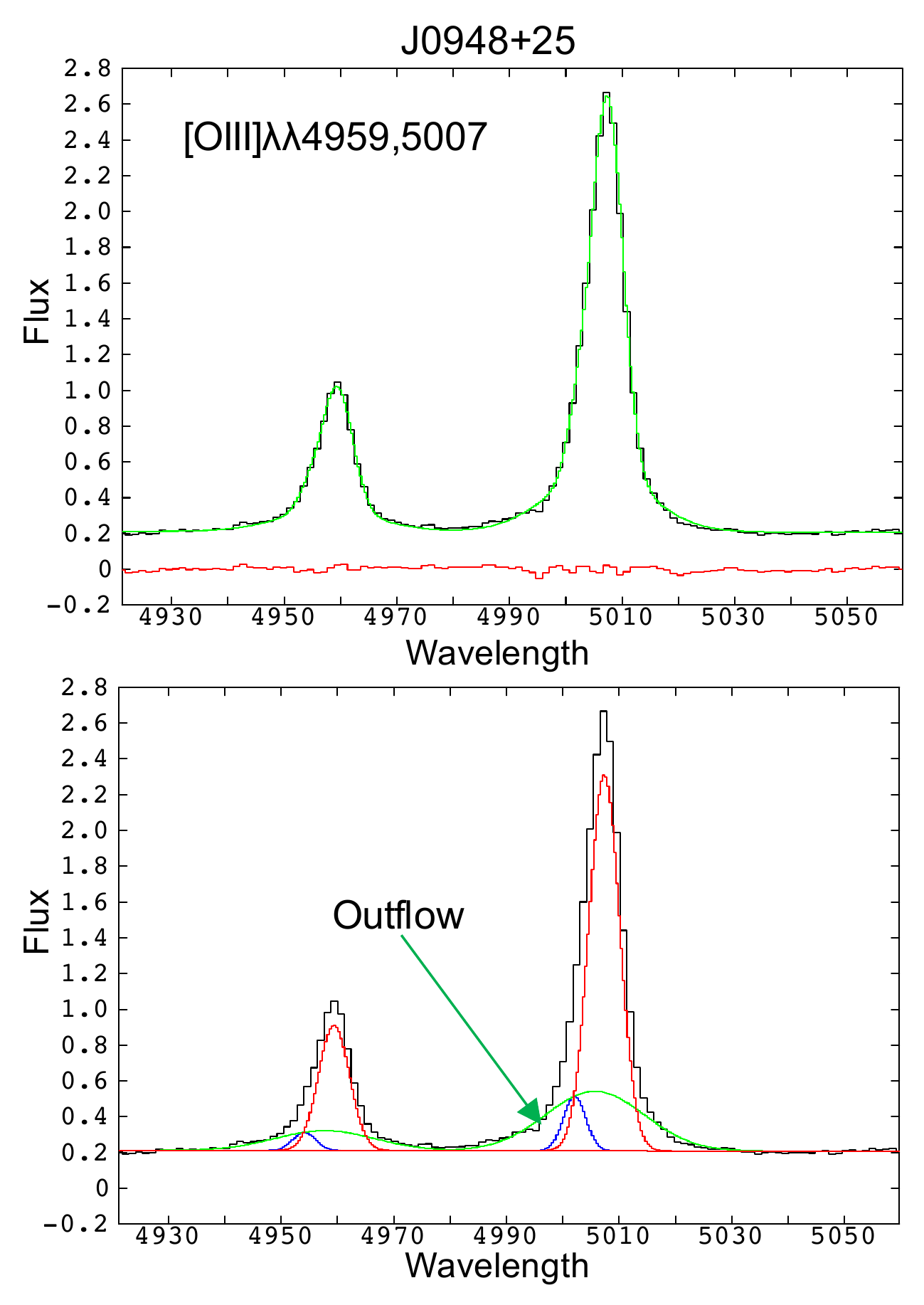}
\caption{[OIII] doublet SDSS  spectrum of J0948+25. Panels and line colour code as in Fig. \ref{outflow0853}.  The broadest of the three kinematic  components isolated in the fit traces an ionised outflow (blue, bottom panel). Wavelength in \AA. Fluxes in units of 10$^{-15}$ erg s$^{-1}$ cm$^{-2}$ \AA$^{-1}$.}
\label{outflow0948}
\end{figure}

An  ionised outflow is identified in the SDSS QSO2 spectrum. It has FWHM=1133$\pm$57 km s$^{-1}$ and  is blueshifted by -77$\pm$16 km s$^{-1}$ relative to the narrow line core (Fig. \ref{outflow0948}).  It contributes 32\% of the total [OIII] flux.

$\bullet$ J1108+06 ($z$=0.182)

   This RQ QSO2 (Fig. \ref{xu}) is a merging system which  likely hosts two AGN separated by 0.7$\arcsec$ (\citealt{Liu2010}).
\cite{Bondi2016} studied the radio and optical morphologies  in detail based on VLA  A-Array L (1.4 GHz), C (5.0 GHz), and X  (8.5 GHz) bands and with the European  VLBI  Network  (EVN)  at  5 GHz. The radio source consists of multiple components located within $\sim$3$\arcsec$ (9 kpc). One of them is probably an AGN radio core. The extended radio emission is  co-spatial with the U-band UV continuum emission  seen in the HST image. They propose that intense star formation is the origin of both the radio and UV continua. The large $q=$1.97$\pm$0.07 is consistent with SF (Table \ref{table-sample}) .

\begin{figure}
\centering
\includegraphics[width=0.40\textwidth]{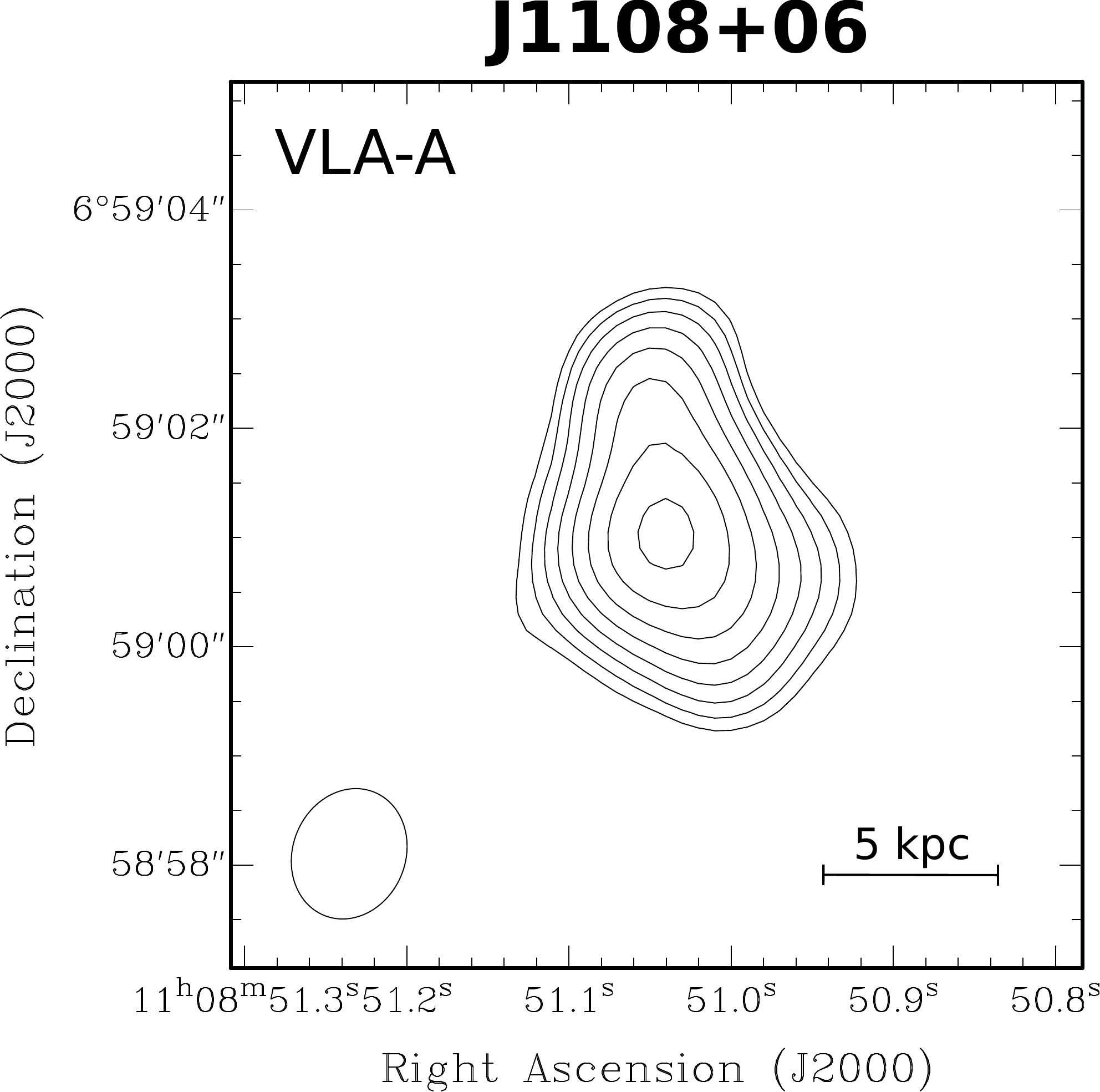}
\caption{A-configuration map of J1108+06. Contour levels start at 300 $\mu$Jy\,beam$^{-1}$ and increase with factor $\sqrt{2}$. This source is unresolved in B-configuration.}
\label{VLA1108}
\centering
\includegraphics[width=0.36\textwidth]{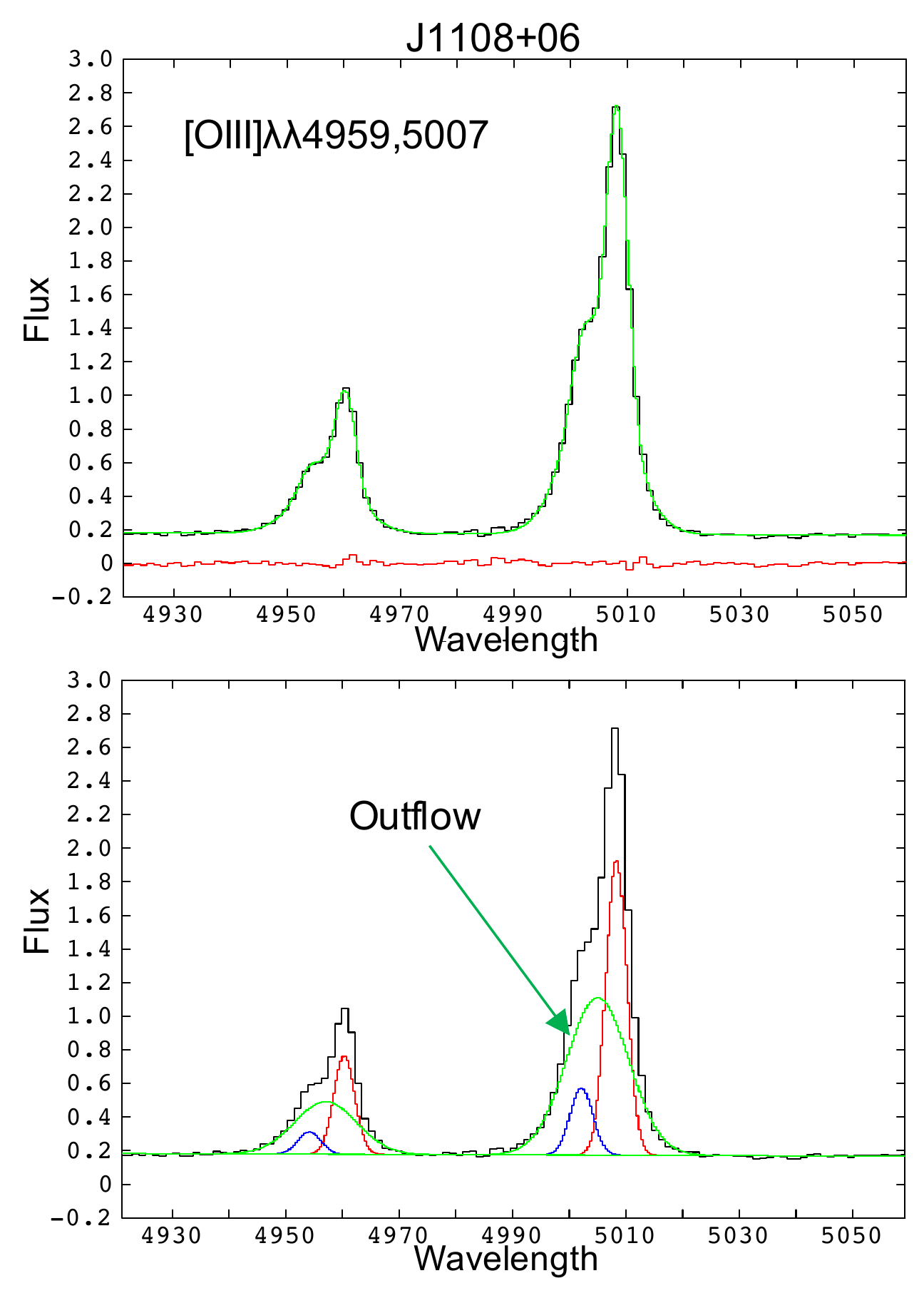}
\caption{[OIII] doublet SDSS  spectrum of J1108+06. The data (black), best fit (green) and residuals (red) are shown in the top panel. The three kinematic components isolated in the fit are shown in the bottom panel with different colours. The broadest component traces a prominent ionised outflow. Wavelength in \AA. Fluxes in units of 10$^{-15}$ erg s$^{-1}$ cm$^{-2}$ \AA$^{-1}$.}
\label{outflow1108}
\end{figure}

We observed this source both with the A (28 minutes) and B   (42 minutes) VLA configurations, for somewhat longer times than  \cite{Bondi2016} (18, 27, and 21 minutes in A-configuration for bands L, C and X respectively). While the source appears unresolved in the B-array map, the A-configuration image shows a $\sim$2$\arcsec$ extension towards the North,  already discussed by those authors.  No additional radio emission  is detected (Fig. \ref{VLA1108}).  No GTC images are available for this object and a radio/optical overlay is not shown.

 We show in Fig. \ref{outflow1108} the SDSS spectrum of J1108+06 in the [OIII]$\lambda\lambda$4949,5007 region and the best fit of the spectral profiles. The  lines consist of two Gaussian narrow components (FWHM=190$\pm$13 and 242$\pm$21 km s$^{-1}$ shifted by 356$\pm$8 km s$^{-1}$). They are likely emitted by the two AGN, since the spatial separation $\sim$0.7$\arcsec$ is significantly smaller than the SDSS 3$\arcsec$ fibre. In addition, a prominent  outflow (49\% of the line flux) is detected (FWHM=813$\pm$25 km s$^{-1}$ and blueshifted by $V_s$=-200$\pm$8 km s$^{-1}$ relative to the dominant narrow component). 

$\bullet$ J1437+30 (B2 1435+30, $z$=0.092)

\begin{figure*}
\centering
\includegraphics[width=0.7\textwidth]{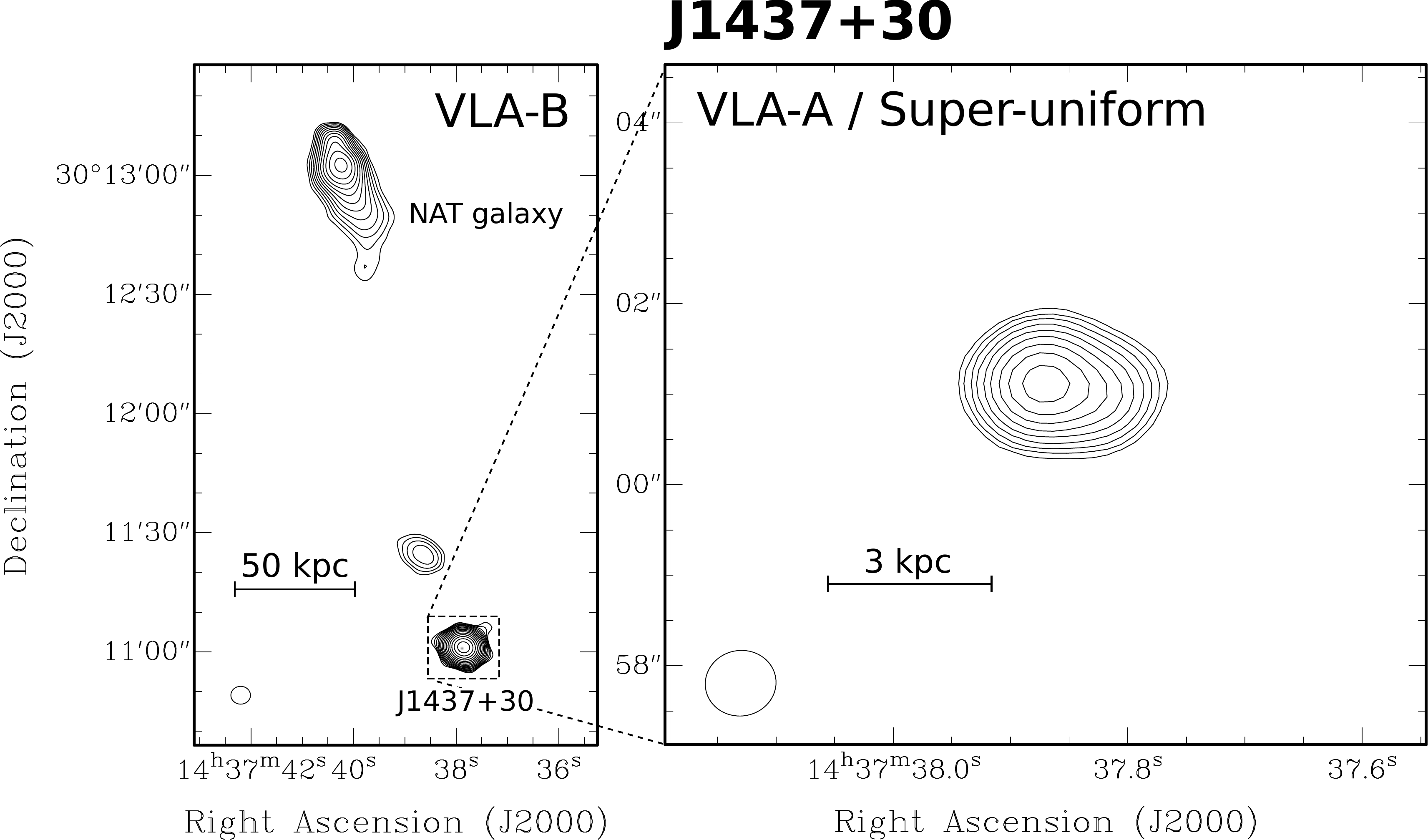}
\caption{VLA radio maps of J1437+30. Same as Fig.\,\ref{VLA0841}. Contour levels start at 350 $\mu$Jy\,beam$^{-1}$ (B-configuration) and 2 \,mJy\,beam$^{-1}$ (A-configuration) and increase with factor $\sqrt{2}$. As shown in Appendix \ref{appendix-tailrg}, source A    is a Narrow-Angle-Tail (NAT), while source B is probably Wide-Angle-Tail galaxy (WAT).  The A-configuration data were imaged using Super-uniform weighting to reveal the marginally resolved radio core in J1437+30.}
\label{VLA1437}
\centering
\includegraphics[width=0.5\textwidth]{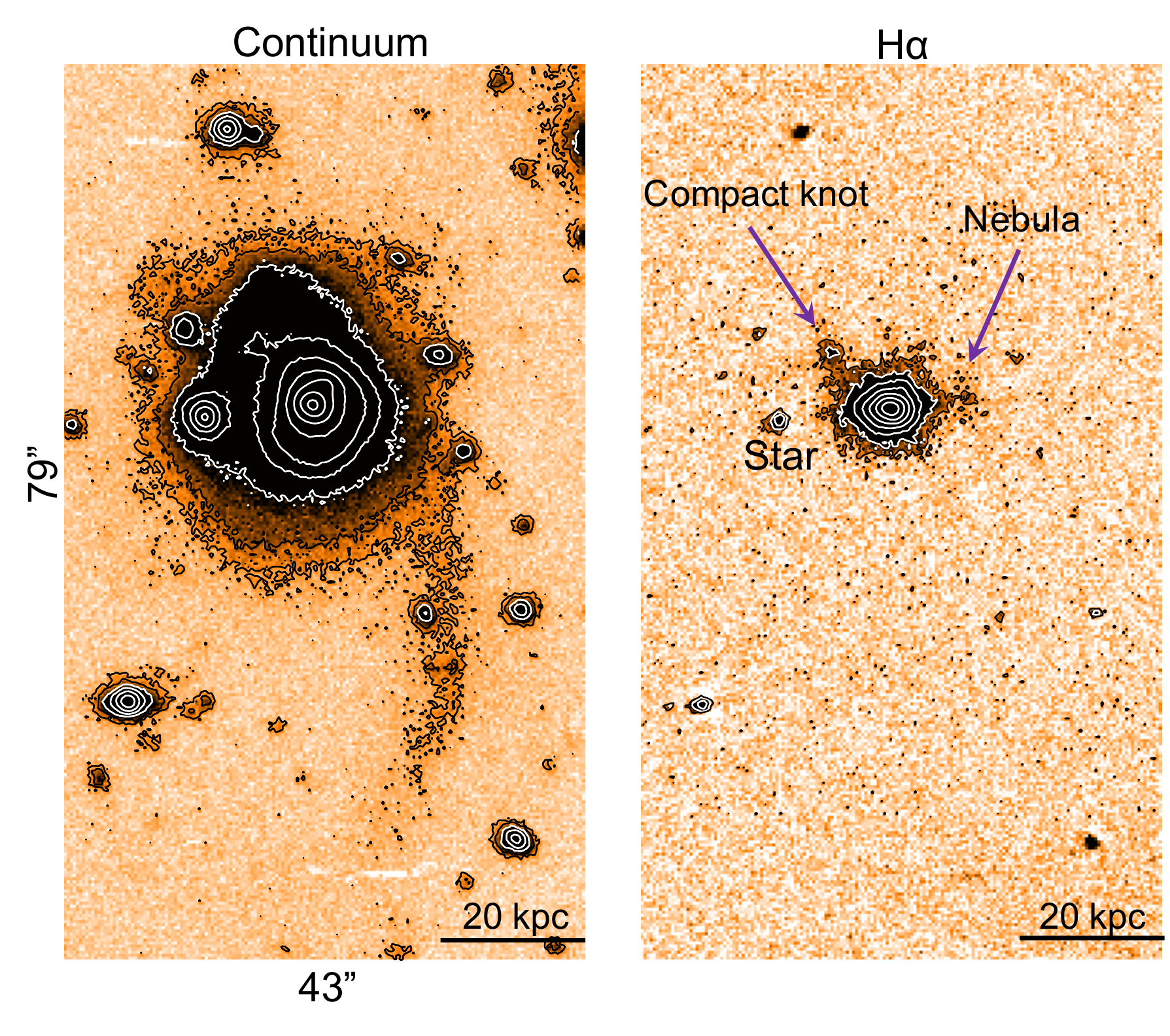}
\caption{J1437+30.  GTC continuum and H$\alpha$ images.  Contour levels in each image start at 3$\sigma$ and increase with factor $\times$3 and 2 For the H$\alpha$  and continuum images respectively. $\sigma$=3.0$\times$10$^{-19}$ erg s$^{-1}$ cm$^{-2}$ pixel$^{-1}$ for the H$\alpha$ image.}
\label{fig1437}
\end{figure*}

This  RQ/RI QSO2 (Fig. \ref{xu}), shows a clear excess or radio emission ($q$=0.43$\pm$0.02,  Table \ref{sample}) and thus has a strong AGN contribution. It is is marginally resolved in the East-West direction at the 1.3$\arcsec$ resolution of our super-uniformly weighted A-configuration data (Fig.\ref{VLA1437})

The continuum and H$\alpha$ GTC images are shown in Fig. \ref{fig1437}.  The optical morphology is strongly distorted due to galactic merger/interaction processes. The QSO2 is associated with a large (30$\arcsec$ or $\sim$51 kpc)  and amorphous  continuum halo,   a  tidal tail that stretches for  more than 60 kpc towards the S and other irregular features. 

At the HST/ACS spatial resolution (HST program 13728; PI: S.B. Kraemer), the NLR  consists of three spatial components    located roughly along the E-W direction (Fig. \ref{overlay1437}; HST program \citealt{Fischer2018}). Using the same terminology as the authors, these  are a compact, bright unresolved core, a cone  to the West of $\sim$2.5 kpc height and a compact arc at $\sim$2.0 kpc to the East.   The overlay with our VLA map is shown in Fig. \ref{overlay1437}. The radio core is located  at the base of the cone. It coincides with the continuum centroid (\citealt{Fischer2018}) and  lies at the expected location of the AGN. The marginal extension of the radio source overlaps with the [OIII] cone, which is the highest  $SB$ emission line structure. The authors measured disturbed kinematics in the cone. All this suggest that the radio source is interacting with this  gas. A higher resolution radio map would be necessary to investigate this, by characterising the radio morphology more accurately.

At the resolution of our GTC image (1.00$\pm$0.05$\arcsec$, Table \ref{log-optical}) and with the overwhelming glare of the  compact [OIII] central source,  the above NLR features are not resolved. The H$\alpha$ emission is dominated by a  compact, barely resolved source with FWHM=1.38$\pm$0.02$\arcsec$ (PA 169$\degr$) and 1.02$\pm$0.01 along the major and minor axis respectively. On the other hand,  lower $SB$  line emission is detected on a significantly larger area than in the HST image.

A  H$\alpha$ knot  is detected at $\sim$11 kpc NE of the QSO2 (Fig. \ref{fig1437}, right panel). The morphology suggests that it is a star forming  object.  Assuming that the [NII] contamination is negligible and ignoring reddening, the implied H$\alpha$ luminosity $L_{\rm H\alpha}$=1.7$\times$10$^{39}$ erg s$^{-1}$ would imply SFR=0.014 M$_{\odot}$ yr$^{-1}$.  Low $SB$ line emission is also detected  up to $\sim$12 kpc to the W of the AGN. It is rather asymmetric, well  in excess above the wings of the central, bright H$\alpha$ source (thus,  seeing smearing is not  responsible).  The QSO2 nucleus is saturated in the long exposures TF  and continuum images. To check whether the extended nebula could instead be due to asymmetric saturation spikes we did several tests using the  continuum and TF  images obtained with short exposures (90  instead of 900 seconds; see Sect. \ref{sec:GTC}).  The H$\alpha$ low $SB$ emission was securely or tentatively detected in all tests. The  $\sim$12 kpc nebula is, thus, real. It is aligned with the  NLR seen in the HST image  and with the radio source. Thus, it is probably ionised by the QSO2 continuum.

 J1437+30 hosts a nuclear ionised outflow (FWHM=1254$\pm$37 and $V_{\rm s}$=+107$\pm$16 km s$^{-1}$)  of radial size $\sim$170 pc (\citealt{Villar2014,Fischer2018}). It contributes $\sim$17\% of the [OIII] nuclear flux.

$\bullet$ J1517+33 ($z$=0.135)

  This RI  QSO2 is associated with the most powerful radio source of our sample and shows the highest radio excess (log($P_{\rm 1.4~GHz}$))=31.8, $q$=0.36$\pm$0.18, Table \ref{sample} and Fig. \ref{xu}).  It has been classified  as a radio galaxy in some works (e.g. \citealt{Rosario2010}). 

These authors obtained   VLA   A-configuration data in four bands: 1.4, 5, 8 and 22 GHz with exposure times at each frequency between five and ten minutes.  The radio source consists of an unresolved flat-spectrum  core   and a steep spectrum bipolar jet extending $\sim$2$\arcsec$ (4.7 kpc) at both sides of the core (see also \citealt{Tingay2011}).  The western jet shows a strong bend of about 75$\degr$ at $\sim$1$\arcsec$ from the core, which is traced by the emission-line gas.  They found a clear similarity in structure between the radio jet
and the ionised gas, which moreover shows highly disturbed kinematics. They propose that the jet is responsible for triggering an outflow  that is interacting with the galaxy ISM.

\begin{figure*}
\centering
\includegraphics[width=0.8\textwidth]{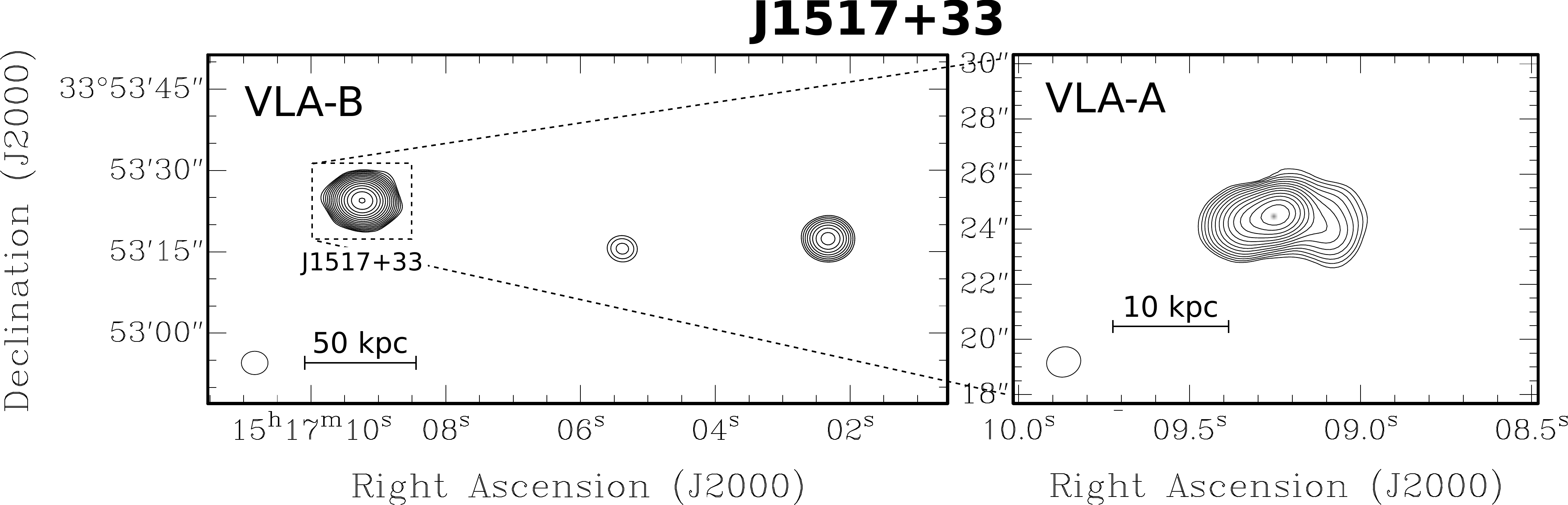}
\caption{VLA radio maps of J1517+33. Same as Fig.\,\ref{VLA0841}. Contour levels start at 800 $\mu$Jy\,beam$^{-1}$ and increase with factor $\sqrt{2}$ for both B- and A-configuration.}
\label{VLA1517}
\end{figure*}

We observed this source with the A  and B VLA configurations for 30 and 60 minutes on source respectively.  The source is unresolved in the B configuration map (Fig. \ref{VLA1517}). The two unresolved knots 0.8$\arcmin$ and 1.4$\arcmin$ to  the East are unrelated   (\citealt{Tingay2011}).
The A configuration map  shows an elongated bright source associated with the QSO2 which extends at both sides of the central radio core for  $\sim$9 kpc in total. The  bent morphology  at $\sim$1$\arcsec$ from the core  identified  at 5 GHz by \cite{Rosario2010}   is also appreciated.  We do not detect additional radio emission. Since no new information is obtained and no GTC images are available, we do not explore further.

\subsubsection{Marginally resolved and  unresolved radio sources}
\label{appendix-unres}

\begin{figure*}
\centering
\includegraphics[width=0.7\textwidth]{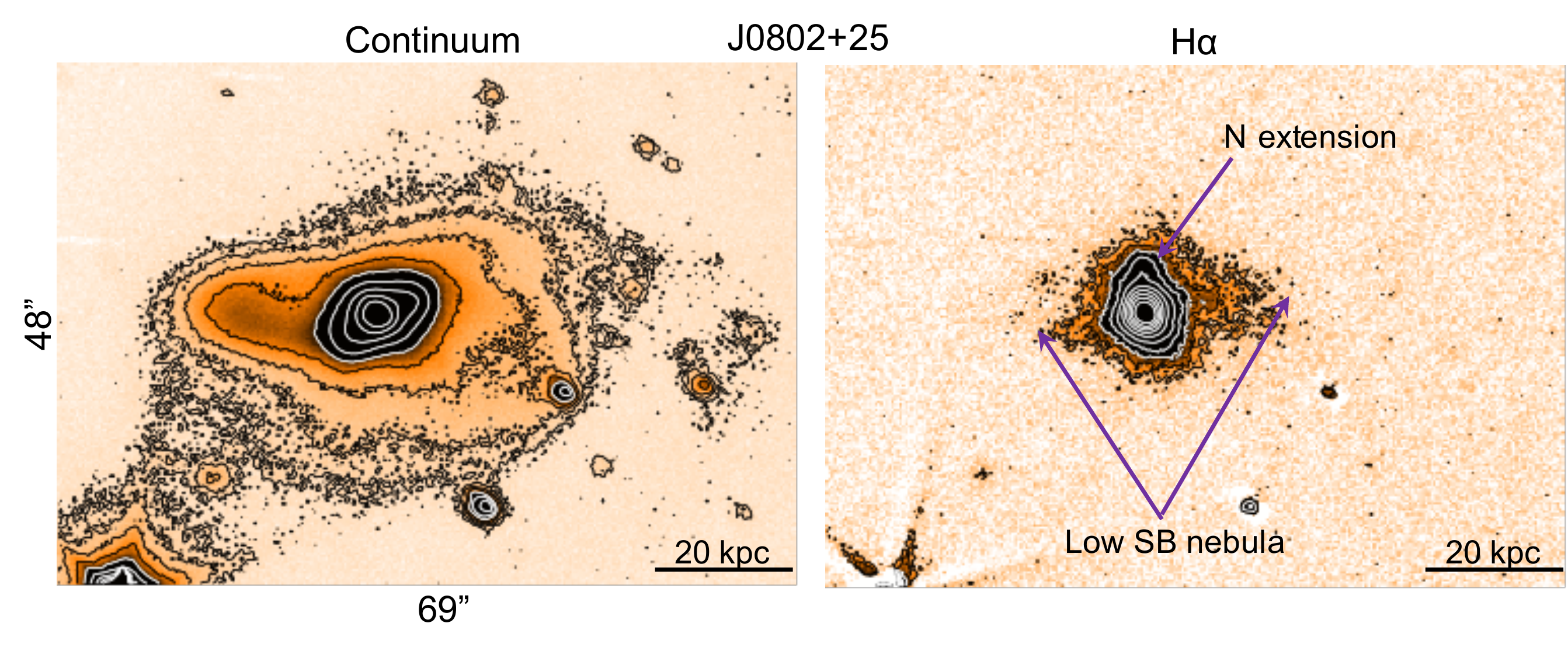}
\caption{J0802+25. GTC continuum and H$\alpha$ images. Contour levels in each image start at 3$\sigma$ and increase with factor $\times$2. For the H$\alpha$ image, $\sigma$=2.7$\times$10$^{-19}$ erg s$^{-1}$ cm$^{-2}$ pixel$^{-1}$.}
\label{fig0802}
\end{figure*}

$\bullet$ J0802+25 ($z$=0.080)  

This RQ QSO2  (Fig. \ref{xu}) has  a significant AGN contribution to the radio emission ($q=$1.39$\pm$0.06, Table \ref{sample}).    Our  VLA maps show that the  source is  unresolved in both the A and B configuration maps.  We do not show them for simplicity.

J0802+25 is a member of an interacting system with a complex optical morphology (Fig. \ref{fig0802}, see also \citealt{Fischer2018}).   A tidal tail and a huge halo are identified in the GTC continuum image extending for $\sim$55$\arcsec$ or $\sim$82 kpc. The major axis of this  halo runs in the E-W direction.   

Fischer et al. (2018) studied the [OIII] morphology and kinematics based on HST/ACS images. [OIII] extends   up to $R_ {\rm max}\sim$2.8 kpc from the AGN towards the North and up to $\sim$1.5 kpc towards the South.  They identified an ionised outflow with radial projected size $R_{\rm out}\sim$0.44 kpc, overlapping with the [OIII] extended emission. Based on its extreme kinematic properties, \cite{Villar2014} proposed that this outflow has been triggered by a so far undetected small scale radio jet (\citealt{Mullaney2013,Molyneux2019}).

Our H$\alpha$ image  (Fig. \ref{fig0802}, right panel)  shows more extended line emission than the [OIII] HST image. H$\alpha$ is detected up to $\sim$6 kpc from the QSO2 nucleus to the North. In addition,  low $SB$ ionised gas   is detected roughly in the E-W direction. It has $d_{\rm max}\sim$38 kpc and  $r_{\rm max}\sim$21 kpc  (Fig. \ref{fig0802}, right panel).     Multiple tests were done to investigate potential artefacts due to saturation spikes using also the short exposure images (Sect. \ref{log-optical}) and/or inaccurate continuum subtraction as well as the potential impact of seeing smearing of the bright central source. We confirm  the detection of the large nebula.  The northern H$\alpha$  extension (also identified by \citealt{Fischer2018}) is likely to be inside the AGN ionisation cone and roughly aligned with it. This is suggested by the higher $SB$ of this gas and the fact that the nuclear ionised outflow is extended in this direction. If such scenario is correct, the large $SB$ nebula lies at least partially outside  the reach of the AGN ionising continuum. Excitation mechanisms unrelated to the nuclear activity (stellar photoionisation or other) must be at work.

\begin{figure*}
\centering
\includegraphics[width=0.7\textwidth]{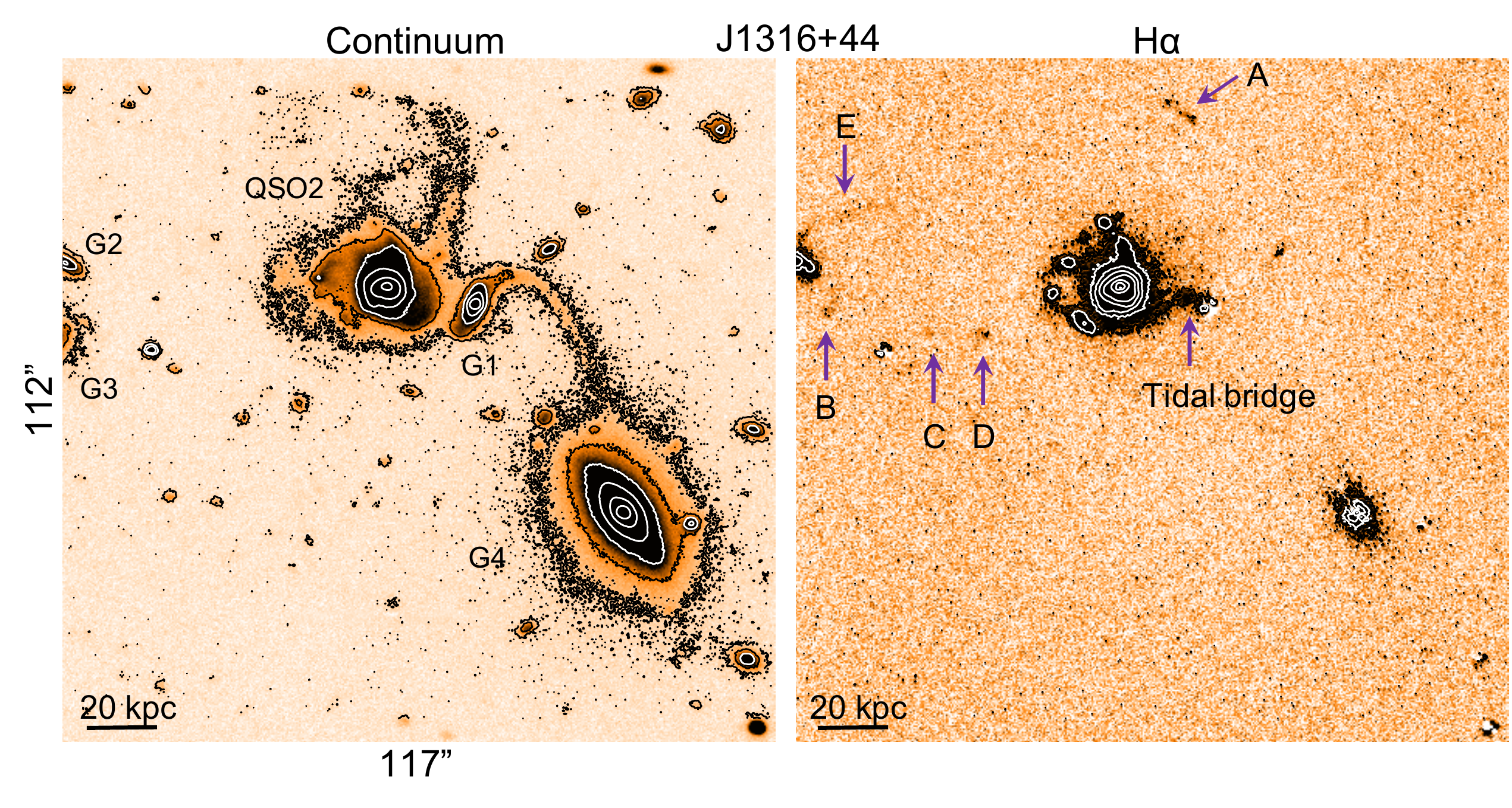}
\caption{J1316+44. GTC continuum and H$\alpha$ images.  G1, G2 and G3 belong to the same group. G4 is at $z$=0.060 and is unrelated. A to E  in the middle panel highlight H$\alpha$ features (see text).   Contour levels in each image start at 3$\sigma$ and increase with factor $\times$3. For the H$\alpha$ image, $\sigma$=2.6$\times$10$^{-19}$ erg s$^{-1}$ cm$^{-2}$ pixel$^{-1}$.}
\label{fig1316}
\end{figure*}

$\bullet$ J1316+44 ($z$=0.091)

The radio emission of this RQ QSO2 (Fig. \ref{xu})  is consistent with  being dominated by star formation  ($q=$2.35$\pm$0.04).  It is unresolved in the B configuration VLA image and appears marginally resolved in the A configuration map along PA $\sim$70$\degr$. The radio maps are not shown for simplicity.

\begin{figure}
\centering
\includegraphics[width=0.36\textwidth]{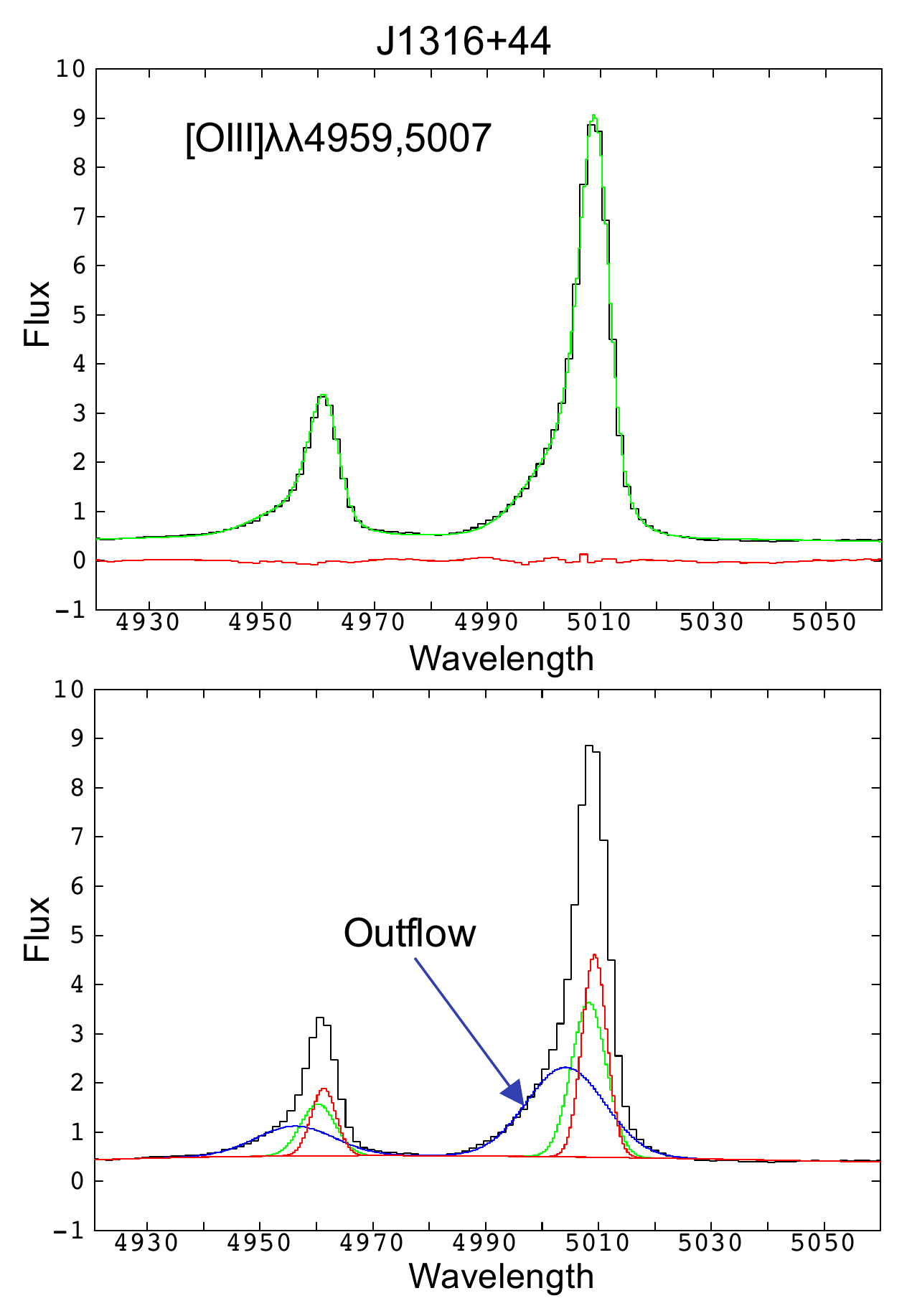}
\caption{[OIII] doublet SDSS  spectrum of J1316+44. Panels and line colour code as in Fig. \ref{outflow0853}.  The broadest of the three kinematic  components isolated in the fit traces an ionised outflow (green, bottom panel). Wavelength in \AA. Fluxes in units of 10$^{-15}$ erg s$^{-1}$ cm$^{-2}$ \AA$^{-1}$.}
\label{outflow1316}
\end{figure}

 Based on its high infrared luminosity, log($L_{\rm IR}/L_{\rm \odot}$)=11.9, J1316+44 is close to the ULIRG regime (Ultraluminous Infrared Galaxy,  log($L_{\rm IR}/L_{\rm \odot})>$12.0). This object  has been classified as a type 1.9 AGN, rather than type 2 (e.g. \citealt{Keel2012}).   Permitted lines such as H$\beta$ and H$\alpha$ show faint very broad wings in the SDSS spectrum originated in  the broad line region (BLR).  An ionised outflow is identified in the SDSS [OIII] doublet  with FWHM 
961$\pm$27 km s$^{-1}$ and blueshifted by -276$\pm$17 km s$^{-1}$ relative to the narrow line core (Fig. \ref{outflow1316}).  It contributes $\sim$40\% of the total line flux.

J1316+44 is a member of a  galaxy group  at $z\sim$0.09.  The GTC continuum image (Fig. \ref{fig1316}, left panel) shows tidal debris spread over a very large area, which extends up to   $\sim$57 kpc to the N of the QSO2 nucleus. The host is interacting with  G1   ($z$=0.091, SDSS  database). The nucleus of this galaxy is not detected in the H$\alpha$ image, which is not surprising, since the SDSS spectrum shows  no emission lines.   It is connected to the QSO2 host by a tidal bridge more clearly  seen in the H$\alpha$ image. The large G4 galaxy to the South is  at significantly lower $z$=0.060 (SDSS spectroscopic $z$).

G2 and G3 at $\sim$90 kpc to the E (right on the edge of the CCD), are  an interacting pair at the same $z$ as the QSO2. The H$\alpha$ image shows prominent emission from G2 and tidal features that point towards the QSO2 (see the tidal tail indicated with  an E'' in Fig. \ref{fig1316}, right). The pair seems to be interacting  also with the QSO2 host.

Multiple regions of star formation in the disk of the QSO2 host are seen in the H$\alpha$ image. Emission line patches are also detected  at different locations far from any galaxy   (purple arrows in Fig. \ref{fig1316}, right). As an example, feature A   is at $\sim$53 kpc N of the QSO2 nucleus.  Features B, C and D may be part of a continuous gaseous $\sim$80 kpc tidal bridge  that connects the QSO2 with the G2-G3 pair. The ionisation of these features could be due to local star formation induced by the galactic interaction across tens  of kpc (\citealt{Weilbacher2003}).  In this scenario, their low line luminosities  $L_{\rm H\alpha}\sim$several$\times$10$^{38}$ erg s$^{-1}$  would imply SFR$\sim$several$\times$10$^{-3}$, consistent with SFR in tidal features of interacting galaxies (e.g. \citealt{Duc2013}).

\section{Serendipitous discovery of two tailed radio galaxies at $z=$0.33.}
\label{appendix-tailrg}

\begin{figure*}
\centering
\includegraphics[width=0.8\textwidth]{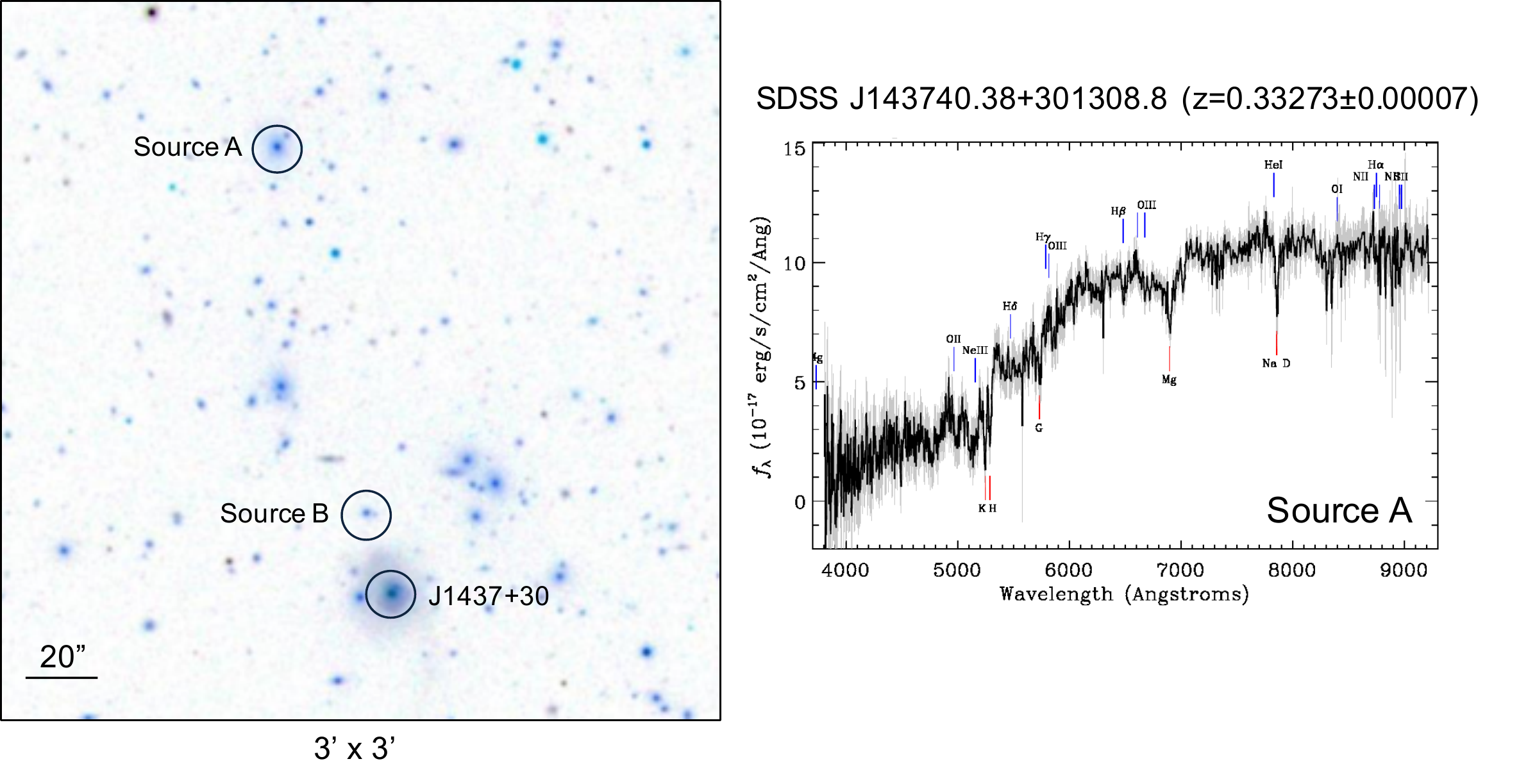}
\caption{The location of the tailed radio sources A and B in the field of J1437+40 are indicated in the SDSS colour  image. The SDSS spectrum of Source A is also shown.}
\label{location}
\end{figure*}

There are two extended radio sources  at $\sim$30$\arcsec$  and 2$\arcmin$ towards  the NE of J1437+30  (Fig. \ref{VLA1437}). Their location in the SDSS colour map is shown in  Fig. \ref{location}.

\begin{figure}
\centering
\includegraphics[width=0.35\textwidth]{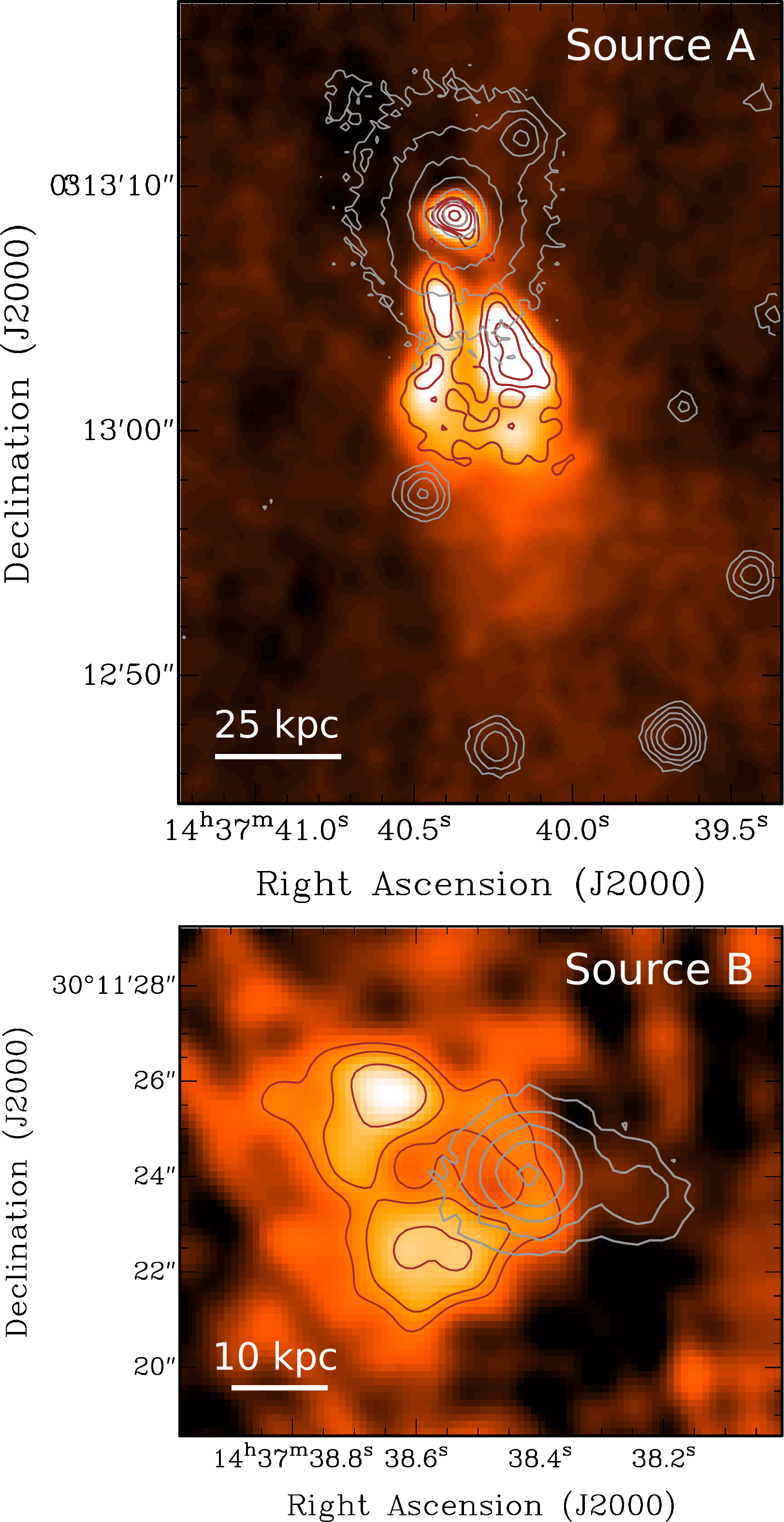}
\caption{Tailed radio galaxies at $z\sim$0.33 in the field of J1437+30. Overlay between the VLA radio maps (coloured image and red contours) and the optical (grey) contours of the SDSS images.}
\label{figtailed}
\end{figure}

Source A, which   is associated with  SDSS J143740.38+301308.8 at $z$=0.333 (Fig. \ref{location}, right), is a tailed FRI radio galaxy (Fig. \ref{figtailed}, top). The radio emission in this type of sources consists in general of a head coincident with the optical galaxy and extensions in one direction in the form of trails or tails that form  an angle with the galaxy apex (\citealt{Ryle1968}).   Depending on the angle, they are classified as wide (WAT) or narrow angle tailed (NAT) radio galaxies.  Source A belongs to the second group. The SDSS optical spectrum  (Fig. \ref{location}, right) shows no emission lines. This is not surprising since Fanaroff–Riley type I (FRI, \citealt{Fanaroff1974})  radio sources often show very weak emission lines (e.g. \citealt{Baum1995,Wills2004}) and the  large size of the SDSS fibre (3$\arcsec$) can contribute to the dilution of any faint nuclear emission features.

The radio source B  is also extended Fig. (\ref{figtailed}, bottom) and overlaps with  the galaxy SDSS J143738.41+301124.2. The SDSS photometric $z_{ph}$=0.366$\pm$0.030 suggests that it belongs to the same system as source A. 
One possible scenario is  that  B is a radio lobe associated with J1437+30. However, based on the multi-component morphology   and  the proximity of SDSS J143738.41+301124.2, we consider more likely that  it not related to the QSO2. The radio morphology is reminiscent of a WAT radio galaxy. The two ``trail'' components are significantly brighter than the poorly defined ``head''. This overlaps precisely with the optical galaxy. Such morphology   is  somewhat peculiar  (e.g. \citealt{Mao2010}), but it is not without precedents (\citealt{Obrien2018}).

It has been suggested that tailed radio galaxies  are members of clusters of galaxies and, as such, they have been proposed to be signposts of  clusters at different $z$ (e.g. \citealt{Miley1972,Giacintucci2009,Mao2010}; but see also \citealt{Obrien2018}). Their peculiar morphologies are thought to be a consequence of the deceleration of  the diffuse radio-emitting plasma  by the intracluster medium as the radio source moves through the gas in the cluster. Sources A and B in the field of J1437+30 are separated by $\sim$510 kpc in projection. Examples of two tailed  radio galaxies in the same cluster have been reported in the literature (e.g. \citealt{Giacintucci2009}). 

Whether sources A and B  belong to a galaxy cluster is beyond the scope of this study, but an overdensity of objects at this $z$ is suggested by the fact that at least three more  galaxies  in the field (C, D and E; Fig. \ref{figtailed}) have similar $z\sim$0.33.

\end{appendix}

\end{document}